\documentclass[a4paper,clock]{article}
\usepackage{graphicx}
\usepackage[dvips]{epsfig}
\usepackage{ amssymb}
\usepackage{bm}
\usepackage{dcolumn}
\catcode`\@=11
\def\marginnote#1{}
\newcount\hour
\newcount\minute
\newtoks\amorpm
\hour=\time\divide\hour by60
\minute=\time{\multiply\hour by60 \global\advance\minute
by-\hour}\edef\standardtime{{\ifnum\hour<12
\global\amorpm={am}%
        \else\global\amorpm={pm}\advance\hour by-12 \fi
        \ifnum\hour=0 \hour=12 \fi
        \number\hour:\ifnum\minute<10
0\fi\number\minute\the\amorpm}}
\edef\militarytime{\number\hour:\ifnum\minute<10
0\fi\number\minute}

\def\draftlabel#1{{\@bsphack\if@filesw {\let\thepage\relax
   \xdef\@gtempa{\write\@auxout{\string
      \newlabel{#1}{{\@currentlabel}{\thepage}}}}}\@gtempa
   \if@nobreak \ifvmode\nobreak\fi\fi\fi\@esphack}
        \gdef\@eqnlabel{#1}}
\def\@eqnlabel{}
\def\@vacuum{}
\def\draftmarginnote#1{\marginpar{\raggedright\scriptsize\tt#1}}
\def\draft{\oddsidemargin -.5truein
        \def\@oddfoot{\sl preliminary draft \hfil
        \rm\thepage\hfil\sl\today\quad\militarytime}
        \let\@evenfoot\@oddfoot \overfullrule 3pt
        \let\label=\draftlabel
        \let\marginnote=\draftmarginnote

\def\@eqnnum{(\theequation)\rlap{\kern\marginparsep\tt\@eqnlabel}%
\global\let\@eqnlabel\@vacuum}  }

\def\numberbysection{\@addtoreset{equation}{section}
        \def\theequation{\thesection.\arabic{equation}}}

\def\underline#1{\relax\ifmmode\@@underline#1\else
 $\@@underline{\hbox{#1}}$\relax\fi}

\catcode`@=12
\relax

\numberbysection
\advance \topmargin by -\headheight
\advance \topmargin by -\headsep

\evensidemargin \oddsidemargin
\def\rf#1{(\ref{#1})}
\def\lab#1{\label{#1}}

\def\br{\begin{eqnarray}}
\def\er{\end{eqnarray}}
\def\be{\begin{equation}}
\def\ee{\end{equation}}

\def\({\left(}
\def\){\right)}

\newcommand{\ct}[1]{\cite{#1}}
\newcommand{\bi}[1]{\bibitem{#1}}

\relax

\def\a{\alpha}

\def\b{\beta}

\def\D{\Delta}

\def\g{\gamma}
\def\G{\Gamma}

\def\l{\lambda}
\def\L{\Lambda}

\def\o{\over}

\def\pa{\partial}

\def\ra{\rightarrow}
\def\s{\sigma}
\def\S{\Sigma}

\def\tp0{\Theta_{+}^{(0)}}
\def\tm0{\Theta_{-}^{(0)}}

\def\cl{{\cal L}}

\def\f#1#2#3 {f^{#1#2}_{#3}}

\def\win1{{\sf w_{1+\infty}}}

\def\Win1{{\sf W_{1+\infty}}}

\def\rlx{\relax\leavevmode}
\def\inbar{\vrule height1.5ex width.4pt depth0pt}
\def\IZ{\rlx\hbox{\sf Z\kern-.4em Z}}
\def\IR{\rlx\hbox{\rm I\kern-.18em R}}
\def\IC{\rlx\hbox{\,$\inbar\kern-.3em{\rm C}$}}
\def\IN{\rlx\hbox{\rm I\kern-.18em N}}
\def\IO{\rlx\hbox{\,$\inbar\kern-.3em{\rm O}$}}
\def\IP{\rlx\hbox{\rm I\kern-.18em P}}
\def\IQ{\rlx\hbox{\,$\inbar\kern-.3em{\rm Q}$}}
\def\IF{\rlx\hbox{\rm I\kern-.18em F}}
\def\IG{\rlx\hbox{\,$\inbar\kern-.3em{\rm G}$}}
\def\IH{\rlx\hbox{\rm I\kern-.18em H}}
\def\II{\rlx\hbox{\rm I\kern-.18em I}}
\def\IK{\rlx\hbox{\rm I\kern-.18em K}}
\def\IL{\rlx\hbox{\rm I\kern-.18em L}}
\def\one{\hbox{{1}\kern-.25em\hbox{l}}}
\def\0#1{\relax\ifmmode\mathaccent"7017{#1}%
B        \else\accent23#1\relax\fi}

\def\PRL#1#2#3{{\sl Phys. Rev. Lett.} {\bf#1} (#2) #3}
\def\NPB#1#2#3{{\sl Nucl. Phys.} {\bf B#1} (#2) #3}

\def\PRD#1#2#3{{\sl Phys. Rev.} {\bf D#1} (#2) #3}

\def\PRA#1#2#3{{\sl Phys. Rev.} {\bf A#1} (#2) #3}
\def\PRB#1#2#3{{\sl Phys. Rev.} {\bf B#1} (#2) #3}

\def\PLA#1#2#3{{\sl Phys. Lett.} {\bf #1A} (#2) #3}

\def\PLB#1#2#3{{\sl Phys. Lett.} {\bf #1B} (#2) #3}
\def\JMP#1#2#3{{\sl J. Math. Phys.} {\bf #1} (#2) #3}

\def\AoP#1#2#3{{\sl Ann. of Phys.} {\bf #1} (#2) #3}

\def\RMP#1#2#3{{\sl Rev. Mod. Phys.} {\bf #1} (#2) #3}
\def\PR#1#2#3{{\sl Phys. Reports} {\bf #1} (#2) #3}

\def\IJMPA#1#2#3{{\sl Int. J. Mod. Phys.} {\bf A#1} (#2) #3}

\def\JPSJ#1#2#3{{\sl J. Phys. Soc. Japan} {\bf #1} (#2) #3}

\def\JHEP#1#2#3{{\sl JHEP} {\bf #1} (#2) #3}

\def\JPAMG#1#2#3{{\sl J. Physics A: Math. Gen.} {\bf A#1} (#2) #3}

\def\NJP#1#2#3{{\sl New J. Phys.} {\bf #1} (#2) #3}

\def\PAN#1#2#3{{\sl Physics of Atomic Nuclei} {\bf #1} (#2) #3}
\def\EPJC#1#2#3{{\sl Eur. Phys. J. C.} {\bf #1} (#2) #3}
\def\JPCM#1#2#3{{\sl J. Phys.: Condens. Matter} {\bf #1} (#2) #3}

\def\SYM#1#2#3{{\sl Symmetry} {\bf #1} (#2) #3}
\def\PhC#1#2#3{{\sl Physica C: Superconductivity and its applications} {\bf #1} (#2) #3}
\def\NRM#1#2#3{{\sl Nat Rev Mater} {\bf #1} (#2) #3}
\def\PZ1#1#2#3{{\sl Phys. Z} {\bf #1} (#2) #3}
\def\PhD#1#2#3{{\sl Physica D: Nonlinear Phenomena} {\bf #1} (#2) #3}
\begin{document}
%\DeclareGraphicsExtensions{.pdf}
\begin{titlepage}
\vspace*{-1cm}

\noindent
%March, 2012 \hfill{IF-P.03/2012}\\
% \hfill{hep-th/12xxx}

\vskip 2cm

\vspace{.2in}
\begin{center}
{\large\bf Majorana zero mode-soliton duality and in-gap and BIC bound states in modified Toda model coupled to fermion}
\end{center}

\vspace{0.1in}

\begin{center}

H. Blas$^{(a)}$, J.J. Monsalve$^{(a)}$ , R. Quica\~no$^{(b)}$ and J. R. V. Pereira$^{(a)}$

$^{(a)}$Instituto de F\'{\i}sica\\
Universidade Federal de Mato Grosso\\
Av. Fernando Correa, $N^{0}$ \, 2367\\
Bairro Boa Esperan\c ca, Cep 78060-900, Cuiab\'a - MT - Brazil \\
$^{(b)}$Instituto de Matem\'atica y Ciencias Afines-IMCA\\
Facultad de Ciencias\\
Universidad Nacional de Ingenier\'{\i}a\\
Av. Tupac Amaru, $N^{0}$ \, 210, Rimac, Lima-Per\'u \\

harold.achic@ufmt.br

\end{center}

\vspace{0.4in}

% \begin{abstract}
A two-dimensional field theory of a fermion chirally coupled to Toda field plus a scalar self-coupling potential is considered. Using techniques of integrable systems we obtain analytical zero modes, in-gap states and bound states in the continuum (BIC) for topological configurations of the  scalar field. Fermion-soliton duality mappings are uncovered for the bound state spectrum, which interpolates the weak and strong coupling sectors of the model and give rise to novel Thirring-like and multi-frequency sine-Gordon models, respectively. The non-perturbative effects of the back-reaction of the fermion bound states on the kink are studied and it is shown that the zero mode would catalyze the emergence of a new kink with lower topological charge and greater slope at the center, in the strong coupling limit of the model. For special topological charges and certain relative phases of the fermion components the kinks can host Majorana zero modes. The Noether, topological and a novel nonlocal charge densities satisfy a formula of the Atiyah-Patodi-Singer-type. Our results may find applications in several branches of non-linear physics, such as confinement in QCD$_2$,  braneworld models, high $T_c$ superconductivity and topological quantum computation. We back up our results with numerical simulations for continuous families of topological sectors. 
%\end{abstract}
\end{titlepage}

\section{Introduction}

The study of fermion-soliton systems have attracted significant interest  in the context of
the localization and bound states
of the fermion, since the interaction of the fermions and the topological solitons are related to interesting physical effects, such as the fermion number fractionization and polarization of the fermionic vacuum \cite{jackiw, goldstone, mackenzie}, string
superconductivity in cosmology \cite{witten0}, and monopole catalysis of the baryon decay \cite{rubakov, callan}. Moreover, in braneworld models, following the original work \cite{rubakov0}, some properties have been studied, such as  the localization of bulk chiral fermions on a brane \cite{randjbar}. Some brane Lagrangians consider the kink soliton
of the $\phi^4$ or sine-Gordon type systems in one dimension (see e.g. \cite{liu}). In this context the
particle transfer and fermion spectrum in braneworld collisions have also been considered \cite{saffin, gibbons}. In addition,
the kink field deforms the energy levels of the fermionic
vacuum leading to bound states and modified continuum states as compared to the free fermion spectrum. These change the zero-point fermion energy, giving rise to the Casimir effect in the presence of the soliton \cite{Shahkarami2}.

In most of the treatments  the back-reaction of the fermion on the kink has been neglected and the kink was taken as prescribed. Since the models of physical interest are in general non-integrable they rarely  possess analytical solutions. So, some recent studies resorted to numerical simulations in order to address the back-reaction of the fermion on the kink by computing self-consistent solutions of the system \cite{perapechka, mohamm}.
    
Many aspects of soliton-fermion systems in two dimensions, such as the quantum corrections, have been studied in
 the literature; however, some important features at the zero order are yet to be investigated. For instance, the effect of the  fermion-scalar interaction term on the topological number of the kink in the strong/weak limits of the model deserves a careful treatment through analytical and numerical methods. So, the dynamics of the fermion-kink coupling sector must be taken into account, since this term might contribute to deform the asymptotic values of the kink in the strong coupling limit. This requires the  study of the self-consistent solutions uncovering the transition from weak to strong coupling sectors of the model. 

In this paper we consider a  modification of the so-called 
$sl(2)$ affine Toda model coupled to matter field (ATM) by introducing an additional potential of self-interacting Toda field. The integrability and the soliton solutions associated to some particular vacua of the ATM model 
have been studied in \ct{matter, npb1}. One of its main properties is that for certain solutions
there is an equivalence between the $U(1)$ Noether current, involving only the Dirac field, and the topological current
associated to the pseudo-scalar soliton. This property was established in
\ct{matter} at the classical level and in \cite{npb1} at the quantum mechanical level.
It implies that the density of the $U(1)$ charge has
to be concentrated in the regions where the pseudo-scalar field has non-vanishing slope. The one-soliton of the
theory is a kink type soliton with fractional topological charge, and therefore the charge density is
concentrated inside the soliton. The  general $sl(n)$ ATM model constitutes an excellent laboratory
to test ideas about confinement \cite{npb1, npb2, prd1, jhep2}, the role of solitons
in quantum field theories \cite{npb1} and duality transformations interchanging
solitons and particles \cite{npb2, aop1, jmp1}.

The ATM model has also been considered in many physical situations related to  fermion-soliton systems. It appears in the study of the dynamical Peierls' energy gap generation in one-dimensional
charge-density wave systems \cite{belve1}, quantum field theory description of tunneling in
the integer quantum Hall effect \cite{mori1}, discussions of fractional charges  induced
in the ground state of a Fermi system by its coupling to a chiral field
\cite{goldstone, keil,stone,gousheh, mackenzie, polychronakos}, the spectrum
and string tension of the low-energy effective Lagrangian of QCD$_2$ \cite{prd1} and  a two dimensional model of high $T_c$ superconductivity \cite{ferraz}. The model has also been considered  in order to discuss the realization of chiral symmetries in $(1+1)$ dimensions \cite{witten}. Numerical self-consistent
solutions have been discussed for the model with additional $\phi^4$ potential \cite{Shahkarami1} and the Casimir energy of the system has been computed for a prescribed  kink \cite{Shahkarami2}.

Here, using some techniques of integrable systems, we perform the study of the soliton and fermion bound state spectra of a modified ATM  model. Although this model is non-integrable in general, its analytic soliton type solutions associated to specific vacua, can be obtained using the Hirota's tau function formalism. The static version of the model is considered and the soliton and fermion bound state spectra is uncovered. In this context, the in-gap and continuum bound states (BIC), as well as the fermion zero-modes, are examined. These soliton solutions allow us to perform a mapping to either modified sine-Gordon or novel Thirring-like  models, in the strong and weak coupling sectors, respectively. We perform this correspondence for each of the sectors, i.e. in-gap, BIC and zero-mode bound states corresponding to a family of soliton profiles of the scalar field. So, one gets extensions and modifications of the duality mappings of the usual ATM model \cite{npb2, aop1}. As a by product of the process, one finds novel scalar and fermionic models, describing the dual sectors of the modified ATM model, respectively. The equivalence between the $U(1)$ and topological charges of the usual ATM model gets an additional contribution as a non-local charge density due to the deformation potential which depends only on the Toda field. 
  
For special topological charges and certain relative phases of the fermion components it is shown that the kinks would host Majorana zero modes. The mapping between the kink and the Majorana zero mode is constructed and shown to describe either a double  sine-Gordon or a novel real fermion Thirring-like  model in the strong and weak coupling sector of the model, respectively. The back-reaction of the Majorana bound state on the kink is shown to catalyze the emergence of a new kink with lower topological charge and greater slope at the center, in the strong coupling limit of the model. However, for the model with vanishing self-coupling  potential and in the intermediate coupling sector the Majorana fermion-soliton system exhibits an equivalence between the topological charge density and the Majorana state density.  In the presence of  the scalar potential this equivalence gets an additional non-local contribution depending on the Toda field. 

The modified ATM model, which incorporates the self-interacting Toda field potential in the dynamics of the system, exhibits a formula of the  Atiyah-Patodi-Singer-type relating the Noether, topological and non-local charge densities. We will verify this type of formula using the analytical kinks and their relevant non-zero and zero mode spinor bound states, respectively. Remarkably, it will be shown that the equivalence between the Noether and topological charges of the undeformed ATM model still holds, such that the proportionality constant factor is redefined  by a term depending on the potential and spinor parameters.  

The choice of complicated background configurations associated to the scalar field renders the problem analytically unsolvable. So, one could resort to certain numerical simulations. However, in the numerical approach some important physical properties and concepts might not be clear as in the analytical approach described above. Nevertheless, in the numerical treatment the scalar field can represent topological kinks with many different shapes and topological charge taking any real value
 hosting fermion bound states. So, in order to check our results we have performed various numerical simulations and they support strongly our analytical approach. In order to perform the numerical simulations we follow the relaxation method which is discussed in the appendix \ref{sec:apprelax}. 

The paper is organized as follows: The section \ref{sec:model} summarizes the main properties of a modified ATM model. In section \ref{sec:topo} the properties of the fermion coupled to topological kinks are examined. The conditions for the threshold, BIC and Majorana bound states are established. A special scalar self-coupling potential is introduced and its vacuum structure is studied in some detail. In section \ref{sec:tau} self-consistent kink and fermion bound state solutions are obtained through the tau function approach. A finite set of topological charges and the associated fermion bound states, as well as the energy spectrum are uncovered. It is established a second order algebraic equation for the bound state energy eigenvalue $E$. In section \ref{sec:backreaction}  we construct a fermion-soliton mapping and  the back-reaction of the fermion on the kink is studied, as well as the strong coupling sector of the model. In the sec. \ref{sec:topnzV} we establish a Atiyah-Patodi-Singer-type formula relating the topological, Noether and non-local charges for potential $V\neq 0$. In sec. \ref{sec:zeromode1} we find the zero modes of the model and the parameters of the Majorana zero modes are obtained. In  sec. \ref{sec:topozmV} it is discussed a formula relating the topological, fermion and non-local charges for zero modes. The results of our numerical simulations are discussed in section \ref{sec:numer1}. In section \ref{sec:discussion} we discuss our main results. The appendices \ref{app1},  \ref{app2}, ...,\ref{app:curr22} present useful equations and identities used in the  tau function approach. The appendix \ref{sec:apprelax} presents the algorithm suitable for solving the model using the relaxation method.  

\section{The model}
\label{sec:model}

We consider a fermion chirally coupled to a scalar field defined by the following Lagrangian\footnote{Our notation: $x_{\pm} = t\pm x $, and so, $\pa_{\pm}=\frac{1}{2} (\pa_t \pm \pa_x)$, and $\pa^2=\pa^2_t -\pa^2_x = 4\pa_{-}\pa_{+}$. We use $\g_0 = \(\begin{array}{cc} 0  & i\\
-i & 0\end{array}\)$, $\g_1 = \(\begin{array}{cc} 0  & -i\\
-i & 0\end{array}\)$, $\g_5 = \g_0 \g_1 = \(\begin{array}{cc} 1  & 0\\
0 & -1\end{array}\)$,  and $\psi = \(\begin{array}{c} \psi_{R}\\
\psi_{L}\end{array}\),\,\, \bar{\psi} = \psi^{\dagger} \g_0$.} 
\br
\cl =  {1\o 2} \pa_{\mu} \Phi \, \pa^{\mu} \Phi
 + i  {\bar{\psi}} \gamma^{\mu} \pa_{\mu} \psi
- M \,{\bar{\psi}} \,
e^{i\beta \, \Phi\,\gamma_5}\, \psi  - V(\Phi),
\label{atm1}
\er
with the potential
\br
\label{pot1}
V(\Phi) \equiv A_1 \cos{(\beta_1 \Phi)} + A_2 \cos{(\beta_2 \Phi)} + A_3 \cos{(\beta_3 \Phi)}.
\er
The parameter $M$ defines a free fermion mass and $\beta$ the coupling strength between the fermion and scalar. The function $V(\Phi)$ defines a multi-frequency scalar potential and the real parameters $\beta_j\,(j=1,2,3)$ define the self-coupling parameters, respectively. 

The model (\ref{atm1}) for vanishing potential $V=0$ belongs to a wide class of integrable soliton
theories, the so-called  $\hat{sl}(2)$ affine Toda systems coupled to matter fields (ATM) \cite{matter}. The integrability properties, the construction of the general solution including the solitonic ones, the soliton-fermion duality properties, as well as its symplectic structures were discussed in \cite{npb1, npb2, aop1}. This model has been shown to describe the low-energy effective Lagrangian of QCD$_2$ with one flavor and $N$ colors \cite{prd1} and the BCS coupling in spinless fermions in a two dimensional model of high T superconductivity in which the solitons play the role of the Cooper pairs \cite{ferraz}. Moreover, the model has earlier been studied as a model for fermion confinement in a chiral invariant theory \cite{chang} and the mechanism of fermion mass generation without spontaneously chiral symmetry breaking in two-dimensions \cite{witten}.

The potential (\ref{pot1})  will destroy the chiral invariance and integrability of the model, and it also would make the model to be  non-renormalizible for a general set of parameters. In fact, the renormalization process would require the divergences to be absorbed into the redefined constants $A_j\, (j=1,2,3)$ (since the addition of new (periodic) counterterms to the action would  modify the three-frequency sine-Gordon model). In this regard, the non-integrable two and three-frequency sine-Gordon models constitute interesting toy models for nonperturbative quantum field theory and they have been regarded as renormalizable field theories in some parameter space \cite{bajnok, toth}.

The solutions to nonlinear differential equations exhibit unexpected states in the quantal Hilbert space with quantum numbers arising from topological properties of the classical field configurations. A model with potential of type $\phi^4$ and kink solutions coupled to heavy fermions has been examined in the semi-classical quantization WKB approach such that a large-N
expansion is employed to treat the scalar-fermion coupling nonperturbatively \cite{naculich}. In another related model without a classical kink, such that a scalar field is chirally coupled to a heavy fermion, it has been reported the creation of a static kink  by quantum corrections \cite{farhi}.

For an arbitrary potential $V$ and $\b \neq 0$ the dynamical model would lose the charge conjugation and energy-reflection
symmetries, resulting in a non-symmetrical energy spectrum
with respect to the $E = 0$ line. We will consider the model (\ref{atm1}) with potential (\ref{pot1}) and show that the dynamics of the  system exhibits kink type and associated spinor bound states such that the charge
conjugation and energy-reflection symmetries hold.

The eqs. of motion  become
\br
\label{521c}
i \g^{\mu} \pa_{\mu} \psi - M  \,
e^{i\beta \Phi\,\gamma_5}\, \psi &=&0,  \\
\partial^2\Phi - M \b  \bar{\psi} \psi \sin{(\b \Phi)} +i  M \b  \bar{\psi} \g_5 \psi \cos{(\b \Phi)}   + V'[\Phi]  &=& 0,\,\,\,\,\,\,\, V'[\Phi]  \equiv \frac{d V(\Phi)}{d\Phi}.\label{525} 
\er

The global $U(1)$ transformation 
\br
\Phi \ra \Phi  \; ;   \qquad 
\psi \ra e^{i \delta_1} \, \psi  \; ; \qquad 
{\psi^{*}_{}} \ra e^{-i \delta_1} \, {\psi^{*}_{}}, 
\lab{globalu1}
\er
gives rise to the next Noether current and the conservation law
\br
\lab{noetheru1}
J^{\mu} = {\bar{\psi}}\, \gamma^{\mu}\, \psi ,\,\,\,
\pa_{\mu}\, J^{\mu} = 0.
\lab{noethersl2}
\er
Next, taking into account the equations of motion (\ref{521c})-(\ref{525}) one can write the anomalous conservation law
\br
\label{quasi1}
\pa_{\mu} J_5^{\mu} &=& -\frac{2}{\b} V'(\Phi).\\
J_5^{\mu}  & \equiv &  \bar \psi \gamma^\mu \gamma_5 \psi 
+ \frac{2}{\b} \pa^{\mu} \Phi  \label{quasi2}
\er
In fact, for vanishing potential, i.e.  $V(\Phi)=0$, the  Lagrangian (\ref{atm1}) is invariant under the global chiral symmetry
\be
\Phi \ra \Phi - \frac{2 \l_1}{\b},\,\,\,\,\,\,\psi \ra e^{i\gamma_5 \lambda_1}\, \psi \; ; \qquad 
{\psi^{*}_{}}\ra e^{-i\gamma_5 \lambda_1}\, {\psi^{*}_{}} ,
\ee
which is associated to the conservation law
\be
\pa_{\mu}J_5^{\mu} =0,
\lab{chiral}
\ee
with the current $J_5^{\mu}$ defined in (\ref{quasi2}). So, a non-vanishing  potential  breaks the global chiral symmetry above and the expression $- \frac{2}{\b}V'(\Phi)$ in the r.h.s. of the quasi-conservation law (\ref{quasi1}) can be dubbed as an anomaly.

Moreover, the Lagrangian \rf{atm1}  for vanishing potential, $V(\Phi)=0$, is invariant under the transformation
 \br
\label{discrete}
\Phi \ra \Phi + \frac{2\pi n}{\b},\,\,\,n \in \IZ,\er
with the fermion field unchanged. The vacua are infinitely degenerate and one can define the topological current and its associated charge  
\br
j^{\mu}_{top} \equiv {\b \o {2\pi }}\epsilon^{\mu\nu} \pa_{\nu} \, \Phi 
\label{topological}\\
Q_{\rm topol.} \equiv \int \, dx \, j_{top}^0 \label{topcharge}
\er
which takes non-zero values depending only on the asymptotic values of $\Phi$, at $x=\pm \infty$. However, for non-vanishing potential the Lagrangian \rf{atm1} will be invariant under (\ref{discrete}) provided that 
\br
\label{bnj}
\b_j = n_j \b,\,\,\,n_j  \in \IZ,\,\,\, j= 1,2,3.
\er
Then, the model (\ref{atm1}) with $V\neq 0$, defined for some set of special coupling parameters $\b_j$ satisfying (\ref{bnj}), will  inherit the degenerate vacua (\ref{discrete}) of the integrable ATM model with $V=0$. 

In the vanishing potential case $V=0$ and defining $J_5^{\mu} \equiv \epsilon^{\mu \nu} \pa_{\nu}\S$ for $\epsilon^{01}=-\epsilon^{10}=1$,  the eq. (\ref{quasi1}) can be rewritten as
\br
\label{free}
\pa^{2} \S  &=& 0,\\
\pa^{\mu} \S  &\equiv&  \bar \psi \gamma^\mu \gamma_5 \psi 
+ \frac{2}{\b} \pa^{\mu} \Phi.
\label{free1}
\er
Therefore, taking into account the trivial solution $\S =0 $ for the free field (\ref{free}) into (\ref{free1}) one can get
\br
\label{index1}
 j_5^{\mu} = - \frac{2}{\b} \pa^\mu
 \Phi\\
 j_5^{\mu} \equiv \bar \psi \gamma^\mu \gamma_5 \psi.
\label{index11} 
\er
The relationship (\ref{index1}) can be regarded as classical versions of a formula of the Atiyah-Patodi-Singer-type (see e.g. \cite{niemi} for a review). For a recent review on the index theorems with massive fermions see also  \cite{fukaya}. Moreover, the relationship (\ref{index1}), upon using the identity $j_5^{\mu}= \epsilon^{\mu \nu }  J_{\nu}$,  can be rewritten as 
\br
\label{NT}
J^{\mu} = - \frac{2\pi}{\b^2} j^{\mu}_{top},
\er
where the currents $J^{\mu}$ and $ j^{\mu}_{top}$ have been defined in (\ref{noetheru1}) and (\ref{topological}), respectively.
This is the equivalence between the Noether (\ref{noetheru1}) and topological (\ref{topological}) currents present in the model, provided that the free massless scalar field $\S$ vanishes. The vanishing of this field has been interpreted as a condition for the
chiral confining property of the spinors inside the solitons of this model \cite{chang}. In \cite{npb1} using bosonization techniques it has been verified that the identity (\ref{NT}) holds at the quantum level, provided that the vanishing of the expectation value of the free field derivative, $< \pa^{\mu}\S> =0$, is imposed. 

The model (\ref{atm1}) with additional mass term and potential $V = \l \Phi^4$ has been
 considered in \cite{Shahkarami1} through a numerical method.  It was one of the first attempts to solve the equations self-consistently and non-perturbatively in which the authors obtained the bound states of the fermion and the shape of the soliton dynamically. In several treatments of this type of models the pseudoscalar field was {\sl a priori} prescribed and the back reaction of the spinor on the soliton has been neglected. For several aspects of the interaction of fermions with kinks, kink-antikink and domain walls,  see  e.g. \cite{gani, yi, brihaye}. The recent studies including the back-reaction were mainly performed using numerical methods \cite{gani1, campos}. In a recent work the analytical scattering states of the fermion on a prescribed sine-Gordon kink has been considered \cite{loginov}. So, the analytical treatment of the  back-reaction  of the fermion bound state on the kink is lacking in the literature. Here, we consider (\ref{atm1}) with $V \neq 0$ and  provide the first analytical results through the tau function approach of soliton theory. 
 
We will see below that the model (\ref{atm1}) with potential $V \neq 0$ would exhibit scalar field solitons approaching wide range of asymptotic values, apart from the values $\b \Phi (\pm \infty) \equiv  2 \pi n, \, n \in \IZ$ as in (\ref{discrete}). In fact, one can impose relevant boundary conditions compatible with the dynamics and the vacuum points defined by the zeroes of the potential $V$. So, as the potential parameters are continuously varied, one can thought that the topologically nontrivial configurations of the scalar field $\Phi$ of the model (\ref{atm1}) evolve continuously and slowly from a topologically trivial configuration to soliton configurations (with $\Phi(\pm \infty) \equiv \pm \phi_0$,\, $\phi_0 \in \IR$), such that  the corresponding spinor bound states assume continuously varying configurations. Below, we will obtain analytical kink and fermion bound state solutions for a finite set of values of $\phi_0$  and resort to numerical simulations to get those solutions for $\phi_0 \in \IR$ belonging to some continuous intervals.  

\section{Topological configurations coupled to the fermion}
\label{sec:topo}

Our aim is to investigate exact analytical solutions of the system of eqs. (\ref{521c})-(\ref{525}) such that nontrivial topological configurations can be achieved by the field $\Phi$, interacting with the bound states formed by the fermion field $\psi$. In the next steps we rewrite the equations of motion of the model in terms of the scalar field $\Phi$ and a set of real fermion field components $\xi_{a}\,\,(a=1,...,4)$. So, let us consider the Ansatz  
\br
\psi_{_{E}} (x,t)&=&  e^{-i E t } \Big[ \begin{array}c
\xi_3 (x) +i\, \xi_4(x) \\
\xi_1(x) + i\,\xi_2(x)
\end{array}\Big]\label{zetas}. 
\er

So, the system  (\ref{521c})-(\ref{525}) becomes
\br\label{5211}
\xi^{'}_1 + E\, \xi_2 - M \xi_4 \sin{\b \Phi}+ M  \xi_3 \cos{\b \Phi}&=&0, \\ \label{5221}
\xi^{'}_2 - E\, \xi_1+ M \xi_3 \sin{ \b \Phi}+ M \xi_4\cos{ \b\Phi}&=&0, \\ \label{5231}
\xi^{'}_3 - E\, \xi_4 + M \xi_2 \sin{\b \Phi}+ M \xi_1\cos{ \b \Phi}&=&0, \\ \label{5241}
\xi^{'}_4 + E\,  \xi_3 - M \xi_1\sin{\b \Phi}+ M \xi_2\cos{\b \Phi}&=&0, \\
\pa_{t}^2\Phi-\pa_x^2\Phi + 2 M \b  \Big[ (\xi_1\xi_3+\xi_2\xi_4)\cos{\b \Phi}- (\xi_1\xi_4-\xi_2\xi_3)\sin{\b \Phi}\Big] + V'[\Phi] &=&0,\label{phizetas1}
\er
where $E$ can be regarded as an eigenvalue of the Hamiltonian in the spinor sector. The self-consistent solutions for a
fermion coupled to static scalar field in the form of a kink (domain wall) have been considered
mainly using numerical methods in the literature.The self-consistent approach to find fermionic bound states must be contrasted to the results in which the scalar field kink is considered as an external field or {\sl a priori} prescribed, as presented for example in \cite{yi} and \cite{brihaye} for of the $\Phi^4$ model and for the case of the kink-antikink of the sine Gordon model, respectively. However, as we will show below, the presence of the fermion bound state will exert a back reaction on the kink by changing its profile and the relevant spectra of the system. Moreover, we will show that the phases of the fermion components  will depend on the parameters of the potential $V$.
  
We will search for analytical solutions to the of the systems of eqs. (\ref{5211})-(\ref{phizetas1}), such that the static version of (\ref{phizetas1}) is considered. In order to tackle this problem we follow the tau function approach developed for the related integrable ATM model in \cite{matter, npb1}. For vanishing potential $V(\Phi)=0$, the integrable ATM model defined in (\ref{atm1}) has been studied in \cite{matter, npb1, aop1, prd1}. The tau function approach can be adapted conveniently to the non-integrable system (\ref{5211})-(\ref{phizetas1}) provided that the potential $V(\Phi)$ (\ref{pot1}) displays the relationships
\br
\label{betas123r}
 \frac{\b_j}{\b} = \nu_j,\,\,\,\,\,\,\,\,\, \nu_j \in \IZ,\er 
between the frequency parameters $\b_j\,(j=1,2,3)$, such that  the set $\{\nu_j\}$ defines a relatively co-prime integers. This process will provide a variety of topologically nontrivial configurations for the scalar field $\Phi$ and the relevant fermionic bound states associated to the field $\xi_a$. 

We will look for solutions to the system (\ref{5211})-(\ref{phizetas1}) such  that they are  invariant under the parity transformation 
\br
\label{par1}
{\cal P}_x: x\rightarrow -x.
\er
So, let us consider certain field configurations satisfying
\br
\label{par11}
{\cal P}(\Phi) &=& - \Phi\\
\label{par12}
{\cal P}(V) &=& V.
\er 
 
We will pursue for some static solutions to the field $\Phi$ representing a kink type solution such that
\br \Phi(x)&=&-\Phi(-x),\label{parity0}\\
 \Phi(\pm \infty)&=&\pm \phi_o,\,\,\,\,\Phi(0)=0,\,\,\,\,\phi_o= const. \label{infty0}.
\er

Under the above assumptions the solutions to the system (\ref{5211})-(\ref{phizetas1}) will be eigenfunctions of the parity operator. Then one must have
\br
\xi_{1}(x)= - \s\, \xi_{4}(-x),\,\,\,\, \xi_{2}(x) = \s\, \xi_{3}(-x),\,\,\,\,\,\s = \pm 1, \label{parity}
\er
where $\s$ are the eigenvalues of the parity operator ${\cal P}_x$. Let us impose the following boundary conditions to the spinors
\br
\label{bcspinors}
\xi_{a}(\pm \infty) \rightarrow  0,\,\,\,a=1,...,4;\,\,\,\,\xi_{1}(0) =  \xi_{0}=const.
\er

Additional consistency conditions must be satisfied by the solutions; in fact, from the equations (\ref{5211})-(\ref{5241}) one gets the relationships
\br
\label{consist1}
\xi_{1}^2+ \xi_{2}^2-\xi_{3}^2-\xi_{4}^2 = c_1= const.
\er
and
\br
\label{consist2}
\pa_x [\xi_{1}^2+ \xi_{2}^2+\xi_{3}^2+\xi_{4}^2] + 4 M  \Big[ (\xi_1\xi_3+\xi_2\xi_4) \cos{(\b \Phi)}- (\xi_1\xi_4-\xi_2\xi_3)\sin{(\b \Phi)}\Big] =0.
\er

Since all the spinor components vanish at $x\rightarrow \pm \infty$, according to the boundary condition (\ref{bcspinors}), from now on we will set $c_1 = 0$ in (\ref{consist1}).  In addition to the above set of conditions we require a finite energy and localized energy density associated to the field configurations.

Useful identities related to (\ref{consist2}), which do not involve the eingenvalue $E$, can be obtained by combining some subset of equations of the system  (\ref{5211})-(\ref{5241}). So, multiplying by $\xi_1$ (\ref{5211}) and (\ref{5221})  by $\xi_2$ and adding the two expressions one has
\br
\label{consist23}
\frac{1}{2} \pa_x [\xi_{1}^2+ \xi_{2}^2] + M  \left\{ [\xi_1\xi_3+\xi_2\xi_4] \cos{(\b \Phi)} -  [\xi_1\xi_4-\xi_2\xi_3]\sin{(\b \Phi)}\right\} =0.
\er
Similarly,  from (\ref{5231})-(\ref{5241}) one can get the next  identity
\br
\label{consist24}
\frac{1}{2} \pa_x [\xi_{3}^2+ \xi_{4}^2] + M  \left\{ [\xi_1\xi_3(x)+\xi_2\xi_4(x)] \cos{(\b \Phi)} -  [\xi_1\xi_4(x)-\xi_2\xi_3(x)]\sin{(\b \Phi)}\right\} =0.
\er

In fact, the identities  (\ref{consist23}) and   (\ref{consist24}) are related to each other due to the parity symmetries (\ref{parity0}) and (\ref{parity}). Moreover, adding the last eqs. (\ref{consist23}) and (\ref{consist24}) one can get the relationship (\ref{consist2}).
 
In addition, the system of equations  (\ref{5211})-(\ref{phizetas1}) exhibits the following discrete symmetry
\br
\label{phim}
\Phi & \rightarrow&   - \Phi ,\,\,\, E  \rightarrow - E\\
\label{xhim1}
\xi_1 &\rightarrow & -i \xi_4,\,\,\,\,\xi_2 \rightarrow  i \xi_3\\
\label{xhim2}
\xi_4 &\rightarrow & \,\,\,\,i \xi_1,\,\,\,\,\xi_3 \rightarrow  -i \xi_2.
\er
This symmetry implies that one can construct two sets of solutions related to each other. In  fact, once we know the set $\{\Phi, \psi_{E}\}$ as a solution of the system, with $\psi_E$ defined in (\ref{zetas}),  one can construct another set, $\{-\Phi, \psi_{-E}\}$, provided that the new components of the spinor $\psi_{-E}$ are constructed according to (\ref{xhim1})-(\ref{xhim2}). Below we will discuss this symmetry in the context of the charge conjugation symmetry and the related formulation for the Majorana fermions in a system of this type.  

Some comments are in order here.  First, the static version of (\ref{phizetas1}) can be written as a suggestive equation involving some current densities. So, taking into account the identity (\ref{consist2}) into the equation  (\ref{phizetas1}) one can write 
\br
\label{xiscurrent}
\pa_x \Big[ J^{0} + \frac{2}{\b} \pa_x \Phi \Big] &=& \frac{2}{\b} V'(\Phi)\\
J^{0} & \equiv  & \xi_1^2 +  \xi_2^2 + \xi_3^2 + \xi_4^2. \label{j0}
\er
Moreover, by integrating (\ref{xiscurrent}) in the whole line and taking into account the relevant b.c.'s one gets
\br
\label{potint}
\int_{-\infty}^{+\infty}  dx\,V'(\Phi) =0.
\er   
Second, taking into account the parity property (\ref{parity0}) and the eq. (\ref{xiscurrent}) one can show that $J^0(x)$ is an even function, i.e. $J^0(-x)=J^0(x)$.

Third, the equation (\ref{xiscurrent}) is invariant under the parity symmetry (\ref{par11})-(\ref{parity}) provided that $V$ is an even function, i.e. $V(-\Phi)=V(\Phi)$. As a realization of this symmetry we will show below the transformation of the kink into an anti-kink, and the related  changes in the  signs of the density charges of the fermion. Moreover, this symmetry implies certain relationship between the topological charge of the kink (anti-kink) and the fermion (anti-fermion) bound state Noether charge in the model.    

Fourth, as mentioned above, for vanishing potential one has the integrable ATM model \cite{npb1, npb2}. It has been shown that in this case  the model exhibits the equivalence of the both topological and Noether currents at the classical and quantum levels, respectively. So, in order to write  an analogous equivalence equation one  integrates the b.h.s. of (\ref{xiscurrent}) in the interval $[-\infty, x]$ in order to get 
\br
\label{integrodif}
J^{0}(x) &=& - \frac{2}{\b} \pa_x \Phi + \frac{2}{\b} \partial^{-1} [\frac{d V(\Phi)}{d \Phi} ].
\er  
This defines a first order integro-differential equation relating the Noether  (\ref{noetheru1}), topological   (\ref{topological}) and a new non-local charge densities, respectively
\br
\label{currents1}
J^{0} (x) &= & (- \frac{4\pi}{\b^2}) \, j^{0}_{top} +  {\cal J}^{0}_{nonloc}\\
\label{nonl}
{\cal J}^{0}_{nonloc} & \equiv &
 \frac{1}{\b} \int^x_{-\infty} \frac{d V(\Phi)}{d \Phi} dx' - \frac{1}{\b} \int^{+\infty}_{x} \frac{d V(\Phi)}{d \Phi} dx'. 
\er

Fifth, remarkably, one can show that the system of first order equations comprising the eqs. (\ref{5211})-(\ref{5241}) and the first order integro-differential eq. (\ref{integrodif}) imply the static version of the second order differential eq. (\ref{phizetas1}). So, a solution of the first-order system of eqs.  (\ref{5211})-(\ref{5241}) and  (\ref{integrodif}) will be solution of the static version of the second order system of eqs. (\ref{5211})-(\ref{phizetas1}).  

Sixth, the strength of the back-reaction of the fermion on the kink  can be controled by the coupling constant $\b$. In fact, by taking the limit $\b \rightarrow 0$ of the integro-differential equation in (\ref{integrodif}) and upon taking the $x-$derivative of the remaining eq.  one is left with an effective equation for the scalar field $\Phi$ alone. In fact,  in this limit the fermion decouples from the scalar as it is clear from the Lagrangian (\ref{atm1}). 

Seventh, the relationship (\ref{currents1}) which relates  the Noether, topological and non-local charge densities can be regarded as another  classical version of a formula of the Atiyah-Patodi-Singer-type for the modified ATM model  incorporating the effect of the potential $V$. In fact,  the formula (\ref{currents1}) becomes a modification of the relationship (\ref{NT}) written for $\mu =0$. Below, we will verify these relationships for the analytical kinks and the relevant non-zero and zero modes in the sec. 6 and sec. 7.2.1., respectively.

Next, we write some identities which will be useful in the construction of the analytic solutions and in order to write a second order algebraic equation for the eigenvalue $E$. A useful  relationship arises for the  charge density $J^0$  once the equation (\ref{consist2}) is integrated in $x$. In fact,  taking into account (\ref{j0}) one can rewrite (\ref{consist2}) as
\br
\label{consist2i}
 J^0= - 4 M \int_{-\infty}^x\, dx\,  \Big[ (\xi_1\xi_3+\xi_2\xi_4) \cos{(\b \Phi)}- (\xi_1\xi_4-\xi_2\xi_3)\sin{(\b \Phi)}\Big].
\er
Taking the derivatives of the system   (\ref{5211})-(\ref{5241}) one can write the identity
\br
\nonumber
(E^2 - M^2) [\xi^2_1+\xi^2_2+\xi^2_3+\xi^2_4] &=& -[ \xi_1 \xi''_1 + \xi_2 \xi''_2+\xi_3 \xi''_3+\xi_4 \xi''_4] +\\
&& 2 M  \b \Phi' \{[\xi_1 \xi_4 - \xi_2 \xi_3] \cos{\b \Phi} + [\xi_1 \xi_3 + \xi_2 \xi_4]\sin{\b \Phi}\} \label{en10}
\er
Let us consider the normalization
\br
\label{norm0}
\int_{-\infty}^{+\infty}  dx\, [\xi^2_1+\xi^2_2+\xi^2_3+\xi^2_4] =  1. 
\er

Integrating in the whole line the expression (\ref{en10}) and taking into account the normalization (\ref{norm0}), as well as the relevant boundary conditions one can write 
\br
\nonumber
(E^2 - M^2) &=& \int_{-\infty}^{\infty} dx \, [ (\xi'_1)^2 +(\xi'_2)^2+(\xi'_3)^2+(\xi'_4)^2] -\\
&& 2 M  \b  \int_{-\infty}^{\infty} dx \, \Phi \frac{d}{dx}\{[\xi_1 \xi_4 - \xi_2 \xi_3] \cos{\b \Phi} + [\xi_1 \xi_3 + \xi_2 \xi_4]\sin{\b \Phi}\}. \label{en2}
\er
In addition, using the eqs. of motion one can write the identity 
\br
\label{id1}
\frac{d}{dx}\{[\xi_1 \xi_4 - \xi_2 \xi_3] \cos{\b \Phi} + [\xi_1 \xi_3 + \xi_2 \xi_4]\sin{\b \Phi}\} &=& - [2E - \b \Phi']\times \\
&& \Big[(\xi_1\xi_3+\xi_2\xi_4) \cos{(\b \Phi)}- (\xi_1\xi_4-\xi_2\xi_3)\sin{(\b \Phi)} \Big].\nonumber\\
&=& \frac{1}{4M} [2E - \b \Phi'] J'_0, \label{id20}
\er
where in order to get the last  line we have used (\ref{consist2}) and the expression of $J^0$ in (\ref{j0}). Then, taking into account the last relationship one can write (\ref{en2})  as
\br
\label{Ephi1}
E^2 &=& M^2 + \int_{-\infty}^{\infty} dx \, \Big\{ (\xi'_1)^2 +(\xi'_2)^2+(\xi'_3)^2+(\xi'_4)^2\Big\}-\b E \int_{-\infty}^{\infty} dx \, \Phi  J'_0 + \frac{ \b^2}{2} \int_{-\infty}^{\infty} dx \,\Phi \Phi' J'_0
\er
This equation provides a second order algebraic equation for the eingenvalue $E$.  We will see below that analogous  second order algebraic equation for the eingenvalue $E$ emerges in the context of the tau function formalism applied to solve the system (\ref{5211})-(\ref{phizetas1}) in order to find kink type and bound state solutions.

\subsection{Threshold states}
\label{sec:threshold}

The threshold or half-bound states are defined as states where the fermion field approaches a constant value at spatial infinity. The solutions $\xi_a$ of this type of state are finite but they do not decay fast enough at infinity to be square integrable \cite{graham, dong1, dong2}. In order to examine these type of states we will write  an equation involving only the bilinears of the $\xi_a\, 's$,  their derivatives and the eigenvalue $E$. In fact, from the system (\ref{5211})-(\ref{5241}) and taking into account the trigonometric identity $\sin^2{\b \Phi}+\cos^2{\b \Phi}=1$ one can write the following relationship
\br
{\cal A}_2 (E^2-M^2) + 2 {\cal A}_1 E  + {\cal A}_0 =0, \label{E2nd20}
\er 
with
\br
\label{aa10}
{\cal A}_2 & \equiv & \xi_{1} ^2+\xi_{2}^2+\xi_{3}^2+\xi_{4}^2,
\\
\label{bb1}
{\cal A}_1 & \equiv & \xi'_{1} \xi_{2}-\xi_{1} \xi'_{2} - \xi'_{3} \xi_{4}+\xi_{3} \xi'_{4},\\
 {\cal A}_0 &\equiv &   (\xi'_1)^2 +(\xi'_2)^2+ (\xi'_3)^2+(\xi'_4)^2. \label{cc1} 
\er  
So, one can analyze (\ref{E2nd20}) for these type of states in order to inspect the values of $E$ between the bound and continuum energy spectrum. So, let us consider
\br
\label{sbc}
\xi_a(\pm \infty) = c^{\pm}_a,\,\,\,\,\xi'_a(\pm \infty) =0,\,\,\,\,\,\,c_a^{\pm} = \mbox{const.},\,\,\,a=1,2,3,4.
\er
Applying the condition (\ref{sbc}) one has that ${\cal A}_0^{\pm} = {\cal A}_1^{\pm} = 0$ in (\ref{bb1})-(\ref{cc1}) and then the equation (\ref{E2nd20}) becomes 
\br
\label{thre0}
{\cal A}_2^{\pm} [E^2_{thr} - M^2 ]= 0.
\er
Since by definition one has ${\cal A}_2^{\pm} \neq 0 $,  the  algebraic equation (\ref{thre0}) provides the  energies of the threshold states 
\br
\label{thre1}
 E_{thr}^{\pm} = \pm M.
\er 
 
\subsection{Bound states in the continuum (BIC)}

The equation (\ref{Ephi1}) can be used to analyze the behavior of the spectra around the threshold states,  $E =   E_{th}^{\pm} $ as defined in (\ref{thre1}). In fact, for kink type solitons in the limit of small soliton width one can consider the function $\Phi = \frac{\phi_o}{2}(H(x)-H(-x)),\,( \phi_o >0)$, where $H(x)$ is the Heaviside step function. Then, substituting this form of kink for $\Phi$ into the equation (\ref{Ephi1}) one can write the next second order algebraic equation for $E$
\br
\label{2ndE}
E^2 - \l_o \,E -(M^2 + I) =0,
\er
where $\l_o \equiv  \phi_o J_{0}(0)$ and $I \equiv   \int_{-\infty}^{\infty} dx \,[ (\xi'_1)^2 +(\xi'_2)^2+(\xi'_3)^2+(\xi'_4)^2]$. Therefore, solving the second order equation (\ref{2ndE}) for the unknown $E$ one has 
\br
\label{Ebic1}
E &=& \pm M \D + \frac{1}{2}\l_o,\,\,\,\,\,\,\,\,\l_o \equiv  \phi_o J_{0}(0)\\
 \D  &\equiv &  \Big\{1+   \frac{\l_0^2 + 4 I}{4 M^2} \Big\}^{1/2},\,\,\,\,\l_o > 0,\,\,\,\,I > 0.  \label{Ebic11}
\er
Notice that, since  $I >0$ then $\D > 1$. Therefore, from the  equation (\ref{Ebic1}) one can see the appearance of bound states in the continuum (BIC), i.e. states with $ E > E^{+}_{th}$, with $ E^{+}_{th}$ being the positive threshold state in (\ref{thre1}) .  In fact, one has a state which takes a value  $E = + M \D + \frac{1}{2}\l_0$ (positive BIC) which is located above the positive threshold  $E^{+}_{th} = + M$ in (\ref{thre1}). Similarly, one has a state with energy $E = - M \D + \frac{1}{2}\l_0$. This will be an in-gap state for $E  > E^{-}_{th}$ provided that $M > \frac{I}{\l_o}$.

Moreover, in order to examine the negative BIC states one must rederive an analogous equation to (\ref{2ndE})  specialized for an anti-kink soliton. So, from  (\ref{Ephi1}) for anti-kink $\Phi = - \frac{\phi_o}{2}(H(x)-H(-x))$ one has
\br
\label{2ndE2}
E^2 + \l_o \,E -(M^2 + I) =0.
\er
Therefore, the solutions become
\br
\label{Ebic111}
E &=& \pm M \D - \frac{1}{2}\l_o, 
\er
with $\D$  and  $\l_o$ defined in (\ref{Ebic1})-(\ref{Ebic11}). Then, one has a negative BIC state with energy value  $E = - M \D - \frac{1}{2}\l_0$, which is located below the negative threshold  $E^{-}_{th} = - M$  in (\ref{thre1}). In addition, one has a state with energy $E = +M \D - \frac{1}{2}\l_0$. This is an in-gap state ($E < +M$) provided that $M > \frac{I}{\l_o}$. Below we will find some positive and negative BIC bound states in the context  of the tau function approach.

\subsection{Majorana zero-mode states}

The system of equations (\ref{5211})-(\ref{5241}) can be written in matrix form as
\br
\label{xiE}
H \xi  &=& E\, \xi,\\
\xi  & \equiv & \(\begin{array}{c} \xi_R \\
\xi_L \end{array} \) = \(\begin{array}{c} \xi_3 + i \xi_4 \\
\xi_1 + i \xi_2 \end{array} \), \er
or in components
\br
\label{RL}
\(\begin{array}{cc} i \frac{d}{dx} & i M e^{-i \b \Phi}\\
 -i M e^{i \b \Phi} & -i \frac{d}{dx}  \end{array} \)  \(\begin{array}{c} \xi_R \\
\xi_L \end{array} \) &=& E 
 \(\begin{array}{c} \xi_R \\
\xi_L  \end{array} \).
\er
We shall assume the operator $H$ to be a hermitian operator with eigenfuntions $\xi$ and a spectrum of real eingenvalues.  One can verify that $H$ satisfies
\br
\label{hc}
 \G^{-1} H \G = H^{\star},\,\,\,\,\,\,  \G \equiv \pm i \g_1.
\er
So, taking into account (\ref{hc}) and the complex conjugate of (\ref{xiE}) one can write the identity
\br
\label{xireal}
\xi  =  \G  \xi^{\star}.
\er 
In terms of the Dirac field $\psi_{_{E}}$ defined in (\ref{zetas}), and  taking into account $ \psi_{_{-E}}^{\star} \equiv  e^{i E t} \xi^{\star}$, the last relationship can be written as 
\br
\label{EmE}
\psi_{_{E}} =  \G  \psi_{_{-E}}^{\star}.
\er
This relationship exhibits the charge conjugation symmetry of the system and then, one expects that the particles and holes will have  identical spectra.  So, one can formulate Majorana fermions for a system of this kind since there exists a
particle-hole symmetry, or, in a relativistic model, a charge conjugation symmetry. 

A Majorana fermion is defined by considering the particle and hole with the same energy as
a single particle. Then, from the relationship (\ref{EmE}) and for zero-mode solutions ($E=0$) one can write
\br
\label{majorana}
\psi_{_{0}} =  \G  \psi_{_{0}}^{\star}.
\er
In fact, the fermions obeying the reality condition (\ref{majorana}) are defined as Majorana fermions. These states are well known to lead to interesting phenomena in condensed matter and we will search for them as fermion bound states. Recently, the model (\ref{RL}) has been considered as the continuum limit of a one-dimensional system of spin-orbit coupled Dirac Hamiltonian and a s-wave superconducting pairing, such that the s-wave pairing is represented by $-i M e^{i \b \Phi}$  \cite{udupa}. The bosonization techniques have been applied to related fermionic systems with finite length which exhibit Majorana zero modes \cite{chua}. So, the complementary descriptions provided by the fermionic and bosonic formulations
of the superconducting phase have been considered. Below, in the context of the tau-function approach, we will perform analogous Majorana zero mode-soliton duality description of the model (\ref{RL}) such that the static scalar field is a self-consistent solution of (\ref{phizetas1}).    

So, setting $E=0$ into the system of eqs. (\ref{5211})-(\ref{5241}) one has that the zero-mode states $\xi_a$ must satisfy
\br
\label{majoxi}
 (\xi'_1)^2 +(\xi'_2)^2+(\xi'_3)^2+(\xi'_4)^2  - M^2 [ (\xi_1)^2 +(\xi_2)^2+(\xi_3)^2+(\xi_4)^2] = 0.
\er
Taking into account the normalization (\ref{norm0}) one can rewrite (\ref{majoxi}) as
\br
\label{majoxi1}
\int_{-\infty}^{+\infty} dx\, [(\xi'_1)^2 +(\xi'_2)^2+(\xi'_3)^2+(\xi'_4)^2  ]= M^2 .\er

Next, imposing the reality condition for $E=0$ (\ref{majorana}) (or in the form  \ref{xireal}) on the spinor $\xi$ one has the following two cases:\\ 
Case I  ($\G = -i \g_1$)
\br
\label{majo1}
\xi_1 = - \xi_3,\,\,\,\, \xi_2 =  \xi_4.
\er 
Case II  ($\G = i \g_1$)
\br
\label{majo2}
\,\,\,\,\,\,\,\,\,\,\xi_1 = \xi_3,\,\,\,\, \,\,\,\xi_2 = - \xi_4.
\er

Below, in the tau function approach, we will find zero-mode bound states satisfying the above conditions (\ref{majoxi1}) and (\ref{majo1})-(\ref{majo2}).

\subsection{A special potential $V(\Phi)$ with $\frac{\b_j}{\b}= j; \, j=1,2,3$}

We will consider a particular potential (\ref{pot1}) with the next relationships imposed to the rational fractions between the coupling constants $\b_j$ in  (\ref{betas123r}) 
\br
\label{betas110}
\b_1 &=& \b ,\,\,\, \b_2= 2 \b,\,\,\, \b_3= 3\b.
\er
Notice that a particular set for the relatively prime integers $\nu_1=1, \nu_2 = 2 $ and  $\nu_3=3$ has been chosen in (\ref{betas123r}). So, the potential has the extrema
\br
\label{cr1}
\Phi_c   =  \frac{\pi}{\b}  m,\,\,m\in \IZ ; \,\,\,\,\,\,\,\, \pm  \frac{\theta_0}{\b}  + \frac{2 \pi}{\b} n,\,\,\,\, n \in \IZ,
\er
where
\br
\label{th00}
\theta_0 \equiv  \frac{1}{\b} \arccos{\Big[ \frac{1}{6} \(-\frac{A_2}{A_3} + \epsilon_o \sqrt{(\frac{A_2}{A_3} )^2- 3 \frac{A_1}{A_3}  + 9}\) } \Big];\,\,\,\,\,\,\,\,\,\,  \epsilon_o \equiv \pm 1,\,\,\,\, \theta_0 \in [-\pi ,+\pi].
\er
The second extrema in (\ref{cr1}) is defined for $A_2^2- 3 A_1 A_3 + 9 A_3^2\geq 0$. From (\ref{th00}) one can get the  next condition for the parameters 
\br
\label{a123}
A_1 = - 4 A_2 \cos{(\b \theta_0)} - 3 [1+ 2 \cos{(2 \b \theta_0)}] A_3, \,\,\,\,\,\,  \,\,\,\,\,A_3 \neq 0.
\er
Consider the case $A_3 \neq 0$ and define 
\br
\label{zz}
z \equiv \frac{1}{6} \frac{A_2}{A_3} + \epsilon_o \sqrt{(\frac{A_2}{A_3} )^2- 3 \frac{A_1}{A_3}  + 9},\,\,\,\,\,\epsilon_o \equiv \pm 1. 
\er
So, one can parametrize the quantities  $(A_1/A_3)$ and   $(A_2/A_3)$ as
\br
\frac{A_1}{A_3}  = 3 - 12 z \cos{(\b \theta_0)} ,\,\,\,\,\frac{A_2}{A_3} = 3 z - 3 \cos{(\b \theta_0)} . 
\er
It is useful to analyze separately the cases $|z| >1$  and $|z| \leq 1$. So, for the case $|z| >1$ and taking into account the above parametrization the potential becomes
\br
\label{pot12}
V_1(\Phi) &=& A_3 \Big\{ [3 - 12 z \cos{(\b \theta_0)}] \cos{(\b \Phi)} + [3 z - 3 \cos{(\b \theta_0)}] \cos{(2 \b \Phi)} +  \cos{(3 \b \Phi)}\Big\} \nonumber - \\
&&\Big\{-\frac{1}{3} A_3 \Big[ 12 z - 3 \cos{(\b \theta_0)} + 6 z \cos{(\b \theta_0)} + \cos{(3 \b \theta_0)} \Big]\Big\},\\
&&|z| > 1, \,\,\, \mbox{sign}(A_3) = \mbox{sign}(z), \nonumber
\er 
where the constant expression has been subtracted in order to define the zero level of the potential at the vacua $\pm  \frac{\theta_0}{\b} $. So,  (\ref{pot12}) describes a family of potentials with kink-type solutions interpolating  the vacua  $\pm \frac{\theta_0}{\b}$. The parameters $\{z,  A_3\}$ must be chosen from either the set $\{z > 1,\, A_3 > 0\}$ or the set $\{z < - 1,\, A_3  < 0\}$, for each value of $\theta_0$ in the interval $\theta_0 \in [-\pi, +\pi]$ . 

For the particular values of $\theta_0 = \pm \frac{\pi}{3},\, \pm \frac{2\pi}{3} $ we show the potentials in the Figs. 1.  These special  values  for $\theta_0 $, as well as the relationship (\ref{a123}),  will be obtained below for kink-type and fermion bound state solutions in the tau function approach.  Notice that the values  $\Phi_c = (3 n \theta_0),\, n \in \IZ$, for these particular $\theta_0's$ correspond to the  first set of the  extrema in (\ref{cr1}) and in this case represents a false vacua, as shown in the Figs. 1. 

For  arbitrary real values of $\theta_0$ in the interval $\theta_0 \in [-\pi, +\pi]$ we will use a numerical simulation in order to find the relevant spectra of kinks and fermion bound states. 

Next, let us define the potential in the region $|z| \leq 1$ such that some points of the extrema in (\ref{cr1}) can be parametrized by $ \frac{1}{\b}\arccos{(-z)}$. So, consider the potential 
\br
\label{pot310}
V_2(\Phi) &=& A_3 \Big\{ [3 - 12 z \cos{(\b \theta_0)}] \cos{(\b \Phi)} + [3 z - 3 \cos{(\b \theta_0)}] \cos{(2 \b \Phi)} +  \cos{(3 \b \Phi)}\Big\} -  c_0,\,\,\,\,\,\, |z| \leq 1,  \nonumber
\er   
where a constant value $c_0$ has been subtracted in order to define the zero level of the potential at the true vacua. The extrema in this parametrization become
\br
\label{cr2}
\Big\{\frac{\pi  k}{\b},\,\,\,\,k \in \IZ \Big\};\,\,\,\,\,\Big\{ \pm \frac{ \theta_0}{\b} + \frac{2\pi n}{\b},\,\,\,\,n \in \IZ\Big\};\,\,\,\,\,\,\,\,\Big\{\pm \frac{\a_0}{\b} + \frac{2\pi m}{\b},\,\,\,\,m \in \IZ\Big\}.
\er
with
\br
\a_0 \equiv  \arccos{(-z)}   .
\er
Some  points  of the extrema in (\ref{cr2}) become a true vacua while the other subset of points become a false vacua.  Let us consider the potential (\ref{pot310}) possessing the true vacua at $\pm \frac{\a_0}{\b}$
\br
\label{pot31}
V_2^{(1)}(\Phi) &=& A_3 \Big\{ [3 - 12 z \cos{(\b \theta_0)}] \cos{(\b \Phi)} + [3 z - 3 \cos{(\b \theta_0)}] \cos{(2 \b \Phi)} +  \cos{(3 \b \Phi)}\Big\}  - \\
&& A_3 \Big\{- 3 z + 2 z^3 + (3+6z^2) \cos{(\b \theta_0)} \Big\},\,\,\,\,\,\,\,|z| \leq 1, \nonumber
\er   
with 
\br
\{A_3 < 0,  \, z  > -\cos{\theta_0}\}\, \,\,\,\, \mbox{or} \, \,\,\,\,\, \{A_3 > 0,  \, z  < - \cos{\theta_0}\}
\er
such that  the constant $c_0$ in (\ref{pot310}) has been assumed to take  a particular form. In the  Fig. 2 (left)  we show an example such that the first two sets of (\ref{cr2}) become a  false vacua while the remaining subset becomes a true vacua with minima $\pm \a_0 + 2\pi m$.

Similarly, the potential with vacua as the first set of points in (\ref{cr2}) for even $k$ can be written as 
\br
\label{pot32}
V_2^{(2)}(\Phi) = A_3 \Big\{ [3 - 12 z \cos{(\b \theta_0)}] \cos{(\b \Phi)} + [3 z - 3 \cos{(\b \theta_0)}] \cos{(2 \b \Phi)} +  \cos{(3 \b \Phi)}\Big\} \nonumber - \\
A_3 \Big\{4+ 3z -3 (1+4 z) \cos{(\b \theta_0)} \Big\},\,\,\,\,\,\,|z| \leq 1,
\er  
such that  
\br
A_3 < 0,  \,\,\,\, z  > \cos{\theta_0}-2,
\er
where a particular form of $c_0$ has been set  to define the zero point energy. In the Fig. 2 (right)  we show another example such that the last two sets of (\ref{cr2}) become a  false vacua while the first subset  becomes a true vacua with minima $\frac{2n \pi }{\b}, \,n \in \IZ$. 

Finally, the potential with vacua as the first set of points in (\ref{cr2}) for odd $k$ can be written as 
\br
\label{pot33}
V_2^{(3)}(\Phi) = A_3 \Big\{ [3 - 12 z \cos{(\b \theta_0)}] \cos{(\b \Phi)} + [3 z - 3 \cos{(\b \theta_0)}] \cos{(2 \b \Phi)} +  \cos{(3 \b \Phi)}\Big\} \nonumber - \\
A_3 \Big\{-4+ 3z -3 (1-4 z) \cos{(\b \theta_0)} \Big\},\,\,\,\,\,\,|z| \leq 1,
\er  
such that  
\br
A_3 > 0,  \,\,\,\, z  < \frac{1}{3}(2- \cos{\theta_0}),
\er
where a convenient $c_0$ has been fixed in order to define the zero point energy. In the Fig. 3 we show an example such that the last two sets of (\ref{cr2}) become a  false vacua while the first subset  becomes a true vacua with minima $\frac{(2n+1) \pi }{\b}, \,n \in \IZ$. 

We will concentrate below, analytically (secs. \ref{sec:tau} and \ref{sec:backreaction}) and numerically (see sec. \ref{sec:numer1}), to the various types of fermion bound states induced by the kink-type solutions associated to the degenerate vacua described by the potential (\ref{pot12}). For another set of parameters such as $A_1 \neq 0,\, \b_1 = \b,\, A_2=A_3=0$ in (\ref{pot1}) we will show that  Majorana bound states are hosted by soliton type solutions interpolating the first type of vacua in (\ref{cr1}) for $m =\pm 1$ (see sections \ref{sec:zeromode1} and \ref{sec:A1n0}).
  
\begin{figure}
\centering
\includegraphics[width=8cm,scale=1, angle=0,height=4.5cm]{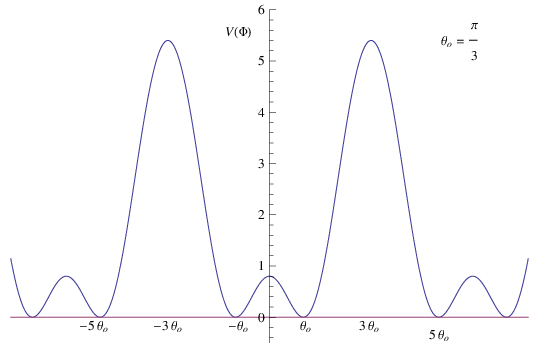}
\includegraphics[width=8cm,scale=1, angle=0,height=4.5cm]{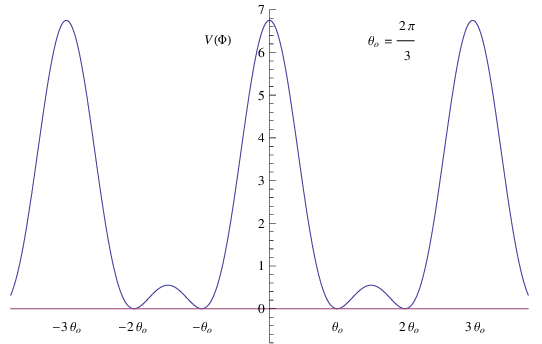}
\parbox{5in}{\caption{The plots of the potential $V_1(\Phi)$ in (\ref{pot12}) for $A_3=0.1, z = 4.5,\,\b =1$. Left Fig. for $\theta_0 = \frac{\pi}{3} $ and vacua $(2n+1)  \theta_o,\,n \in \IZ$.  Right Fig. for  $\theta_0 =\frac{2\pi}{3} $ and vacua $n  \theta_o,\,n \in \IZ$.}}
\end{figure}

\begin{figure}
\centering
\includegraphics[width=8cm,scale=1, angle=0,height=4.5cm]{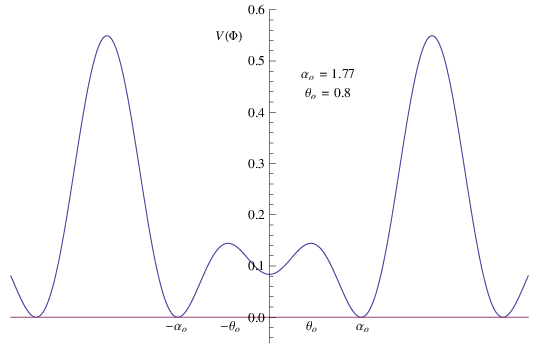}
\parbox{5in}{\caption{Left, the potential $V_2^{(1)}(\Phi)$ in (\ref{pot31})  for $A_3 = - 0.1, z = 0.2,\,\b =1,\,\theta_0 = 0.8$ and vacua   $\a_o = \pm 1.77 $.}}
\end{figure}
 
\begin{figure}
\centering
\includegraphics[width=8cm,scale=1, angle=0,height=4.5cm]{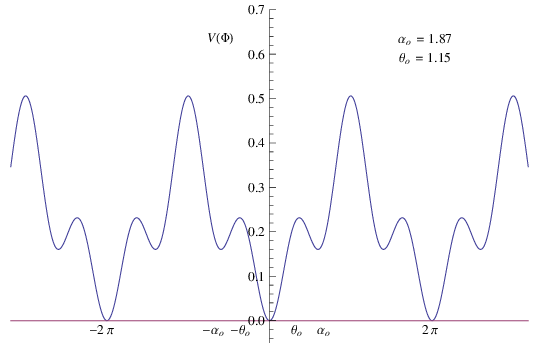}
\includegraphics[width=8cm,scale=1, angle=0,height=4.5cm]{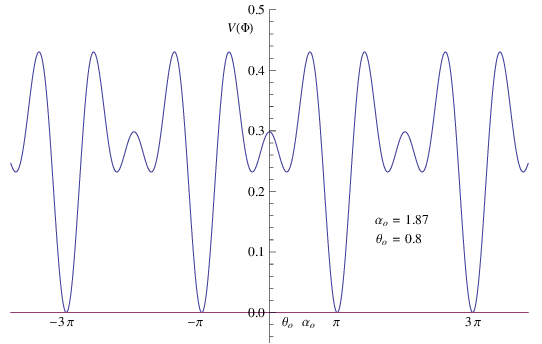}
\parbox{5in}{\caption{Left, the potential $V_2^{(2)}(\Phi)$ in (\ref{pot32}) for $A_3 = - 0.1, z = 0.3,\,\b =1,\,\theta_0 = 1.15,\, \a_o = 1.88$ and vacua $\pm 2 \pi n,\,n\in \IZ$. Right,  the potential $V_2^{(3)}(\Phi)$ in (\ref{pot33}) for $A_3 =  0.1, z = 0.3,\,\b =1,\,\theta_0 = 0.8,\, \a_o = 1.88$ and vacua $\pm (2n+1) \pi ,\,n\in \IZ$.  }}
\end{figure}

\section{Soliton solutions and the tau function approach}
\label{sec:tau}

In this section we solve the system of equations (\ref{5211})-(\ref{phizetas1}) analytically and determine a class of  kink and  fermion bound states with eigenvalue $E$. In order to perform this task we will follow the tau function approach and write the fields $\xi_a$ and  $\Phi$ in terms of the so-called tau functions. Even though the tau function and Hirota methods have been applied to integrable systems, those methods can also be applied to some non-integrable models in order to find 1-soliton type solutions as in the generalized sine-Gordon models \cite{jhep2}  and, sometimes, in order to construct the two-soliton solutions, as in the non-integrable modified regularized long wave equation (mRLW) \cite{jhep20, kdvnucl}.

So, let us consider the scalar and the fermion fields of the model  to depend on the tau functions $\{\tau_{1},\,\tau_{0},\, \tau_{\xi}^{a},\,\widetilde{\tau}_{\xi}^{a}\}$ through the Ansatz
\br
e^{i \b \Phi} &=& e^{-i \theta_{0}} \, \frac{\tau_{1}}{\tau_{0}},\,\,\,\, \label{tau00}\\
\xi_1 &=&  \,[\frac{ \tau_{\xi }^{1}}{\tau_0}] \, e^{-i \theta_{1}}+ \, [\frac{ \widetilde{\tau}_{\xi }^{1}}{\tau_1}] \, e^{i \theta_{1}} \label{tau1}\\
\xi_{2}&=& -i\{ \, [\frac{\tau_{\xi}^{2}}{\tau_0}] \, e^{-i \theta_{2}} - \, [\frac{ \widetilde{\tau}_{\xi}^{2}}{\tau_1}]  \, e^{i \theta_{2}}\} \label{tau2}\\
\xi_{3} &=& \, [\frac{ \tau_{\xi}^{3}}{\tau_0}] \, e^{-i \theta_{3}} + \, [\frac{\widetilde{\tau}_{\xi}^{3}}{\tau_1}] \, e^{i \theta_{3}}\label{tau3} \\
\xi_{4} &=& i\{  \,[\frac{ \tau_{\xi}^{4}}{\tau_0}] \, e^{-i \theta_{4}} -  \, [\frac{\widetilde{\tau}_{\xi }^{4}}{\tau_1}] \, e^{i \theta_{4}}\},\label{tau4}
\er
where $\theta_{0},\,\theta_{a} \, (a=1,...,4)$ are real parameters.  

We will concentrate below on the analytical kink and fermionic bound state profiles for the case $\b_n = n \b\,(n=1,2,3)$ in (\ref{betas110})  and the topological sector defined by  the charge $Q_{topol} = \frac{\theta_0}{\pi}$.
 
Next, we substitute the Ansatz (\ref{tau00})-(\ref{tau4}) into the system (\ref{5211})-(\ref{phizetas1}) for the fermion components coupled to the field $\Phi$ and write the differential equations  in terms of the  relevant tau functions. The four eqs.  (\ref{5211})-(\ref{5241}) in polynomial form in terms of the tau functions are provided in the appendix \ref{app1}, see eqs.(\ref{exi1})-(\ref{exi4}), respectively. 

Notice that for the chosen particular potential with the coupling constants defined in  (\ref{betas110}) the derivative of the potential $V'(\Phi)$ terms in the eq. (\ref{phizetas1}) can be written as integer powers of the tau functions. So, in the appendix \ref{app2} we present the scalar field equation (\ref{phizetas1}) written in the form of  a polynomial equation in integer powers of the tau functions (see  eq. (\ref{secorder})).

In addition, the consistency conditions   (\ref{consist1}) (for $c_1 =0$)  and (\ref{consist2}) in terms of the tau functions are provided in the appendix  \ref{app3}, see eqs. (\ref{consist11})-(\ref{consist22}), respectively.

\subsection{Tau functions for kink-type solitons and fermion bound states}
\label{subsec:n1}

A self-consistent solution to the  system of equations (\ref{5211})-(\ref{phizetas1}) for the topological sectors defined by the parameter $\theta_0$ will be found for the following tau functions
\br
\label{t01}
\tau_{0} &=& 1+ e^{- i \theta_0} e^{ 2 \kappa x },\,\,\,\tau_{1}= 1  + e^{ i \theta_0} e^{ 2 \kappa x },\\
\label{tauxi1}
\tau_{\xi}^{1} &=& \rho_1 e^{ \kappa x },\,\, \tau_{\xi}^{2} = \rho_2 e^{ \kappa x },\,\,\tau_{\xi}^{3} = \rho_3 e^{ \kappa x },\,\, \tau_{\xi}^{4}= \rho_4 e^{ \kappa x },\\\label{tauxi2}
\widetilde{\tau}_{\xi}^{1} &=& \rho_1 e^{ \kappa x },\,\, \widetilde{\tau}_{\xi}^{2} = \rho_2 e^{ \kappa x },\,\,\widetilde{\tau}_{\xi}^{3} = \rho_3 e^{ \kappa x },\,\,\widetilde{\tau}_{\xi}^{4} = \rho_4 e^{ \kappa x },\er
where $\kappa, \theta_0, \rho_a \, (a=1,...,4)$ are real parameters.  

In general, the tau functions become linear combinations of the monomials of the types $1, e^{\kappa x}, e^{2\kappa x},... (\kappa \in \IR)$. So, substituting these tau functions into the relevant equations in the appendices \ref{app1}, \ref{app2} and \ref{app3} one gets some polynomial equations of the form $c_0^{l} + c_1^l e^{\kappa x} + c_2^l e^{2\kappa x}+ ....+ c_n^l e^{n_k \kappa x} =0,\,(n_l \in \IN,\,\,\, l=1,...,7)$. So, by setting to zero the coefficients (i.e. $c_n^l \equiv 0$)  we will get some parameter relationships and determine the relevant eigenvalues $E$ depending on the set of parameters $\{M, \kappa, \theta_0, \s\}$. Next, we will implement this process below.
 
Firstly,  it is convenient to get the simplest forms of the parameter relationships through the auxiliary equations such as the parity eq.  (\ref{parity}) and the consistency equations  (\ref{consist1})  and (\ref{consist2}), respectively.  So, imposing the parity conditions (\ref{parity}) to the $\xi_{a}$ fields defined in (\ref{tau1})-(\ref{tau4}) for the tau functions (\ref{t01})-(\ref{tauxi2}) provide the next  parameter relationships
\br
\label{barb1}
 \rho_4   &=& \s \rho_1,\,\,\,\,\,\,\,\,\,\,\,\,\,\,\,\theta_4 =- \theta_1 + \theta_0- \frac{1}{2}\pi\\
 \rho_3  &=&  -\s \rho_2,\,\,\,\,\,\,\,\,\,\,\,\theta_3 = -\theta_2 + \theta_0 + \frac{1}{2}\pi .\label{barb2}
\er
Additional parameter relationships can be obtained from the equations (\ref{consist1})  and (\ref{consist2}) written in terms of the tau functions (\ref{consist11}) and  (\ref{consist22}), respectively. So, from (\ref{consist11}) upon substituting the tau functions (\ref{t01})-(\ref{tauxi2}) and taking into account the relationships (\ref{barb1})-(\ref{barb2}) one gets
\br
\label{rho12}
\rho_2^2 &=&   \rho_1^2  \csc{(2\theta_2-\theta_0)} \sin{(2\theta_1-\theta_0)}.  
\er
Next,  instead of replacing the tau functions (\ref{t01})-(\ref{tauxi2}) directly into (\ref{consist22}) (which corresponds to (\ref{consist2})) it is more convenient to use its simplified version as in (\ref{consist23}). In order to get other simple expressions for  the parameters relationships it is convenient  to use the  integrated form of (\ref{consist23}). So, upon substituting the tau function expressions (\ref{t01})-(\ref{tauxi2}) into (\ref{consist23})  one can integrate it  in the interval $[-\infty, x]$ and taking into account the boundary conditions (\ref{infty0}) and (\ref{bcspinors}) the outcome is a first order polynomial in $\mbox{sech}(2 \kappa x)$. So,  setting to zero its zero'th and first order term coefficients and after some algebraic manipulations taking into account  (\ref{rho12}) one  gets, respectively  
\br
\label{km}
\kappa = \s M \frac{\sin{(\theta_1-\theta_2)}}{\cos{(\theta_1+\theta_2)} - 2 \sin{\theta_1}\cos{\theta_2} \cot{\theta_0} }
\er
and
\br \label{rho12l}
\rho_2  &=&  \frac{\rho_1}{4}  \csc{(2\theta_2-\theta_0)} \times \\
  && \Big[\frac{\cos{(2\theta_2)} - \cos{(2 \theta_1)} + 4 \cos{[2(\theta_1-\theta_2)]} - \cos{[2(\theta_1-\theta_0)]}+ \cos{[2(\theta_2-\theta_0)]} - 4  \cos^2{(\theta_1+\theta_2-\theta_0)}}{2 \cos{\theta_2} \sin{\theta_1}\cos{\theta_0} - \cos{(\theta_1+\theta_2)}  \sin{\theta_0}}\Big].\nonumber
\er

Next, taking into account (\ref{rho12}) and (\ref{rho12l}) one can get an equation relating only the angles $\{\theta_0, \theta_1, \theta_2\}$
\br
\nonumber
\cos{[2(\theta_1-\theta_2+\theta_0)]}+\cos{[2(\theta_1-\theta_2-\theta_0)]} - 4 \cos{(2\theta_1)} + 4 \cos{(2\theta_2)} + 18 \cos{[2(\theta_1-\theta_2)]} - \\
4 \cos{[2(\theta_1-\theta_0)]} + 4 \cos{[2(\theta_2-\theta_0)]} - 6\{1+ 2 \cos{[2(\theta_1+\theta_2-\theta_0)]}\} -2 \cos{(2\theta_0)} =0.\label{thetas120}
\er
 
Now, let us consider the equations of appendices \ref{app1} and \ref{app2}. Substituting the expressions  (\ref{tau00})-(\ref{tau4}), given in terms of the tau functions (\ref{t01})-(\ref{tauxi2}), into the system of equations  (\ref{exi1})-(\ref{exi2}) one gets a $8$th order  polynomial in powers of the exponential $e^{\kappa x}$. The even order terms vanish identically and the odd order terms, i.e. the first, third, fifth and seventh orders,  provide non-vanishing relationships. So, from (\ref{exi1}) one has 
\br
\label{eq11}
\kappa \rho_1 \cos{\theta_1}  - E  \rho_2 \sin{\theta_2} - M  \s [ \rho_2 \sin{(\theta_2-\theta_0)}  \cos{\theta_0} +\rho_1 \sin{\theta_0} \cos{(\theta_1-\theta_0)}] =0,\\ 
\label{eq12}
\cos{\theta_0} [\kappa \rho_1 \cos{\theta_1}  - 3 \rho_2 (E + M\s)\sin{\theta_2} ]+\sin{\theta_0} [3 \kappa \rho_1 \sin{\theta_1} - M \rho_1 \s \cos{\theta_1} + \rho_2 (E+2 M \s) \cos{\theta_2}]=0,\\ 
\nonumber
4 \kappa  \rho_1 \cos{\theta_1} +\rho_2 ( 4 E + 5 M \s) \sin{\theta_2} + \rho_2 ( 2 E + M \s) \sin{(\theta_2-2\theta_0)} - \\
2 \rho_1 [\kappa \cos{(\theta_1-2\theta_0)} + M \s \cos{(\theta_1-\theta_0)} \sin{\theta_0}]=0,
\label{eq13}\\
\cos{\theta_0} [\kappa \rho_1 \cos{\theta_1}  + \rho_2 (E+M\s)\sin{\theta_2}] - \sin{\theta_0} [ M \rho_1 \s \cos{\theta_1} + E \rho_2  \cos{\theta_2} - \kappa \rho_1 \sin{\theta_1}] =0. \label{eq14}
\er
Similarly, from (\ref{exi2}) one has
\br
\label{eq21}
2 \kappa \rho_2 \sin{\theta_2}  +  \rho_1 (2E + M \s) \cos{\theta_1} +M  \s \rho_1 \cos{(\theta_1-2\theta_0)} -2 M \rho_ 2 \s \sin{\theta_0} \sin{(\theta_2-\theta_0)}=0,\\ 
\label{eq22}
\cos{\theta_0} [\kappa \rho_2 \sin{\theta_2}  + 3 \rho_1 (E + M\s)\cos{\theta_1} ]+\sin{\theta_0} [\rho_1 (E+2 M \s) \sin{\theta_1} -3 \kappa \rho_2 \cos{\theta_2} - M \rho_2 \s \sin{\theta_2} ]=0,\\ 
\nonumber
-4 \kappa  \rho_2 \sin{\theta_2} +\rho_1 ( 4 E + 5 M \s) \cos{\theta_1} + \rho_1 ( 2 E + M \s) \cos{(\theta_1-2\theta_0)} + \\
\rho_2 [2 \kappa \sin{(\theta_2-2\theta_0)} - M \s \cos{\theta_2} + M \s  \cos{(\theta_2-2\theta_0)}]=0,
\label{eq23}\\
\cos{\theta_0} [\rho_1 (E+M\s)\cos{\theta_1}-\kappa \rho_2 \sin{\theta_2} ] + \sin{\theta_0} [ M \rho_2 \s \sin{\theta_2} + E \rho_1  \sin{\theta_1} + \kappa \rho_2 \cos{\theta_2}] =0. \label{eq24}
\er

Next, we consider the problem of finding the bound state energy $E$.  The four eqs. (\ref{eq11})-(\ref{eq14}) can be written as an homogeneous linear system of equations for the unknown independent variables $\{\cos{\theta_1},  \sin{\theta_1}, \cos{\theta_2}, \sin{\theta_2}\}$.  So, in order to find a non-trivial solution, one imposes the condition of vanishing determinant for the $4 \times 4$ matrix formed by the coefficients in that linear system of equations, and one gets a quadratic algebraic equation for the bound state energy $E$  
 \br
\nonumber
[ 4 \kappa^2 - M^2 \sin^2{(\theta_o)}] E^2+   M \s [ 4  \kappa^2 -  2 M^2 \sin^2{\theta_o} - \kappa M \s  \sin{(2 \theta_o)} ] E  - \\
M^2 \sin{ \theta_o} [ 2 \kappa M \s \cos{\theta_o}-(\kappa^2-M^2) \sin{\theta_o}] = 0 \label{quadratic1}
 \er

Following similar reasoning as above for the system of equations (\ref{eq21})-(\ref{eq24}) one gets the same algebraic equation (\ref{quadratic1}). So, the equation (\ref{quadratic1})  is a second order polynomial in the variable $E$, and its exact solutions become
\br
\label{energy11}
E^{\s} = - \s M + \frac{M}{4} \Big\{ \frac{ 2 [\s \pm \cos{\theta_o}] + \frac{M}{2\kappa} (1\pm 2 \s) \sin{\theta_o} }{1-(\frac{M}{2 \kappa})^2 \sin^2{\theta_0}} \Big\}  
\er

Notice that the eigenvalue $E$ depends on  the parameters $\s, \, M, \theta_0$ and $\kappa$. The fraction $\frac{M}{2\kappa}$, which appear in the expression for $E^{\s}$ in (\ref{energy11}),  can be determined  from (\ref{km}). So, one has 
\br
\label{mk1}
\frac{M}{2\kappa} = - \frac{\s}{2} \cot{\theta_0} + \frac{1}{2} \s \csc{\theta_0}\, [\frac{\sin{(\theta_0-2 \theta_{+})}}{\sin{(2\theta_{-})}}],
\er
where 
\br
\label{thpm1}
\theta^{\pm} \equiv \frac{\theta_{1} \pm \theta_{2}}{2}.
\er
Since the set of angles $\{\theta_{1} ,\theta_{2},\theta_{0}\}$ satisfy the relationship (\ref{thetas120}) one can write this eq. as 
\br
\nonumber
6 - 18 \cos{(4 \theta^{-})} + 4 \cos{(2 \theta_0 )} \sin^2{(2\theta^{-})} + 12[1-2 \sin^2{(\theta_{0}-2\theta^{+})}] + 16 \cos{\theta_{0}}
\sin{(2 \theta^{-})}  \sin{(\theta_{0}-2\theta^{+})} =0.
\er
The last  equation  is a second order polynomial in the `variable'  $\sin{(\theta_{0}- 2\theta^{+})}$. So, one can write the two solutions in the form
\br
\label{fr11}
\sin{(\theta_{0}-2\theta^{+})}  = \frac{1}{3} \Big[\cos{\theta_0} + \epsilon_0 \sqrt{2} \sqrt{7+\cos{(2 \theta_0)}} \Big] \, \sin{(2\theta^{-})},\,\,\,\,\,\,\,\epsilon_0 \equiv  \pm 1,
\er
where $\theta^{\pm}$ are defined in (\ref{thpm1}).

Therefore, substituting the expression for $[\frac{\sin{(\theta_{0}- 2\theta^{+})}}{\sin{(2\theta^{-})}}]$ from (\ref{fr11}) into the equation (\ref{mk1}) one can get  
\br
\label{mk2}
\frac{M}{2\kappa}  = -\frac{\s}{3}\Big[\cot{\theta_0} + \epsilon_2 \frac{ \sqrt{2}}{2} \csc{\theta_0} \sqrt{7 +\cos{(2 \theta_0)}}\Big],\,\,\,\epsilon_2 = \pm 1.
\er
Remarkably, the last equation shows that $\frac{M}{2\kappa}$ depends only on the parameters $\{\s, \theta_0, \epsilon_2\}$.  
Substituting the expression for $\frac{M}{2\kappa}$ provided in (\ref{mk2})  into  (\ref{energy11}) one can get the eigenvalues  $E^{(\s)}$ 
\br
E^{\s}(\theta_0)  = - \s M \left\{1- \frac{1}{8} \Big[\frac{36 - 6(1-4 \s \epsilon_1) \cos{(\theta_o)} - 3\sqrt{2}(1+2 \s \epsilon_1) \epsilon_2 \sqrt{7+\cos{(2 \theta_o)}} }{5-\cos{(2 \theta_o)} -  \epsilon_2 \sqrt{2} \cos{\theta_o} \sqrt{7+\cos{(2 \theta_o)}}}\Big]\right\},\,\,\,\,\,\,\,\,\,\epsilon_1 = \pm 1.
\label{esp2}
\er
This is our final expression for the eigenvalue $E$ depending on the set of parameters  $\{M , \s, \theta_0\}$ and the signs $\epsilon_a = \pm 1\,(a=1,2)$. Notice that $E^{\s}(-\theta_0) =E^{\s}(\theta_0)$ as  $\kappa$ changes sign under $\theta_0 \rightarrow - \theta_0$. In fact, the change of sign for $\theta_0$ implies different asymptotic values for $\Phi$ at $x = \pm \infty$ and the interchange soliton/anti-soliton.

In addition, combining  the relationship for $\frac{M}{2\kappa}$ in (\ref{km}) and  the above equation (\ref{mk2}) one gets again the equation (\ref{fr11}). In fact, once a particular value for $\theta_0$ is known one has that the equation  (\ref{fr11}) defines the allowed set of angles $\{ \theta_1 , \theta_2\}$. It is known that one-dimensional bound-states are non-degenerate \cite{landau}; so, in our case the  bound-states with equal set of $\{ \theta_1 , \theta_2\}$ must  be labeled by their relevant parities. The solutions of (\ref{fr11}) for each $\theta_0$ will define a continuous family of parameters $\{ \theta_1 , \theta_2\}$ and the associated parameter $z$ (\ref{zz}) of the potential $V_1(\Phi)$ in (\ref{pot12}) will depend on these parameters through $A_j (j=1,2,3)$, as we will show below.
 
Next, let us analyze the second order equation (\ref{phizetas1}). So, substituting the expressions  (\ref{tau00})-(\ref{tau4}), given in terms of the tau functions (\ref{t01})-(\ref{tauxi2}), into the equation (\ref{phizetas1}) one gets a $12^{th}$ order polynomial in powers of the exponential $e^{\kappa x}$.  The terms with odd order power and the terms of order $6, 8, 10$\, and \,$12$ vanish identically, provided that the previous parameters relationships are assumed. The terms with the even power $0, 2$\, and \, $4$ provide certain non-trivial relationships between the parameters. So, the zero'th, second and fourth order power terms furnish the following relationships, respectively
\br
\label{aa1}
A_1&=&-4 A_2 \cos{\theta_0} - 3[1 +2 \cos{(2\theta_0)}] A_3,\\
\label{aa2} 
A_2 &=&  \frac{\kappa^2}{\b^2  \sin^2{(\theta_0)}} - \\
&& M \sigma \csc^2{\theta_0}  \Big\{  \rho_1^2   \sin{\theta_1} \sin{(\theta_1-\theta_0)} +  \rho_2^2  \cos{\theta_2} \cos{(\theta_2-\theta_0)} -  \rho_1  \rho_2  \cos{(\theta_1-\theta_2)} \cos{\theta_0} \Big\},  \nonumber\\
A_3 &=& \frac{1}{6} M \sigma  \Big[ \rho_1^2+ \rho_2^2 - 2 \rho_1 \rho_2  \cos{(\theta_1 - \theta_2)}\Big]  \csc^2{(\theta_0)}.  \label{aa3} 
\er
 
In fact, the equation (\ref{aa1}) is precisely the relationship (\ref{a123}) which has earlier been defined in order to introduce a particular vacua (\ref{cr1}) of the model.

\subsubsection{The tau function approach and the $\theta_o$ values}
\label{sec:theta0}

Below, we will find the values of the parameter $\theta_o$. We proceed by establishing additional parameter relationships from (\ref{currents1})-(\ref{nonl}) and (\ref{consist2i}) for the above solutions (\ref{tau00})-(\ref{tau4}). So, let us write the relevant expressions for the charge densities of the  topological and non-local currents as 
\br
\label{nonl11}
 {\cal J}^{0}_{nonl} (x) =\frac{a_0 + a_1  \cosh{(2 \kappa x)} + a_2  \cosh{(4 \kappa x)} }{4\kappa [\cos{\theta_0} + \cosh{(2 \kappa x)}]^2}  - \frac{a_2}{2 \kappa},
\er
with
\br
\nonumber
a_0 &\equiv&  16 A_2 \cos{\theta_0} \sin^3{(\theta_0)} + A_3  \{(45+\log{8}) \sin{\theta_0}  - 36 \sin{(2 \theta_0)} + 3[7+ \log{16}] \sin{(3 \theta_0)} -\\
&& 9 \sin{(4 \theta_0)} + \log{8} \sin{(5 \theta_0)}\},\label{a00}\\
\nonumber
a_1 & \equiv & 12 (A_2-3 A_3)\sin{\theta_0} + 4 A_3 (15+ \log{8}) \sin{(2\theta_0)} - 4 (A_2 + 9 A_3) \sin{(3 \theta_0)} + \\&&
6 A_3 (1+\log{4}) \sin{(4 \theta_0)},\label{a11}\\
a_2 &\equiv& 9 A_3 [\sin{\theta_0} - 2 \sin{(2 \theta_0)} ].\label{a22}
\er
These parameters satisfy the relationship 
\br
\nonumber
(4 \cos{\theta_0}-1) [\frac{a_0 -a_1 \cos{\theta_0}}{a_2} ]&=& \frac{2}{3}[\log{2} - 3\cos{\theta_0}+(5+\log{2}) \cos{(2\theta_0)} -3 \cos{(3\theta_0)}+ \\
&&
(1+\log{2})\cos{(4\theta_0)} ].  \label{a012rs}
\er
Notice that the non-local current  (\ref{nonl11}) satisfies the boundary condition
\br
 {\cal J}^{0}_{nonl} (\pm \infty)   = 0. 
\er 
Next, the topological charge density $ j^0_{top}$ can be written as
\br
\nonumber
 - \frac{4\pi}{\b^2} \, j^0_{top}  &=&- \frac{2}{\b} \, \pa_x \Phi\\
 &=& - \frac{4 \kappa}{\b^2} \, [\frac{\sin{\theta_0}}{\cos{\theta_0} + \cosh{(2 \kappa x)}}].\label{jtop1}
\er
By integrating the r.h.s. of  (\ref{consist2i}) one can get  the charge density $J^{0} $ written as
\br
J^{0} & = &\frac{4}{[\cos{\theta_0} + \cosh{(2 \kappa x)}]^2} [b_0 + b_1\cosh{(2 \kappa x)}], 
\label{density1}\er
where the expressions for $b_0$ and $b_1$ are provided in the appendix  \ref{app:b0b1}.

Next, let us consider the current charge densities (\ref{nonl11}), (\ref{jtop1}) and (\ref{density1}) and substitute them into the relationship (\ref{currents1}). So, one gets the next first order polynomial in $ \cosh{(2 \kappa x)}$
\br
\label{c01}
c_0 + c_1  \cosh{(2 \kappa x)} &\equiv & 0,
\er 
where an overall factor $[\cos{\theta_0} + \cosh{(2 \kappa x)}]^{-2}$ has been omitted and the coefficients  $c_0, c_1$ are provided in the appendix \ref{app:curr22}. 
  
So, the coefficients in the identity (\ref{c01}) must satisfy $c_0=c_1=0$. Next, take into account the relationships between  $\rho_1^2$ and $\rho_2^2$ in  (\ref{rho12}),  $\kappa$ and $M$ in  (\ref{km}) and $\rho_1$ and $\rho_2$ in (\ref{rho12l}). So,  taking into account the form of $c_0$ in (\ref{c00}) and setting $c_0 =0$ according to  the identity (\ref{c01}), one can get the following equation 
\br
\label{products1}
[3 + 5 \cos{(2\theta_0)} + \cos{(4\theta_0)} ]  \,  \csc{(2 \theta_2-\theta_0)} \, \csc^2{\theta_0} \, \sin{[2 (\theta_1-\theta_2)]} = 0.
\er
Similarly, setting $c_1=0$ in (\ref{c01}) and taking into account (\ref{c11}) one can get
\br
\label{products2}
[1+2  \cos{(2 \theta_0)}] \, \cos{(\theta_1-\theta_2)} \cot{\theta_0} \csc{(2 \theta_2-\theta_0)} \sin{(\theta_1-\theta_2)}^2 = 0.
\er
Notice that  the equation for $\kappa$ (\ref{km}) implies  
\br
\label{cond11}
 [\sin{\theta_0} \cos{(\theta_1+\theta_2)} - 2\cos{\theta_0} \sin{\theta_1} \cos{\theta_2} ] \neq 0.
\er 
Moreover, in order to define real values for $\rho_1$  and   $\rho_2$ in  (\ref{rho12}) one must have
\br
\label{cond22}
\theta_1 - \theta_2 \neq \frac{n\pi}{2}.
\er
Then the equations (\ref{products1})-(\ref{products2}) must hold provided that (\ref{cond11}) and (\ref{cond22}) are satisfied.   
So, the factors involving the angle $\theta_0$ in (\ref{products1}), $[3 + 5 \cos{(2\theta_0)} + \cos{(4\theta_0)} ] $  and $[1+2  \cos{(2 \theta_0)}]$ in (\ref{products2}) provide, upon equating them to zero,  the common solutions
\br
\label{th0}
\theta_0  = \pm \frac{\pi}{3} + \pi n,\,\,\,\,\,n\in \IZ.
\er
Notice that $|\theta_0|\leq \pi$ in the kink(anti-kink) solution (\ref{kink1}) and this parameter determines the asymptotic value of the kink (anti-kink) at $x= \pm \infty$. Therefore,  one can choose the set
\br
\label{thetas0}
\theta_0  =   \{ \pm \frac{\pi}{3} ,\,\, \pm \frac{2\pi}{3}\}.
\er
These are the possible  values of the parameter  $\theta_0$ defined in (\ref{tau00}) and the tau functions $\tau_{0}$ and $\tau_{1}$  in (\ref{t01}) . So, for the $\theta_0's$ in (\ref{thetas0}) and taking into account the expressions (\ref{mk2}) and \ref{esp2}) one can compute $\kappa$ and the energy  eigenvalues $E$, respectively. In addition, the $\theta_0$ values and the eq. (\ref{thetas120}) will determine a relationship between the phases $\{\theta_1, \theta_2\}$ of the spinor components in (\ref{tau1})-(\ref{tau2}). The other phases  $\{\theta_3, \theta_4\}$ in (\ref{tau3})-(\ref{tau4}) can be determined from (\ref{barb1})-(\ref{barb2}). 

In the Figs. 4-7  we present the relevant values of the set $\{\frac{\kappa}{M},\, \frac{E}{M}\}$ for the angles provided in (\ref{thetas0}) and a given parity $\s$. A careful observation of the tables and points for the both parities allows one to deduce that the discrete points with coordinates $(\kappa\, ,\, E)$ (for a fixed value of $M>0$) satisfy the symmetry
\br
\label{Eksign}
\kappa \rightarrow -\kappa\,\, \mbox{and} \,\, \s \rightarrow -\s \, \mbox{implies} \,\,\,\,  E \rightarrow -E.  
\er  

In the case $\theta_0 = \pi/3$ (tables and points in the Figs. 4 and 5) a careful  examination allow us to say that all of the states become in-gap states such that $|\frac{E}{M}| < 1$ and satisfy  the symmetry (\ref{Eksign}). 
 
In the case $\theta_0 = 2 \pi/3$ (see the tables in Figs. 6 and 7)  one has the both type of states, in-gap and BIC states. Remarkably, one notices the appearance of some bound states in the continuum (BIC) with energies $|\frac{E}{M}| > 1$; i.e. $E = \pm 1.01 M , \pm  1.1 M$. These bound states remain localized even though they coexist with the  continuous spectrum of radiating waves that can carry energy away,  i.e. for energies above or below the threshold ($|E|>M$) as defined in (\ref{thre1}). These types of bound states were
first predicted by von Neumann and Wigner \cite{wigner} in 3D linear Schr\"odinger equation with a localized potential.
 For an account of the developments in the BIC theory and the explanation of the physical mechanisms responsible for  their appearance in  different physical areas see e.g. \cite{BIC1}.  In the context of soliton theory the BIC states appeared as embedded solitons which can be robust despite being in resonance with the linear spectrum. For different mechanisms underlying
the existence of these solitons  see e.g.  \cite{embeddsol, malomed}). 
\begin{figure}
\centering
\includegraphics{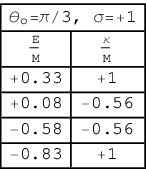}
\includegraphics[width=4cm,scale=1, angle=0,height=4.5cm]{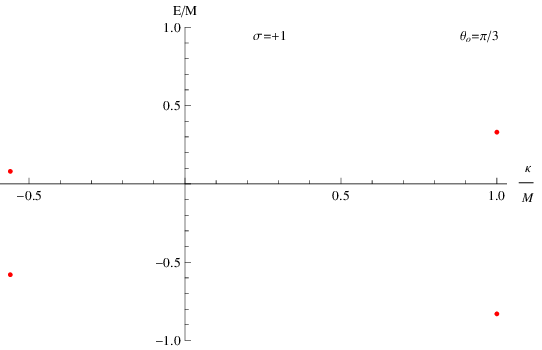}
\parbox{5in}{\caption{The left table shows the $E/M$ and $\kappa/M$ values for  $\theta_0 = \pi/3, \s = 1$.The right Fig. shows  the points with coordinates $(\kappa/M\,,\, E/M )$.}}
\end{figure}

\begin{figure}
\centering
\includegraphics{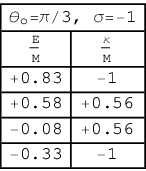}  
\includegraphics[width=4cm,scale=1, angle=0,height=4.5cm]{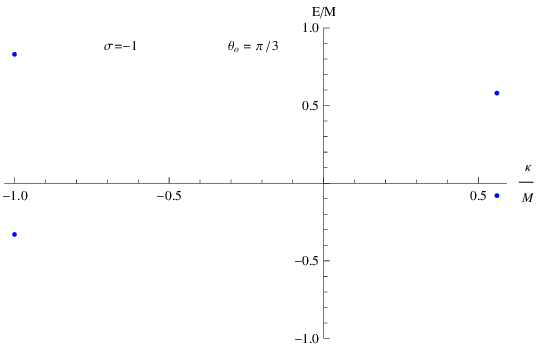}
\parbox{5in}{\caption{The left table shows the $E/M$ and $\kappa/M$ values for  $\theta_0 = \pi/3, \s = -1$.The right Fig. shows  the points with coordinates $(\kappa/M\,,\, E/M )$.}}
\end{figure}

\begin{figure}
\centering
\includegraphics{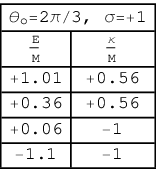}  
\includegraphics[width=4cm,scale=1, angle=0,height=5cm]{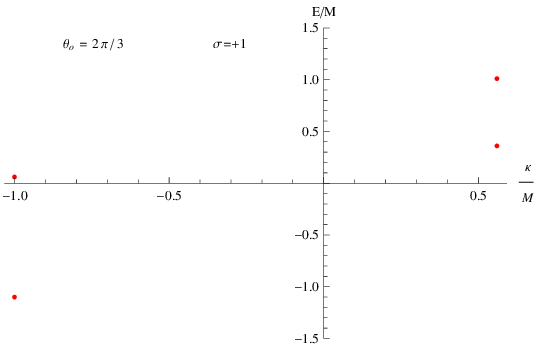}
\parbox{5in}{\caption{The left table shows the $E/M$ and $\kappa/M$ values for  $\theta_0 = 2\pi/3, \s = 1$.The right Fig. shows  the points with coordinates $(\kappa/M\,,\, E/M )$. Notice the BIC states for $E = +1.01 M , - 1.1 M$.}}
\end{figure}

\begin{figure}
\centering
\includegraphics{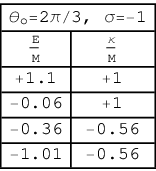}  
\includegraphics[width=4cm,scale=1, angle=0,height=5cm]{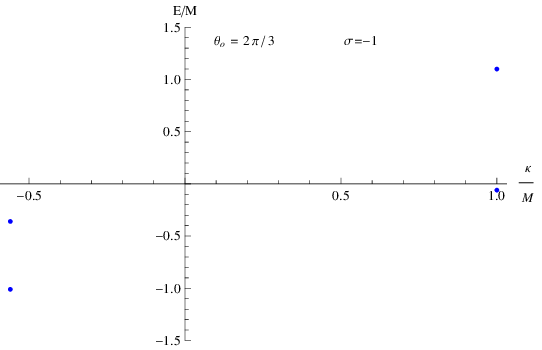}
\parbox{5in}{\caption{The left table shows the $E/M$ and $\kappa/M$ values for  $\theta_0 = 2\pi/3, \s =-1$.The right Fig. shows the points with coordinates $(\kappa/M\,,\, E/M )$. Notice the BIC states for $E = -1.01 M , + 1.1 M$.}}
\end{figure}

Let us write the explicit forms of the fermion bound states and  the kink solutions. The fermion components become
\br
\label{spinors12}
\xi_{a} = (-i)^{a-1} \, \rho_a\, e^{\kappa x -i \theta_a} \Big[ \frac{1}{1+e^{2 \kappa x- i \theta_0}}+ (-1)^{a-1} \frac{e^{2 i \theta_a}}{1+e^{2 \kappa x+i \theta_0}}\Big],\,\,\,\,a=1,2.
\er
The remaining components $\xi_{a}\,(a=3,4)$ can be obtained from the above expressions (\ref{spinors12}) taking into account the parity condition relationships (\ref{parity}). Notice that the expressions (\ref{spinors12}) define real  $\xi_{a}$ functions. 

Next, taking into  account  the relationship  (\ref{tau00}) and the tau functions (\ref{t01}), as wel as the conditions (\ref{parity0})-(\ref{infty0}), one can write the kink as
\br
\label{kink1}
\Phi(x) &=& \frac{2}{\b}  \arctan{\Big[ \tan{(\frac{\theta_0}{2})}  \,\, \tanh{(\kappa x)}\Big]},\,\,\,\, \,\,\,\,\,\,\,\,\,\,\,\, \theta_0 \in  \{ \pm \frac{\pi}{3} ,\,\, \pm \frac{2\pi}{3}\}.
\er
The asymptotic values of the field are $ \Phi (\pm \infty)\equiv \pm \frac{\theta_0}{\b}$.  

In addition, a condition for the spinor  will be given by setting the spinor component $\xi_1$ to a constant at $x=0$, i.e. 
\br
\label{cond1}
\xi_1(0)\equiv \xi_0 = \rho_1 \cos{\theta_1} [ 1+ \tan{\theta_1} \tan{(\frac{\theta_0}{2})}],
\er
where $\xi_0$ is a real constant.

\begin{figure}
\centering
\includegraphics{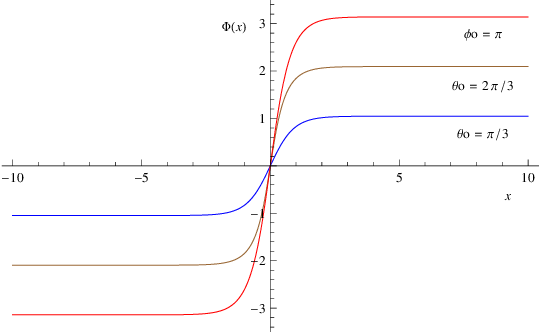}
\parbox{5in}{\caption{The analytical kinks $\Phi(x)$ for the set of parameters $\kappa=1, \b=1$,  $\theta_0 = \{2\pi/3, \pi/3\},\, \phi_0=\pi$. The kink with $\phi_0 = \pi$ value will host a Majorana fermion.}}
\end{figure}

\begin{figure}
\centering
\includegraphics[width=4cm,scale=1, angle=0,height=4cm]{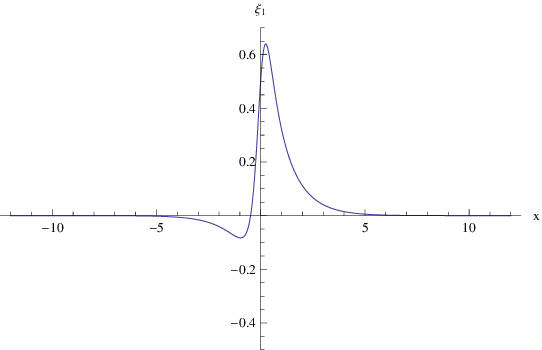}
\includegraphics[width=4cm,scale=1, angle=0,height=4cm]{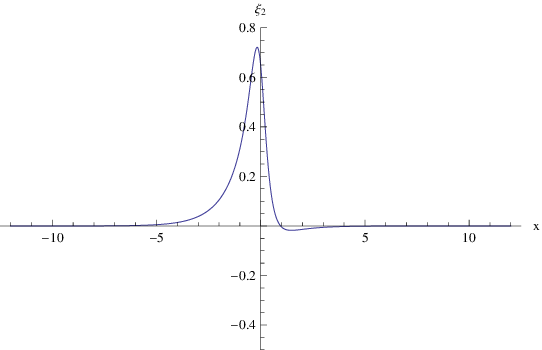}
\includegraphics[width=4cm,scale=1, angle=0,height=4cm]{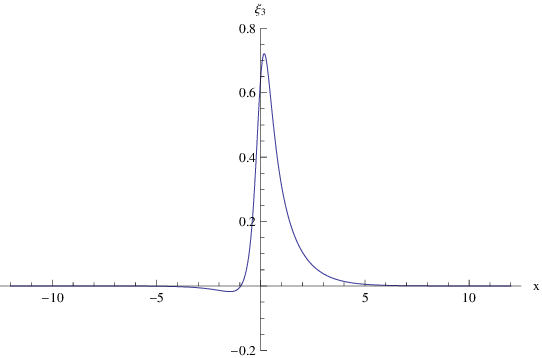}
\includegraphics[width=4cm,scale=1, angle=0,height=4cm]{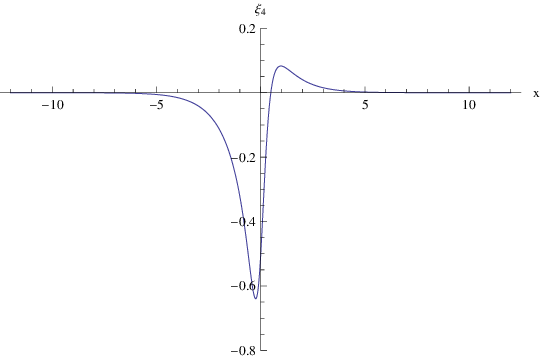}
\parbox{5in}{\caption{The analytical bound state of the fermion as a function of $x$ for the {\bf positive parity} $\s=1$. It is plotted for $\rho_1 = 0.41, \rho_2 = -0.65,\, \theta_1 = 0.138,\, \theta_2 = 0.85,\,\kappa =1, \theta_0 = 2 \pi/3$. Notice the spinor components relationships $\xi_{1}(-x)= - \xi_{4}(x)$ and $\xi_{2}(-x)= + \xi_{3}(x)$}}
\end{figure}

\begin{figure}
\centering
\includegraphics[width=4cm,scale=1, angle=0,height=4cm]{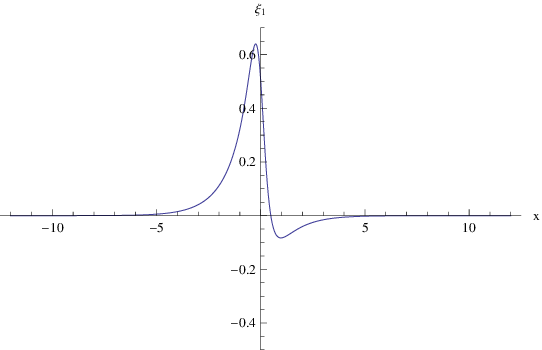}
\includegraphics[width=4cm,scale=1, angle=0,height=4cm]{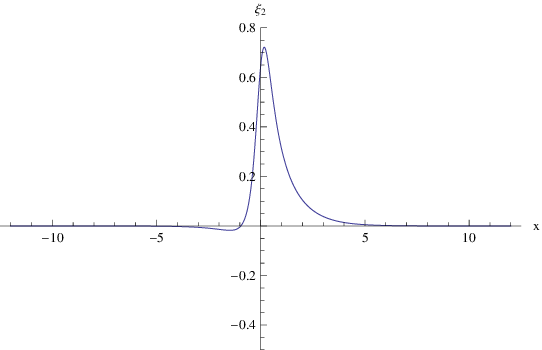}
\includegraphics[width=4cm,scale=1, angle=0,height=4cm]{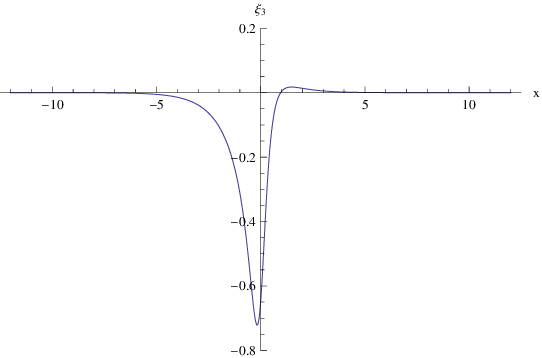}
\includegraphics[width=4cm,scale=1, angle=0,height=4cm]{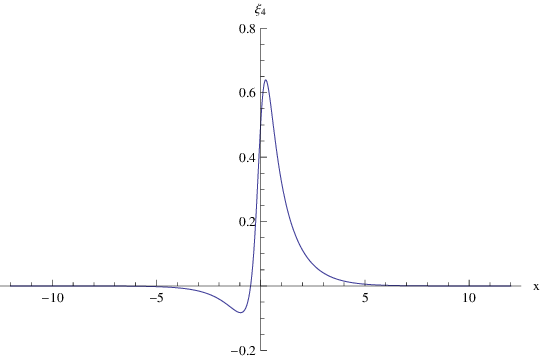}
\parbox{5in}{\caption{ The analytical bound state of the fermion as a function of $x$ for the {\bf negative parity} $\s=-1$. It is plotted for $\rho_1 = 0.41, \rho_2 = -0.64,\, \theta_1 = 0.14,\, \theta_2 = 0.85,\,\kappa =1, \theta_0 = 2 \pi/3$. Notice the spinor components relationships $\xi_{1}(-x)=+\xi_{4}(x)$ and $\xi_{2}(-x)= -\xi_{3}(x)$}}
\end{figure}

The above (\ref{kink1}) kink-type  solutions  for the values $\{\theta_0 =\,\pi/3, \, 2\pi/3\}$ are plotted in the Fig. 8. For comparison we also plotted  the sine-Gordon  type kink, $\frac{4}{\b} \arctan{ \Big[ e^{2 \kappa x  }\Big]} - \frac{\pi}{\b}$,  with asymptotic values $\phi_o = \pm \pi/\b $. The sine-Gordon  type kink  also appear as a solution of the model for another set of tau functions and this kink will host a Majorana fermion, as we will show below in the section \ref{sec:zeromode1} .

One can define the topological charge as
\br
\label{topol1}
Q_{topol} &\equiv &\frac{\b}{2 \pi}\,\int_{-\infty}^{+\infty} \pa_x \Phi\\
\label{topol11}
&=& \frac{\b}{2 \pi}[\Phi(+\infty)-\Phi(-\infty)].
\er
In the case at hand it becomes  $Q_{topol}=\frac{\theta_0}{\pi}$.  So, our analytical solutions are associated to the non-integer topological charge values  
\br
\label{topcharges1}
Q_{topol} =\pm \frac{1}{3},\, \pm \frac{2}{3},\er
where the $+$ sign corresponds to solitons and the $-$ sign for antisolitons.  

The fermion bound state corresponding to the soliton ($\theta_0 > 0$) can be normalized as follows 
\br
\label{cond31}
\frac{1}{{\cal N}}\int_{-\infty}^{+\infty}  dx\,  J^{0} &=& 1 \\
{\cal N} & \equiv & \frac{2}{\kappa} \Big[\rho_1^2 \cos{(2\theta_1 -\theta_0)}-\rho_2^2 \cos{( 2\theta_2 -\theta_0)} + \theta_0 (\rho_1^2+\rho_2^2) \csc{\theta_0}\Big]\label{cond32}.
\er
where the charge density  $ J^{0} = \sum_{a=1}^{4} \xi_{a}^2 $ as defined in (\ref{j0}) becomes 
\br
\nonumber
J^{0} & = &\frac{4}{[\cos{\theta_0} + \cosh{(2 \kappa x)}]^2} \Big\{ \rho_1^2 \cos{(\theta_1-\theta_0)} \cos{\theta_1} + \rho_2^2 \sin{(\theta_2-\theta_0)} \sin{\theta_2}  +\\
&&\frac{1}{2} [\rho_1^2 + \rho_2^2 + \rho_1^2 \cos{(2\theta_1-\theta_0)} \cos{\theta_0} - \rho_2^2 \cos{(2\theta_2-\theta_0)} \cos{\theta_0} ] \cosh{(2 \kappa x)} \Big\}.
\label{density}\er
The normalization condition ${\cal N} =1$ in (\ref{cond31}) and the eq.  (\ref{rho12}) can be used to determine the particular values of $\{\rho_1,  \rho_2\}$, once the angles $\{ \theta_1 , \theta_2\}$ are fixed from the curves defined by (\ref{fr11}) for each $\theta_0$. 

Regarding the full set of parameters of the system of equations (\ref{5211})-(\ref{phizetas1}) and the solution (\ref{spinors12})-(\ref{kink1}) one has  $17$ parameters, i.e. $\{\rho_a,\, \theta_a, \, A_j,\, \theta_0,\,\kappa, E,\,M,\,\beta, \sigma\}\, \mbox{for}\, a=1,2,3,4\, \mbox{and}\, j=1,2,3$.  One can consider the parameters $\b, M$ (which appear in the Lagrangian) and $\rho_1$ as free parameters. In fact, $\rho_1$ can be fixed from the condition (\ref{cond1}) once a particular value for $\xi_0$ is chosen (this value can be set according to the normalization condition (\ref{cond31})-(\ref{cond32})). In addition,  once a particular value for $\theta_0$ is chosen from the set (\ref{thetas0}) and a parity eigenvalue $\s$ is determined for a particular solution one can get the parameter $\kappa$ and the eigenvalue $E$, as functions of $M$,  from the eqs.(\ref{mk2}) and \ref{esp2}), respectively. The $\rho_2$ can be  determined from (\ref{rho12}), and the set $\{\theta_3, \theta_4\}$ and $\{\rho_3, \rho_4\}$ can be obtained from (\ref{barb1})-(\ref{barb2}). So, the values for $ A_j$ can be determined from (\ref{aa1})-(\ref{aa3}).

Moreover,  the equation (\ref{fr11}) defines a family of curves in the plane $\theta_1\times \theta_2$ for each $\theta_0$ of the set (\ref{thetas0}). Notice that a choice of a particular point $(\theta_1 , \theta_2)$ in a given curve defines a parameter value $z$ in (\ref{zz}) corresponding  to the potential of type $V_1(\Phi)$ in (\ref{pot12}). In fact, the parameter $z$ depends, through the parameters $A_j\,( j=1,2,3)$ in (\ref{aa1})-(\ref{aa3}),  on the angles $\theta_1, \theta_2, \theta_0$ and the parameters $M, \s, \b,\kappa$.  

\begin{figure}
\centering
\includegraphics[width=4cm,scale=1, angle=0,height=6cm]{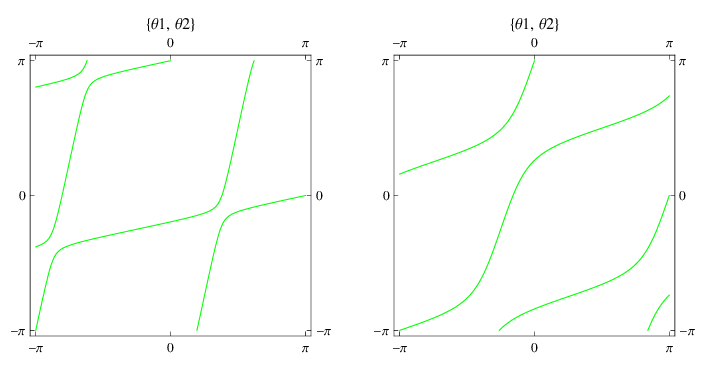} 
\parbox{5in}{\caption{The Figs. show the family of curves in the plane $\theta_1 \times \theta_2$ defined by (\ref{fr11}) for $\theta_0 = \frac{2\pi}{3}$ and the cases $\epsilon_0 = +1$(left figure) and $\epsilon_0 = -1$(right figure).}}
\end{figure}

In the Figs. 9 and 10 we present, for some set of parameters, the plots of the solutions  $\xi_a,\,\,a=1,...4$ defined in (\ref{tau1})-(\ref{tau4}) considering the tau functions (\ref{t01})-(\ref{tauxi2}) for the both signs of the parity eigenvalues $\s= \pm 1$, respectively. Their relevant parameters satisfy the relationships discussed above,  therefore the properties  of the solitons in Fig. 8, such as their asymptotic values, width and slope, will always depend on the relevant spinor parameters, and vice versa.

In the Fig. 11 we show a family of curves in the plane $\theta_1 \times \theta_2$ as defined through the eq. (\ref{fr11}) for $\theta_0 = \frac{2\pi}{3}$ and the cases $\epsilon_0 = +1$(left figure) and $\epsilon_0 = -1$(right figure).

\section{Back-reaction of spinor on the kink and strong coupling sector}
\label{sec:backreaction}

Fermions bounded by kinks were previously considered by some authors in the literature. However, in most  of the studies the effect of the  back-reaction of the fermions on the soliton were considered to be nearly zero or completely neglected.  Some  attempts  to take into account the back-reaction of the fermion on the kink have been performed mainly by numerical simulations \cite{perapechka, Shahkarami1}, although self-consistent analytical solutions, to our knowledge,  are still missing. The main reason for that is the lack of integrability of the models considered and the non-existence of general exact methods to solve the highly nonlinear and non-integrable models, as well as the integral-differential equations to solve the self-consistent system of equations.

Next, we examine the back-reaction of the Dirac fermion on the kink using the exact solutions obtained above. So, from (\ref{spinors12}) and (\ref{kink1}) one can construct the following mappings between the spinor components and scalar field
\br
\label{map11}
\xi_1 &=& \frac{2 \rho_1 }{\sin{\theta_0}} \sqrt{ \sin{(\frac{\theta_0 -\b \Phi}{2})} \sin{(\frac{\b \Phi +\theta_0 }{2})} } \, \,\,\,\cos{(\theta_1 - \frac{\theta_0 +\b \Phi}{2})},\\
\label{map12}
\xi_2 &=& - \frac{2  \rho_2}{\sin{\theta_0}} \sqrt{\sin{(\frac{\theta_0 -\b \Phi}{2})} \sin{(\frac{\b \Phi +\theta_0 }{2})}}  \, \sin{(\theta_2 - \frac{\theta_0 + \b \Phi}{2})},\\
\label{map13}
\xi_3 &=&- \frac{2 \s  \rho_2}{\sin{\theta_0}} \sqrt{\sin{(\frac{\theta_0 -\b \Phi}{2})} \sin{(\frac{\b \Phi +\theta_0 }{2})}}  \, \sin{(\theta_2 - \frac{\theta_0 -\b \Phi}{2})},\\
\label{map14}
\xi_4 &=& - \frac{2 \s \rho_1 }{\sin{\theta_0}} \sqrt{ \sin{(\frac{\theta_0 -\b \Phi}{2})} \sin{(\frac{\b \Phi +\theta_0 }{2})} } \, \,\cos{(\theta_1 - \frac{\theta_0 -\b \Phi}{2})}.
\er
Notice that these spinor components satisfy the parity symmetry defined in (\ref{parity}) provided that the kink satisfies (\ref{parity0}). In fact, the kink  presented in (\ref{kink1}) satisfies  (\ref{parity0}). 

Moreover, using the above mappings one can get the following identity
\br
\label{id2}
 2 M \b  \Big[ (\xi_1\xi_3+\xi_2\xi_4)\cos{\b \Phi}- (\xi_1\xi_4-\xi_2\xi_3)\sin{\b \Phi}\Big] =\b B_1 \sin{(\beta \Phi)} + 2 \b B_2 \sin{(2 \beta \Phi)} + 3 \b B_3 \sin{(3 \beta \Phi)},
\er
with
\br
\nonumber
B_1 &=& \frac{M \s}{2 \sin^2{\theta_0} } \Big\{ \rho_1^2+ \rho_2^2 - 2\rho_1 \rho_2  \cos{(\theta_1 - \theta_2)} - 4 \rho_1^2 \cos{(2\theta_1 - \theta_0)} \cos{\theta_0}+ 4 \rho_2^2 \cos{(2\theta_2 - \theta_0)} \cos{\theta_0}\Big\}\\
B_2 &=& -\frac{M \s}{2 \sin^2{\theta_0} } \Big\{ [ \rho_1^2+ \rho_2^2 - 2\rho_1 \rho_2  \cos{(\theta_1 - \theta_2)} ] \cos{\theta_0} -  \rho_1^2 \cos{(2\theta_1 - \theta_0)}  + \rho_2^2 \cos{(2\theta_2 - \theta_0)}  \Big\}\\
\label{bb3}
B_3 &=& \frac{M \s}{6 \sin^2{\theta_0} } \Big\{ \rho_1^2+ \rho_2^2 - 2\rho_1 \rho_2  \cos{(\theta_1 - \theta_2)} \Big\}.
\er
One can verify the next relationship
\br
\label{B123}
B_1 = - 4 B_2 \cos{\theta_0} - 3 [1+ 2 \cos{(2 \theta_0)}] \, B_3.
\er
Notice that the r.h.s. of the identity (\ref{id2}) depends only on the scalar field. We proceed by constructing a static equation of motion involving only the scalar field $\Phi$. This follows by making use of the identity (\ref{id2}) into the static version of the equation (\ref{phizetas1}). In fact,  the identity (\ref{id2}) defines an equivalence of the spinor-scalar interactions terms (l.h.s of (\ref{id2})) to a sum of multi-frequency sine terms (r.h.s. of (\ref{id2})). So, substituting the l.h.s of (\ref{id2}) into the relevant terms of the static version of the equation of motion  (\ref{phizetas1}) one can get
\br
\label{multiSG}
\Phi'' + \b C_1 \sin{(\b \Phi)} +  2\b C_2 \sin{(2 \b \Phi)}+3 \b C_3 \sin{(3\b \Phi)} =0,
\er
where 
\br
\label{cjj}
C_j = A_j - B_j, \,\,\,\,\, j=1,2,3.
\er
Inspecting the form of the parameters $A_3$ in (\ref{aa3}) and $B_3$ in (\ref{bb3}) one concludes that $B_3= A_3$, then 
\br
\label{c3}
C_3 =0.
\er  
 So, (\ref{multiSG}) defines a static equation of motion for a double SG model, such that an effective potential can be defined as
\br
\label{Veff}
V_{DSG} = C_1 \cos{(\b \Phi)} +  C_2 \cos{(2 \b \Phi)}. 
\er
The equation (\ref{multiSG}) can be regarded as an effective model for the scalar field $\Phi_{DSG}$ which incorporates the back-reaction of the fermion on the kink from the original system (\ref{5211})-(\ref{phizetas1}). The effective model (\ref{multiSG}) carries the effect of the spinor on the kink  through the parameters $C_j$ which depend on $B_j$ (\ref{cjj}). In fact, the  $B_j$ depend on the spinor parameters $\{\rho_{1,2}, \theta_{1,2},\s, \theta_0, \kappa\}$. The parameters $C_j$ also incorporate, through the terms $A_i$, the kink parameters $\{\kappa, \theta_0\}$, as well as the parameters $\{\b, M\}$ which appear in the Lagrangian.  

Taking into account (\ref{aa1}) and (\ref{B123}) one can show the next relationship
\br
\label{C123}
C_1 = - 4 C_2 \cos{\theta_0}.
\er 
Next, let us examine the vacua of the potential $V_{DSG} $ in (\ref{Veff}). Since the parameters satisfy (\ref{C123}) one gets the critical points of the potential (\ref{Veff}) to be
\br
\label{vcdsg}
\Phi_{DSG}^{vac}  =   \frac{\pi}{\b}  m,\,\,\pm \frac{\theta_0}{\b} + \frac{2 \pi}{\b} n;\,\,\,\, \, m, n  \in \IZ.
\er
Remarkably,  these vacua take the same values as the ones we have defined corresponding to the special potential in (\ref{cr1}).

The model (\ref{multiSG}) with potential  (\ref{Veff}) and the parameters relationships (\ref{c3}) and (\ref{C123}) possesses the following two types of kink solutions.
 
{\bf Type A}

By direct substitution in the static equation  of motion (\ref{multiSG}) for  $C_3 =0$ and $C_1 = - 4 C_2 \cos{\theta_0}$, one can verify the solution  
\br
\label{kinkA}
\Phi_{DSG}^{A} = \frac{2}{\b} \arctan{\Big[\tan{(\frac{\theta_0}{2})} \tanh{(\g x)}\Big]}, 
\er
with
\br
\g = \sqrt{\b C_2} \sin{\theta_0},\,\,\, C_2 >0.
\er

Notice that  the solution (\ref{kinkA}) resembles to the one in (\ref{kink1}). However, we will show below that the $\theta_0$ dependence of their relevant widths $\frac{1}{\kappa}$ and $\frac{1}{ \g}$, respectively for (\ref{kink1}) and (\ref{kinkA}), are completely different in the both cases. In fact, the solution (\ref{kink1}) has been self-consistently obtained  for the soliton-fermion system (\ref{5211})-(\ref{phizetas1}) with potential $V$ defined for $\b_j = j\, \b\, (j=1,2,3)$) such that the parameter $\theta_0$ takes the particular values of $\theta_0 = \pm \frac{\pi}{3},\, \pm \frac{2\pi}{3} $. 

So, in the model (\ref{5211})-(\ref{phizetas1})  one can examine the degree of the back-reaction of the spinor components  on the soliton by comparing the widths of the both solitons $\Phi$ in (\ref{kink1}) ($\sim \frac{1}{\kappa}$)  and  $\Phi^{A}_{DSG}$ in (\ref{kinkA}) ($\sim \frac{1}{\g}$), respectively. In doing so, one can argue that the soliton (\ref{kinkA}) of the model (\ref{multiSG})-(\ref{c3}) carries the spinor back-reaction of the soliton-fermion system (\ref{5211})-(\ref{phizetas1}) through the parameter $\g$, which defines its with as $\sim \frac{1}{\g}$. 

\begin{figure}
\centering
\includegraphics[width=4cm,scale=1, angle=0,height=4.5cm]{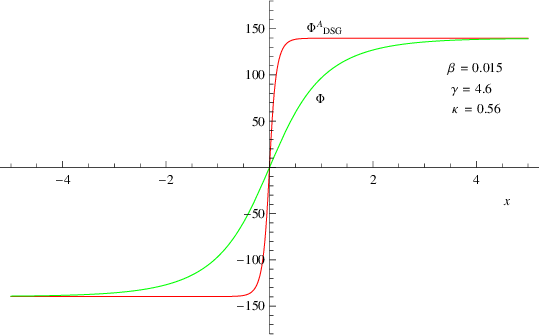}
\includegraphics[width=4cm,scale=1, angle=0,height=4.5cm]{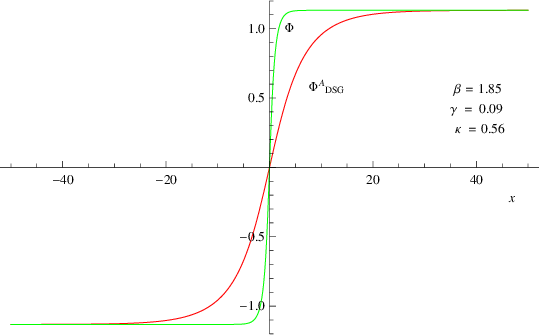}
\parbox{5in}{\caption{The kinks $\Phi$ (green) and $\Phi_{DSG}^{A}$ (red) as a function of $x$ for weak and large coupling parameter $\b$. Left: For $\b=0.015,\,\g=4.6$ . Right:  For $\b = 1.85,\,\g =0.09$. In the both cases  $\theta_0 = 2\pi/3,  M=1, \kappa =0.56, \rho_1 = 0.41, \rho_2 = -0.65, \theta_1 = 0.138, \theta_2 = 0.85, \s =1$.}}
\end{figure}

The changes in the local features of the kink, once the asymptotic value $\theta_0 = |\Phi(\pm \infty)|$ is fixed, will be considered below in order to study the back-reaction  of the fermion on the kink; so, let us compute the slope of the kink $\Phi^{A}_{DSG}$ at the origin $x=0$ 
\br
\label{slope1}
\mu^A =\frac{2}{\beta}  \g\, \tan{(\theta_0/2)},
\er
 where 
\br
 \label{g11}
\g  &=& \frac{2 \sqrt{2}}{\b^{3/2}} \Big\{ 2 \kappa^2 + M \b^2 \s (2 \b -1) \times \\
&& \Big[\( \rho_1^2 + \rho_2^2- 2 \rho_1 \rho_2 \cos{(\theta_1-\theta_2)}\) \cos{\theta_0} -  \rho_1^2 \cos{(2\theta_1-\theta_0)} +  \rho_2^2 \cos{(2\theta_2-\theta_0)}  \Big]\Big\} ^{1/2},\nonumber
\er 
such that $\kappa$, in terms of $M, \theta$ and  $\s$,  provided in  (\ref{mk2}) must be taken into account.

On the other hand, the slope of the kink $\Phi$ (\ref{kink1}) at the origin $x=0$ becomes
\br
\label{slope0}
\mu=\frac{2}{\beta}  \kappa \, \tan{(\theta_0/2)}.
\er
In (\ref{mk2})  one has $\kappa = f(\theta_0, \s) M$ with $f(\theta_0,\s)$ being a proportionality factor; so,  the slope $\mu$ of the kink $\Phi$ defined in (\ref{slope0}) is directly proportional to the  parameter $M$ which appears in the Lagrangian (\ref{atm1}) as the fermion mass in the free fermion limit of the model ($\b \rightarrow 0$).  Whereas, $\g$ in (\ref{g11}) carries the effect of all the parameters of the model, including the spinor parameters. Therefore,  one can argue that the slope $\mu^A$ in (\ref{slope1}) measures  the degree of back-reaction of the spinor components on the kink. Notice that the slopes satisfy 
\br
\label{ratio1}
\frac{\mu^A}{\mu} = \frac{\gamma}{\kappa}.
\er
Let us analyze the strong coupling sector ($\b \rightarrow $ large) of the model by considering the kink solutions. So, for finite $M$ (remember that $\kappa \sim M$)  and fixed $f(\theta_0, \s)$ the slope of the kink $\Phi$ (\ref{slope0}) in the limit $\b \rightarrow$ large,   tends to zero i.e.  $\mu \rightarrow 0$. However, it is possible to obtain a kink  $\Phi^{A}_{DSG}$ with finite slope in  (\ref{slope1}) by maintaining $\frac{\g}{\b}=$ finite in the limit $\b \rightarrow $ large.  In fact, taking into account (\ref{g11}) and the relationship between the parameters $\rho_1$ and $\rho_2$ in (\ref{rho12}), in the limit ($\b \rightarrow $ large) one can consider $\rho_1 \rightarrow $ large, in order to get 
\br
\label{strong22}
\mu^A &=&  [  \frac{\g}{\b} ] 2 \tan{\frac{\theta_0}{2}} \\
& \sim & \sqrt{M}\,  \frac{\rho_1}{\b} = \mbox{finite},\label{strong221}
\er     
where a set of values for $\{\theta_1, \theta_2, \theta_0\}$ and the parity $\s$ have been fixed. 

In view of the above discussions, one can argue that the double SG  model  (\ref{multiSG}) describes the strong coupling sector of the model defined by the Lagrangian (\ref{atm1}) for kink solitons of the type (\ref{kinkA}). In order to see qualitatively the back-reaction of the fermion $\psi$ on the soliton we plot the solitons of types $\Phi $ and $\Phi_{DSG}^{A} $ in the Fig.  12  for weak  and  strong  coupling constant parameter $\b$ and  the particular value $\theta_0 = \frac{2 \pi}{3}$. 
 
{\bf Type B}

The model  (\ref{multiSG})-(\ref{c3}) also exhibits the next  kink and bounce type solutions \cite{jhep2}. Consider
\br
\Phi_{DSG}^{B} = \frac{2}{\b} \arctan{\Big[  (\frac{1}{d}) \frac{1+ h \exp{(2 k x)}}{\exp{(k x)}} \Big]} .
\er
The kink type solution arises for the relationships
\br
 C_2 = -\frac{1}{4 \b}(\frac{k^2}{1+\cos{\theta_0}}),\,\,\,\,\, h =  -\frac{1}{4}(\frac{d^2 \cos{\theta_0}}{1+\cos{\theta_0}}),\,\,\,d=2 \sqrt{1+ \sec{\theta_0}}.
\er
For the solution $\theta_0 = \frac{\pi}{3}$ in (\ref{thetas0}) one gets  $d =  2 \sqrt{3}$. Then, one has the kink-type solution
\br
\Phi_{DSG}^{B} = \frac{2}{\b} \arctan{\Big[   \frac{1}{\sqrt{3}}  \sinh{k x}\Big]}.
\er  
This solutions interpolates the values $\pm \pi$ of the set of points in the vacua (\ref{vcdsg}). 

Moreover, for $\theta_0 =\frac{2\pi}{3}$ one has $d =  2$ and the next bounce-like solution
\br
\Phi_{DSG}^{bounce}  = \frac{2}{\b} \cosh{(k x)}.
\er

\section{Atiyah-Patodi-Singer-type formula and non-zero modes}
\label{sec:topnzV}

The one and two soliton solutions of the ATM model (eqs. of motion (\ref{521c})-(\ref{525}) with $V=0$) have been  obtained in \cite{npb1} and  verified that they satisfy the Noether and topological currents equivalence (\ref{NT}). However, their space-time dependence were given through the exponentials of type $\exp{[\g_j (x- v_j t + c_j)]}$, such that $j=1$ ($j=1,2$) for one soliton (for two-solitons). Therefore, in order to have static solutions one needs to set $v_j =0$ and this procedure will spoil the appearance of the phase factor $e^{-i E t}$ with $E \neq 0$, which is needed to define localized fermionic modes with
non-zero eigenvalues as in the Anzats performed in (\ref{zetas}). So, the N-soliton(antisoliton) ($N=1,2$) bound state spinor solutions obtained in \cite{npb1} correspond to zero-mode states ($E=0$), with  the scalar N-soliton (N-antisoliton) possessing the topological charges $Q_{top} = \pm 1$ and $Q_{top} = \pm 2$, for $N=1$ and $N=2$, respectively. 

Next, let us examine the currents relationship (\ref{currents1}) for the modified ATM model, i.e.  for $V\neq 0$, and for spinor bound states with non-zero eingenvalues, i.e.  $E\neq 0$. Let us introduce a potential function $\L$ for the current component ${\cal J}^{0}_{nonl}$ in (\ref{nonl11}), such that 
\br
\label{nonl111}
\pa_x \L \equiv {\cal J}^{0}_{nonl} (x).
\er 
For the kink sector $\Phi(x)=\Phi_{kink}(x)$ with topological charges $\pm \frac{\theta_0}{\pi}$ obtained in sec. 4, one can write 
\br
\label{potch11}
\L(x) &=& \zeta \, \Phi(x) \pm  \nu  \frac{\sqrt{16 \kappa^2 - 3 [ \pa_x \Phi - 2 \sqrt{3}\kappa/3]^2}}{(1+2 \cos{\theta_0}) \pa_x \Phi + 2 \sqrt{3}\kappa}\\
\zeta & \equiv & \frac{\b }{16 \kappa^2} (\csc{\theta_0})^3 [2 a_1 -(2 a_0+3 a_2) \cos{\theta_0} + a_2 \cos{(3 \theta_0)}]\label{zeta1}\\
\nu &\equiv& \frac{1}{8 \kappa^2}  (\csc{\theta_0})^2 [a_0 - a_1 \cos{\theta_0} + a_2 \cos{(2 \theta_0)}]. 
\er
Taking into account the asymptotic values $\pa_x \Phi(\pm \infty)=0$, from (\ref{potch11}) one can get
\br
\label{del1}
 \D \L &\equiv & \L(+\infty)-\L(-\infty) \\
\label{del2}
          &=& \zeta \, [\Phi(+\infty)-\Phi(-\infty)]\\
           &=&\frac{2\pi}{\b} \, \zeta \,Q_{topol},\label{del3}
\er
where the relationships (\ref{topol1})-(\ref{topol11}) have been used in the last line. Notice that the parameter $\zeta$ (\ref{zeta1}) depends on the variables $\theta_0, \kappa, \b$ and the set  $\{a_0, a_1, a_2\}$ defined in  (\ref{a00})-(\ref{a22}). Since these variables depend on  the potential variables $A_1, A_2$, and $ A_3$ defined in (\ref{aa1})-(\ref{aa3}) one can argue that the parameter $\zeta$ carries the effect of the potential $V$ and the spinor parameters. So, one concludes that $\D \L $ will provide a correction to the $U(1)$ charge due to the  combined effects of the potential $V$ and the spinor parameters, which are interrelated in a nontrivial way. In order to see this, let us rewrite the charge densities equivalence relationship (\ref{currents1}) as
\br
\label{jj00}
J^{0}(x) = - \frac{2}{\b} \pa_x \Phi(x) + \pa_{x} \L(x),
\er
where $j^0_{top}(x)$ in  (\ref{topological}) and ${\cal J}^{0}_{nonl} (x)$ in (\ref{nonl111}) have been used in order to write the first and second terms in the r.h.s. of the equation (\ref{jj00}), respectively.  So, from (\ref{jj00}) the $U(1)$ charge becomes
\br
\label{equivch1}
{\cal Q} &\equiv & \int_{-\infty}^{+\infty}  dx\, J^{0}(x)\\
\nonumber
               &=& -\frac{2}{\b}  [\Phi(+\infty)-\Phi(-\infty)]  + [\L(+\infty)-\L(-\infty)] \\
                &=&  \frac{2\pi}{\b} \(\frac{2}{\b} - \zeta \)  |Q_{topol}|,
\label{equivch2}
\er
where the relationships (\ref{del1})-(\ref{del3}) have been used. So, one can argue that  in the modified ATM model the equivalence between the Noether and topological charges still holds provided that the factor of proportionality is shifted as  $ \(\frac{2}{\b} - \zeta \) $, in which the effect of the potential is carried by the parameter $\zeta$. Actually, for $\zeta=0$ one recovers the ATM model relationship.  Then, we have verified analytically  a classical version of a formula of the Atiyah-Patodi-Singer-type (\ref{jj00}) for the modified ATM model  incorporating the effect of the potential $V$ and for the non-zero mode bound states.

In the next section we will pursue localized fermionic modes with zero eigenvalues for the both cases separately, i.e.   $V = 0$ and $V \neq 0$. We will show below that among those zero-modes there is a subset of solutions satisfying the Majorana condition (\ref{majorana}) provided that $V=0$.   

\section{Zero-mode bound states}
\label{sec:zeromode1}

Regarding the zero modes, notice that the both conditions (\ref{majo1}) and (\ref{majo2}) in order to have Majorana fields satisfy trivially the relationship (\ref{consist1}). Then,  we do not  have at our disposal a relationship between the parameters analogous to the equation (\ref{rho12}),  which has been used in the calculations to arrive at the fermion spectra for $E\neq 0$. Consequently, the constructions above performed to arrive at the equations (\ref{km})-(\ref{thetas120}) and (\ref{products1})-(\ref{products2}) do not hold anymore. So, one must search for new bound state solutions to the whole system of equations   (\ref{5211})-(\ref{phizetas1}) specialized to the zero-modes $E=0$.
 
So, from this point forward in this section we set $E=0$ in the system of equations (\ref{5211})-(\ref{5241}). So,   let us set   $E=0$ in the equivalent system of equations (\ref{RL}). So, one has in components
\br
\label{ma10}
\frac{d }{dx}  \xi_{0R}&=& -M e^{- i \b \Phi} \xi_{0L} ,\\
\frac{d }{dx}  \xi_{0L} &=& -M e^{i \b \Phi} \xi_{0R},
\label{ma1c0}
\er  
where $\xi_{0R} = \xi_3 + i \xi_4$ and $\xi_{0L} = \xi_1 + i \xi_2$.  Notice that the system (\ref{ma10})-(\ref{ma1c0}) describes a Majorana fermion coupled to a scalar field, provided that either the identification (\ref{majo1}) or the one in (\ref{majo2}) is taking into account. In fact, the case with $\b = 0$ describes a free massive Majorana system \cite{fabrizio, gogolin}. Recently, the model (\ref{ma10})-(\ref{ma1c0}) has been considered as the simplest continuum model of a one-dimensional superconducting fermionic symmetry-protected topological (SPT) phase in condensed matter such that the mean-field superconducting (SC) pairing potential is represented by the field $\Phi$  \cite{chua, udupa}.  

Moreover, the static  second order equation derived from (\ref{phizetas1}) for the kink $\Phi$ and fermion components $\xi_0^{T} = \( \xi_{0R},\, \xi_{0L}\)$ becomes  
\br
\label{sec20}
-\frac{d^2 \Phi}{dx^2}  + M \b [ \xi^{\star}_{0R} \xi_{0L} e^{i \b \Phi}  + \xi^{\star}_{0L} \xi_{0R} e^{-i \b \Phi}] + V'(\Phi) =0.
\er
Then, in order to find the kinks and zero-mode fermion bound states  one must solve the system of equations (\ref{ma10})-(\ref{sec20}).  However, in the following we will construct those solutions by solving a system of first order integro-differential equations. Notice that  the integro-differential equation (\ref{integrodif}) has been written for any bounded static configuration associated to an arbitrary value of the parameter $E$, and in particular, it must hold for the zero-modes with $E=0$. So, let us consider the identity (\ref{integrodif}) written for the field $\xi_0$ and the kink $\Phi$
\br
\label{id00}
\frac{1}{\b} \frac{d \Phi}{dx} + \frac{1}{2} (\xi^{\star}_{0R} \xi_{0R} + \xi^{\star}_{0L} \xi_{0L}) = \int_{-\infty}^{x} d\hat{x} \, V'(\Phi).
\er 
Remarkably, one can show by direct computation that the system of first order equations (\ref{ma10})-(\ref{ma1c0}) together with the first order integro-differential eq. (\ref{id00}) imply the second order differential eq. (\ref{sec20}). So, a solution of the first-order system of eqs. comprising (\ref{ma10})-(\ref{ma1c0}) and  (\ref{id00}) will also be a solution of the system of eqs. (\ref{ma10})-(\ref{sec20}).
This property has also been used in the preceding sections  in order to find  the fermion-kink solitons for the non-vanishing spectrum, i.e. for $E\neq 0$.

Moreover, the identity (\ref{majoxi}) for $E=0$ and upon integration in the whole line and written for the $\xi_0$ spinor components becomes 
\br
\label{nor11}
 \int_{-\infty}^{+\infty} dx  [   (\xi^{\star}_{0R})^{'} \xi'_{0R}  +  (\xi^{\star}_{0L})^{'} \xi'_{0L} ] - M^2 \int_{-\infty}^{+\infty} dx \(   \xi^{\star}_{0R} \xi_{0R}  +  \xi^{\star}_{0L} \xi_{0L} \)  =0.
\er  
The bound state solutions must  satisfy the last identity, as we will show below.
    
Next, in order to find the zero-mode fermion bound states coupled to  kink-type  solitons we resort to the tau function approach applied to the system of first order equations (\ref{ma10})-(\ref{ma1c0}) and the first order integro-differential equation  (\ref{id00}). Let us assume the following Ansatz 
\br
\label{t010}
e^{i \b \Phi/2}  &=& e^{i \a_0} \, \frac{\tau_0}{\tau_1},\\
\label{tr0}
\xi_{0R}  &=& \frac{\tau_R}{\tau_0} ,\\
\label{tl0} 
\xi_{0L}  &=& \frac{\tau_L}{\tau_1}.
\er
So,  replacing the relationships (\ref{t010})-(\ref{tl0})  into the equation  (\ref{ma10}) and (\ref{ma1c0}), respectively,  one can write the following equations in terms of the tau functions
\br
\label{tr10}
e^{-2 i \a_0} M \tau_1 \tau_L - \tau_R \tau'_0 + \tau_0 \tau'_R &=&0,\\
\label{tl10}
e^{2 i \a_0} M \tau_0 \tau_R - \tau_L \tau'_1 + \tau_1 \tau'_L &=&0.
\er 

Let us choose the parameters $\b_j$ in the potential (\ref{pot1}) to be $\b_j = \b\, j , (j=1,2,3)$. We will see that this choice of the $\b_j$ parameters provides a bound state solution. Likewise, substituting the tau functions  in (\ref{t010})-(\ref{tl0})  into (\ref{id00})  one gets 
\br
\nonumber
- \frac{\b^2 \tau_L \tau_R  + 2 i \tau_1 \tau'_0 - 2i \tau_0 \tau'_1 }{\b^2\tau_0 \tau_1} + \frac{i \b}{2} \int^{x}_{-\infty}d\hat{x}\, 
\(\frac{e^{2 i \a_0} \, \tau_0^4 - e^{-2 i \a_0} \tau_1^4}{\tau_0^6 \tau_1^6}\) \times \\
\Big[ 3 A_3 e^{4 i \a_0} \tau_0^8 + 2 A_2 e^{2 i \a_0} \tau_0^6 \tau_1^2 + (A_1+3A_3)   \tau_0^4 \tau_1^4+2 A_2 e^{-2 i \a_0} \tau_0^2 \tau_1^6 + 3 A_3 e^{-4 i \a_0} \tau_1^8    \Big]=0.\label{id10}
\er
Notice that the non-local terms above will vanish for $A_j =0,\, j=1,2,3$, which corresponds to $V=0$. Below, we will search for special tau functions such that the whole non-local term above, once the $x-$integral is performed, maintain similar tau function structure as the first local term in (\ref{id10}), i.e. a rational expression of bi-linear terms in the tau function variables. 

So, let us consider the following tau functions, 
\br
\label{tau01}
\tau_0 &=& 1+ i \, \epsilon \, e^{2 \kappa x},\,\,\,\,\,
\tau_1 = 1 - i \, \epsilon \, e^{2 \kappa x},\,\,\,\,\,\,\,\,\,\epsilon = \pm 1,\\
\label{taurl}
\tau_L &=& \zeta_1  e^{i \a_1} e^{\kappa x},\,\,\,\,\,\,\,\tau_R = \zeta_2  e^{i \a_2} e^{\kappa x},
\er
 where the parameters $\kappa, \zeta_a, \a_a\, (a=1,2)$ are real numbers.  

Then, for the above tau functions the fields take the forms  
\br
\label{k10i}
\Phi &=& \frac{4}{\b} \arctan{ \Big[ \epsilon \, e^{2 \kappa x  }\Big]} + \frac{2 \a_o}{\b},
\\
\label{mr0}
\xi_{0R} &=& \frac{1}{\sqrt{2}} \zeta_2   \Big[\mbox{sech}(2 \kappa x)\Big]^{1/2}\, e^{ i \Theta_R},\,\,\,\,
\Theta_R(x) = - \arctan{ \Big[\( \frac{    \sinh{[\kappa x- \frac{1}{2} \epsilon \log{(\tan{\a_2})}]  }    }{   \cosh{[\kappa x + \frac{1}{2} \epsilon \log{(\tan{\a_2})}]} }   \)^{\epsilon}\Big] }\\
\label{ml0}
\xi_{0L} &=& \frac{1}{\sqrt{2}} \zeta_1   \Big[\mbox{sech}(2 \kappa x)\Big]^{1/2}\, e^{i \Theta_L},\,\,\,\,
\Theta_L(x) = - \arctan{ \Big[ \( \frac{    \sinh{[\kappa x- \frac{1}{2} \epsilon \log{(\tan{\a_1})}]  }    }{   \cosh{[\kappa x +\frac{1}{2} \epsilon  \log{(\tan{\a_1})}]} }   \)^{\epsilon} \Big]}
\er
Notice that $\Phi$ in (\ref{k10i}) represents a kink type static soliton interpolating the asymptotic values $\frac{2}{\b}(\a_0 + \epsilon \pi)$ and $\frac{2 \a_0}{\b}$. Whereas, the fermion components (\ref{mr0})-(\ref{ml0}) define a zero-mode fermion bound state, i.e. a solution for $E=0$. 

Regarding the parity properties of this solution, the parameters must satisfy certain relationships implied by (\ref{parity0})-(\ref{parity}).  So, from the kink solution (\ref{k10i}) one has that 
\br
\label{a0}
\a_o  =  \epsilon \frac{\pi}{2},\,\,\,\,\epsilon = \pm 1,
\er
in order to satisfy the parity condition (\ref{parity0}). In fact, for $\epsilon = -1\, (\epsilon = +1)$ one has a kink (antikink) solution. This kink $\Phi$ resembles the sine-Gordon static soliton; however,  the kink in (\ref{k10i}) interpolates two vacua  defined by  
\br 
\label{phi0}
\b \Phi (\pm \infty)= \pm \phi_0,\,\,\,\,\,\,\,\, \phi_0 \equiv \pi,
\er
in order to satisfy $\Phi (-x) = - \Phi(x)$ in  (\ref{parity0}).  So, following the definition in (\ref{topol11}) one gets the topological charges 
\br
\label{topo11}
Q_{topol} = \pm 1.
\er

Moreover, the parity conditions  (\ref{parity}) for the fermion components furnish  the  next parameter relationships
\br
\label{a2m1}
\a_2 &=& \a_1 + n \pi, \,\,\,\ n \in \IZ,\\
\zeta_2 &=&  [\frac{\s}{\cos{(n \pi)}}]\,\zeta_1.
\label{r12s}
\er

The tau functions (\ref{tau01})-(\ref{taurl}) will be substituted into the above system of equations (\ref{tr10})-(\ref{id10}) and the condition (\ref{nor11}) must be imposed in order to find the parameters relationships and study the properties of the solutions for $\Phi$ and $\xi_0$. 

Firstly, let us substitute the tau functions (\ref{tau01})-(\ref{taurl}) into the pair of equations (\ref{tr10})-(\ref{tl10}). So, one gets  the next relationships
\br
\label{a01}
\a_o &=& - \frac{\pi}{2} (m+n),\,\,\,\, m \in \IZ.\\
\label{km1}
\kappa &=& - \epsilon  M.
\er   
By comparison of the above result (\ref{a01}) with (\ref{a0}) one gets the next condition for the integers $m$ and $n$
\br
m+n  =  - \epsilon.
\er 
For later convenience let us write the slope of the kink (anti-kink) (\ref{k10i}) at $x=0$ for the parameter $\kappa $ defined in (\ref{km1}). So, one has 
\br
\label{slkinkmajo}
\mu^{K} \equiv - \epsilon \frac{4 M}{\b}.
\er

Notice that the identity (\ref{nor11}) is a condition for the zero-mode solutions and it does not depend explicitly on the form of the potential $V$. So, one can derive a general relationship for the parameters by substituting (\ref{tr0})-(\ref{tl0}) in the form  (\ref{tau01})-(\ref{taurl})  into   (\ref{nor11}). The relevant terms of (\ref{nor11}) become
\br
 \int_{-\infty}^{+\infty} dx  [   (\xi^{\star}_{0R})^{'} \xi'_{0R}  +  (\xi^{\star}_{0L})^{'} \xi'_{0L} ] &=& \frac{\pi}{4}\, \kappa\, [\zeta_1^2 + \zeta_2^2]\\
\int_{-\infty}^{+\infty} dx \(   \xi^{\star}_{0R} \xi_{0R}  +  \xi^{\star}_{0L} \xi_{0L} \)&=& \frac{\pi}{4} \, \frac{1}{\kappa} \, [\zeta_1^2 + \zeta_2^2]
\er
Substituting the above quantities into (\ref{nor11}) one gets again (\ref{km1}).

So,  the kink and anti-kink of type (\ref{k10i}) host the zero-mode bound states (\ref{mr0})-(\ref{ml0}). In order to compare qualitatively the kink  (\ref{k10i}) to the ones of the section \ref{sec:theta0} we have plotted the kink (\ref{k10i}) in the Fig. 8. Notice that its asymptotic values are $\pm \pi$  as in (\ref{phi0}), whereas the kinks of  sec. \ref{sec:theta0} interpolate the asymptotic values $\theta_0 = \pm \frac{\pi}{3}, \pm \frac{2\pi}{3}$, respectively.
 
Below, in order to verify the equation (\ref{id10}) let us consider  the two cases:  For vanishing, $V=0$, and non-vanishing, $V \neq 0$, potentials, respectively. 

\subsection{Zero-mode bound states for $V=0 $}

In this case let us consider  $A_j =0  \rightarrow V=0$. So, by substituting (\ref{tr0})-(\ref{tl0}) in terms of the tau functions  (\ref{tau01})-(\ref{taurl})  into   (\ref{id10}) with $A_j=0$ one derives the following  expression for  the parameter $\kappa$
\br
\label{k10}
\kappa = - \epsilon \, \b^2  \(\frac{\zeta_1^2+\zeta_2^2}{16}\).
\er
Next, let us examine the Majorana zero-modes for $V=0$.

\subsubsection{Majorana zero-modes for $V=0$}
 \label{sec:majoV0}

In order to have  a Majorana  spinor $\xi$ it has been imposed the reality condition  (\ref{majorana}). It has provided the both cases (\ref{majo1})-(\ref{majo2}). So, we consider below each case separately.

{\bf Case I.}
The  conditions (\ref{majo1}) provides the next relationships
\br
\label{r12a}
 \zeta_2 = \(\frac{\sin{\a_1}}{\sin{\a_2}} \)\zeta_1\\
\a_1 + \a_2 = q \pi ,\,\,\,\,\,\,  q \in \IZ. 
\label{a2p1}
\er
Then, from (\ref{a2m1}) and (\ref{a2p1}) one can get
\br
\label{a12}
\a_1 = \frac{\pi}{2} (q-n),\,\,\,\,\a_2 = \frac{\pi}{2} (q+n).
\er
Inspecting the relationship (\ref{r12a}) one must have $\sin{\a_1} \neq 0$ and $\sin{\a_2} \neq 0$; therefore, in (\ref{a12}) the integers $n$  and  $q$ must satisfy
\br
\label{c1q}
q & \equiv & s+r +1\\
n & \equiv & s-r ,\,\,\,\,\,\,\, s, r \in \IZ. \label{c1n}
\er 
In fact, these identities make the quantities $( q\pm n )$ to be odd integers for any set of integers $ s, r \in \IZ$, implying $\sin{(\a_j)} \neq 0 \,(j=1,2)$.
 
{\bf Case II.}
The  conditions (\ref{majo2}) provides the next relationships
\br
\label{r12aii}
 \zeta_2 = \(\frac{\cos{\a_1}}{\cos{\a_2}} \)\zeta_1\\
\a_1 + \a_2 = p \pi ,\,\,\,\,\,\,  p \in \IZ. 
\label{a2p1ii}
\er
Then, from (\ref{a2m1}) and (\ref{a2p1ii}) one can get
\br
\label{a12ii}
\a_1 = \frac{\pi}{2} (p-n),\,\,\,\,\a_2 = \frac{\pi}{2} (p+n).
\er
Likewise, inspecting the relationship (\ref{r12a}) one must have $\cos{\a_1} \neq 0$ and $\cos{\a_2} \neq 0$; therefore, in (\ref{a12ii}) the integers $n$  and  $p$ must satisfy
\br
\label{c2p}
p & \equiv & s+r\\
n & \equiv & s-r ,\,\,\,\,\,\,\, s, r \in \IZ.\label{c2n}
\er 
In fact, these identities make the quantities $( p \pm n )$ to be even integers for arbitrary set of integers $\{s, r\}$, implying $\cos{(\a_j)} \neq 0 \,(j=1,2)$.

In summary, in order to have  a Majorana spinor $\xi$, in the both cases I and II above, the parameters must satisfy 
\br
\label{zeta122}
\zeta_1^2 = \zeta_2^2 =  \frac{8M}{\b^2}.
\er

In the Fig. 13 we plot some examples of the Majorana bound state inside the kink for the both cases I and II, respectively. 

\begin{figure}
\centering
\includegraphics[width=4cm,scale=1, angle=0,height=4.5cm]{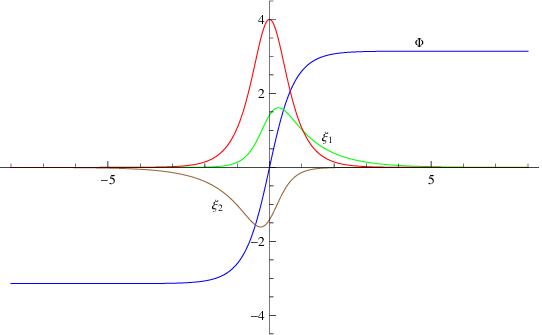}
\includegraphics[width=4cm,scale=1, angle=0,height=4.5cm]{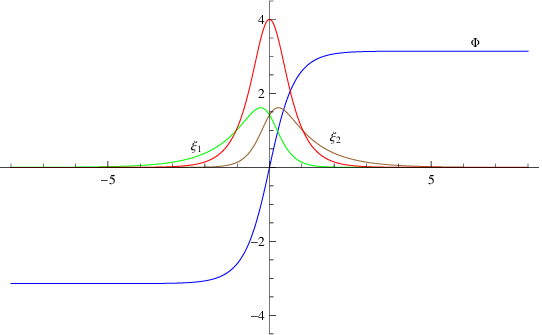}
\parbox{5in}{\caption{The kink $\Phi$ (blue) and the Majorana zero-mode components $\xi_{1}$ (green) and  $\xi_{2}$ (brown) as  functions of $x$. Left (Case I): For $\b=1,\,\a_0 = -\frac{\pi}{2},\,\zeta_1 = - \zeta_2=-\sqrt{8}, \kappa=M=1, \a_1 = \pi/2$. Right (Case II):  For $\b = 1,\,\a_0= -\frac{\pi}{2},\,\zeta_1 =  \zeta_2 = -\sqrt{8}, \kappa=M=1, \a_1 = \pi$. The red lines represent the Majorana density $(\xi_1^2+ \xi_2^2)$ vs $x$, in each case. }}
\end{figure}

\subsubsection{Fermion back-reaction and soliton-particle(Majorana) duality}
\label{duality11}

Here we follow an analogous approach considered in the sine-Gordon/massive Thirring mapping in order to exhibit the soliton/particle duality in the affine Toda model coupled to a Dirac fermion \cite{npb2, orfa1}. So, in our case the zero-mode fermion components can be written as functionals of the scalar field for the fermion and kink solutions of the type (\ref{mr0})-(\ref{ml0}) and (\ref{k10i}), respectively. 

Then, from (\ref{mr0})-(\ref{ml0}) and (\ref{k10i}) one has the next two cases, either for the soliton or the anti-soliton field.

1. Majorana coupled to a soliton: $\a_0 = - \frac{\pi}{2},\,\,\,\epsilon =1$
\br
\label{ms2}
\xi_{0L}  &=& \frac{1}{2} \{\tan{[\frac{1}{4}(\b \Phi + \pi)]}\}^{1/2}  \zeta_1 e^{i \a_1}  [1 + i \, e^{\frac{i}{2} \b \Phi}]\\
\xi_{0R}  &=& \frac{1}{2} \{\tan{[\frac{1}{4}(\b \Phi + \pi)]}\}^{1/2}  \zeta_1 e^{i \a_2}  [1 - i \, e^{-\frac{i}{2} \b \Phi}],\,\,\,\,\,\,\,\,\,\zeta_1^2 = \frac{8M}{\b^2}.
\label{ms22}
\er

2. Majorana coupled to an anti-soliton: $\a_0 = \frac{\pi}{2},\,\,\,\epsilon = -1$
\br
\label{ms1}
\xi_{0L}  &=& \frac{1}{2} \{\cot{[\frac{1}{4}(\b \Phi + \pi)]}\}^{1/2}  \zeta_1 e^{i \a_1}  [1 - i \, e^{\frac{i}{2} \b \Phi}]\\
\xi_{0R}  &=& \frac{1}{2} \{\cot{[\frac{1}{4}(\b \Phi + \pi)]}\}^{1/2}  \zeta_1 e^{i \a_2}  [1 + i \, e^{-\frac{i}{2} \b \Phi}],\,\,\,\,\,\,\,\,\,\,\zeta_1^2 = \frac{8M}{\b^2}.
\label{ms11}
\er
 
Notice that these mappings between the Majorana components and  the scalar field soliton are non-perturbative by construction. Even though they are classical, they resemble the bosonization formulas at the quantum level \cite{fabrizio}. In fact, following the ideas of  \cite{npb2, orfa1}, this mapping can be regarded as  the strong coupling (DSG) $\leftrightarrow$ weak coupling (Majorana fermion model) sectors of the  affine Toda coupled to Majorana fermion.  

In the case of the affine Toda model coupled to (charged) Dirac field (eq. (\ref{atm1})  with $V=0$) it has been obtained an analogous classical mapping  relating  the particle (fermion) to a soliton (boson field) such that they  describe the strong (sine-Gordon) and weak  (massive Thirring) coupling sectors of the model, respectively \cite{npb2}. At the quantum level, the bosonized Toda model coupled to Dirac field  has been discussed in relation to a confinement mechanism in the context of the equivalence between the sine-Gordon and massive Thirring models  \cite{npb1}.
    
Remarkably, one can use the above mappings in order to decouple the strong and weak  coupling  sectors of the model at the level of the equations of motion. First, let us substitute the mappings (\ref{ms2})-(\ref{ms22}) and (\ref{ms1})-(\ref{ms11}) into the soliton(anti-soliton)-fermion interaction terms of the model in (\ref{sec20}). So, one has
\br
\label{compot}
\xi^{\star}_{0R} \xi_{0L} e^{i \b \Phi}  + \xi^{\star}_{0L} \xi_{0R} e^{-i \b \Phi} &\equiv& - \epsilon \zeta_1 \zeta_2 \cos{(\a_1 - \a_2)} \Big[ \sin{(\b \Phi)}  +   \sin{(2 \b \Phi)} \Big], \,\,\,\,\, \epsilon = \mp 1.\\
&=& - \epsilon \zeta_1^2 (-1)^{N} \Big[ \sin{(\b \Phi)}  +   \sin{(2 \b \Phi)} \Big],\,\,\,\,\,\,N \equiv r+n-s, \label{compot1}
\er
where the last relationship holds for the both cases I and II defining the Majorana fermions in the section \ref{sec:majoV0}. Notice that the r.h.s. of (\ref{compot1}) is non-vanishing   for any set of values of the integers $r,n,s$ in the both cases I and II, (\ref{c1q})-(\ref{c1n}) and (\ref{c2p})-(\ref{c2n}), respectively. 

Since the r.h.s. of (\ref{compot1}) is non-vanishing one can conclude that the back-reaction of the Majorana fermion on the soliton is non-vanishing and depends on the special values of the fermion parameters  $\zeta_1^2=\zeta_2^2 \equiv \frac{8 M}{\b^2}$ and the non-vanishing factor $(-1)^{N},\, N \in \IZ$, determined by the relationship between the phases $\a_1$ and $\a_2$ of the Majorana fermion components in the cases I and II presented in the sec.  \ref{sec:majoV0}. Recently, for the scattering of the iso-doublet femions on an {\sl external} sine-Gordon field and considering a small fermion-soliton coupling, it has been reported a vanishing of the back-reaction of the Majorana fermion on the soliton \cite{loginov}.   

Therefore, substituting the identity (\ref{compot1}) into (\ref{sec20}) (with $V=0$) one has 
\br
\label{sec201}
-\frac{d^2 \Phi}{dx^2}  - 8   (-1)^{N}  \epsilon \frac{M^2}{\b}\Big[ \sin{(\b \Phi)}  +   \sin{(2 \b \Phi)} \Big]=0.
\er
This is the static equation of the double sine-Gordon model (DSG) defined for the scalar field $\Phi$ with the relevant potential 
\br
\label{potdsg1}
V_{DSG}(\Phi)  =   8   (-1)^{N} \epsilon \frac{M^2}{\b} \Big[ \cos{(\b \Phi)}  + \frac{1}{2}  \cos{(2 \b \Phi)} - \cos{(\frac{2\pi}{3})}  - \frac{1}{2}  \cos{(\frac{4\pi}{3})} \Big].
\er
The Fig. 14 shows the plot of this potential with vacua $\b \Phi = \frac{2 \pi}{3}\, n\,(n \in \IZ)$. In fact, the kink type solutions of the DSG interpolate these vacuum points. Notice that the points $\b \Phi = n \pi, (n \in \IZ )$ of the potential (\ref{potdsg1}) are critical points which do not support stable soliton type solutions.  On the other hand, the soliton  (\ref{k10i})  hosts  the Majorana fermion and it interpolates the vacuum values  $\Phi = \pm \pi$. So, the vacua of the effective potential (\ref{potdsg1}) do not support kink-type solitons of the type (\ref{k10i}).
  
In analogy to the mapping between the strong/weak coupling sectors of the affine Toda model as the soliton/particle correspondence \cite{npb1} and the discussion in section \ref{sec:backreaction} of the present paper on the strong coupling sector of the model (\ref{atm1}), one can think of the DSG model (\ref{sec201}) as describing the strong coupling sector of the model (\ref{atm1}) with $V=0$ with equations of motion (\ref{ma10})-(\ref{sec20}) and (\ref{sec20}) in the zero-mode sector.

It is interesting to analyze the kink-like solutions of the DSG model. The model  (\ref{sec201}) exhibits a kink solution of the type  
\br
\label{kink12}
\Phi_{DSG}(x) &=& \frac{2}{\b}  \arctan{\Big[\tan{(\frac{\pi}{3})}  \,\, \tanh{(\sqrt{3} M x)}\Big]},  
\er
provided that $\epsilon =1 $ and $N$ is an even integer in the potential (\ref{potdsg1}).
In fact, this kink interpolates the vacua $\pm 2\pi/3$ and it resembles to one of the kinks plotted in the Fig. 8 with slope at $x=0$ provided by 
\br
\label{sldsg1}
\mu^{DSG} = 6 \, \frac{M}{\b}.
\er 
Notice that the kink (\ref{k10i}) which hosts the Majorana zero-mode interpolates a different  vacua, i.e.  $\pm \pi $. However, at the origin $x=0$ one can compute the ratio between the slopes of the DSG kink (\ref{sldsg1}) and the slope of the SG-type  kink  in (\ref{slkinkmajo}). So, one has  
\br
\label{ratio2}
\frac{\mu^{DSG}}{\mu^{K}} = \frac{3}{2}.
\er
Remarkably, in the strong coupling sector of the model the back-reaction effect of the zero mode (\ref{mr0})-(\ref{ml0}) on the kink (\ref{k10i}) reflects by shifting its vacuum points and rising the kink slope according to (\ref{ratio2}).    
  
Moreover, the DSG model  (\ref{sec201}) possesses a bounce-like solution which  interpolates between the vacuum value $\pi$ and the point $\Phi_{DSG} = \frac{\pi}{2}$ and then it comes back. Since $\pi$ is a false vacuum position this type of solution is not related to any stable particle in the quantum theory (see \cite{jhep2} and references therein). 
 
 The DSG model has been mapped to various spin models at the quantum level, and in particular, its relation with the deformed quantum Ashkin-Teller model has been considered \cite{fabrizio}. In these developments the bosonized Majorana fermion field plays a fundamental role. 

Moreover, the sine-Gordon type models play an important role in superconductors, especially multi-band superconductors including layered high-temperature superconductors. The vortex fluxes and the associated kink-like  solutions of the double SG appear  in the study of multi-band superconductors in the Ginzburg-Landau  theory approach to superconductivity (see e.g. \cite{yanasigawa, aguirre} and references therein).

Next, from the expressions (\ref{t010})-(\ref{tl0}) and (\ref{tau01})-(\ref{taurl}) in order to achieve the Majorana fermion sector corresponding to the weak coupling sector of the model one can get the relationship 
\br
\label{fermioni}
e^{i \b \Phi}  = - \(\frac{\xi_{0L}}{\xi_{0R}}\)^2.
\er
So, substituting  (\ref{fermioni}) into  (\ref{ma10})-(\ref{ma1c0}) one gets the non-linear field equations for the Majorana field  components
\br
\label{maeq1}
\frac{d }{dx}  \xi_{0R}&=& \mp M \frac{\xi_{0R}^3}{|\xi_{0L}|^2},\\
\frac{d }{dx}  \xi_{0L} &=& \mp M \frac{\xi_{0L}^3}{|\xi_{0R}|^2},
\label{maeq2}
\er
where the $\mp$ signs stand for the case I and II, respectively. This system of equations can be regarded as describing the weak coupling sector of the model  (\ref{ma10})-(\ref{sec20}).

Moreover, the equation (\ref{id00}) for $V=0$ can be written as
\br
\label{topden}
\frac{1}{\b} \frac{d \Phi}{dx} = - \frac{1}{2} (\xi^{\star}_{0R} \xi_{0R} + \xi^{\star}_{0L} \xi_{0L})
\er
In fact, the mapping (\ref{ms1})-(\ref{ms11}) [or  (\ref{ms2})-(\ref{ms22})] satisfies the above relationship. This is reminiscent of the topological and Noether currents equivalence for the charged fermion \cite{npb1}. However, in the real field Majorana fermion case the equivalence  (\ref{topden}) can be regarded as an equivalence of the topological charge density to the Majorana state density.  

Our result above is analogous to the mapping of the Liouville equation for a real scalar plus an auxiliary field to a relativistic  non-linear and non-local model for real Majorana type spinor fields in 1+1 dimensions in the context of the dual $\s $ model \cite{pashaev1}. Moreover, it is known the equivalence of a gapless Majorana fermion and the critical point of a real scalar field in $(1+1)d$. The diccionary being  ${\cal L}_{f}^{(1+1)d} = \bar{\xi}\g_\mu \pa_\mu \xi  \leftrightarrow {\cal L}_{b}^{(1+1)d} = (\pa_\mu \phi)^2 + \phi^4$, which holds in the infrared region (see \cite{max} and references therein). In fact, this relationship can be understood as a bosonization of $(1+1)d$ free Majorana fermions. In this context, the Lagrangian formulation and  the bosonization of the self-interacting Majorana  fermion system  (\ref{maeq1})-(\ref{maeq2}) deserves a further study.

On the other hand, Majorana bound states in topological superconductors are pursued due to their potential applications in topological quantum computation. Since there are  other entities as topological bound states in such systems, a so-called Majorana polarization quantity $P_M$ has been defined in order to
discriminate between proper Majorana fermions from  other bound states \cite{vigliotti}. Analogously, let us define the quantity
\br
\label{polm}
P_{M} = \frac{{\cal N}}{{\cal D }},
\er
where
\br
\label{ndm}
{\cal N} \equiv \int_{-\infty}^{+\infty} dx \, 2 \,\xi_1 \xi_2;\,\,\,\,\,\,{\cal D} \equiv \int_{-\infty}^{+\infty} dx \,  [\xi_1^2+ \xi_2^2].
\er

For the above soliton (antisoliton) solutions and the host Majoranas one finds
\br
{\cal N} = \mp  \frac{1}{4} \zeta_1^2  \int_{-\infty}^{+\infty} dx  [1 + \cos{(\b \Phi)} ],\,\,\,\,{\cal D } = \frac{1}{4} \zeta_1^2  \int_{-\infty}^{+\infty} dx  \cos{(\frac{\b \Phi}{2})},
\er
where the $\mp$ signs corresponds to the cases I and II, respectively.
Since $ \Phi(\pm \infty) = \pm \pi/\b $ one has
\br
\label{polm1}
P_M = \mp \frac{\pi}{4} \approx  \mp  0.7854.
\er
\begin{figure}
\centering
\includegraphics{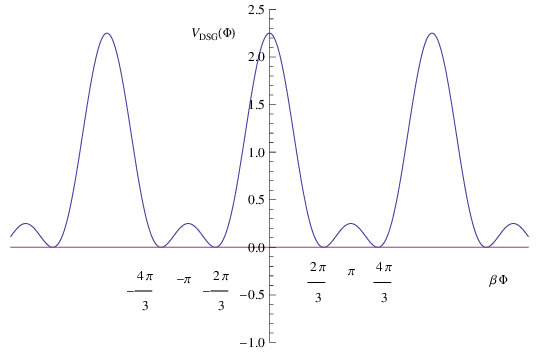}
\parbox{5in}{\caption{The plot of the DSG potential $V(\Phi)$ in (\ref{potdsg1}). Notice the minima for $\b \Phi = \frac{2 \pi}{3}\, n\,(n \in \IZ)$.  The values $\Phi = \pm \pi$ correspond to unstable  critical points.}}
\end{figure}
In order to compare these values of $P_M$  with the behavior of the Dirac spinor obtained  in the above sections let us define the quantity 
\br
\label{pold}
P_{D} = \frac{{\cal N}_D}{{\cal D }_D},
\er
where
\br
\label{ndd}
{\cal N}_D \equiv \int_{-\infty}^{+\infty} dx \, \,(\xi_1 \xi_2 - \xi_3 \xi_4);\,\,\,\,\,\,{\cal D}_D \equiv  \frac{1}{2} \int_{-\infty}^{+\infty} dx \,   [\xi_1^2+ \xi_2^2+\xi_3^2+ \xi_4^2],
\er
where the Dirac spinor components have been considered.
 
Notice that the definition (\ref{pold})-(\ref{ndd}) reduces to the relationship (\ref{polm})-(\ref{ndm}) once the identification (\ref{majo2}) (or (\ref{majo1}) with ${\cal N}_D \rightarrow - {\cal N}_D$ ) is taken into account in order to define a Majorana fermion.   
Computing the above integrals for the explicit Dirac spinor solutions and setting $\theta_1=\theta_2=\theta_o$ one has 
\br
\label{pold1}
P_{D} = - \frac{(\sin{\theta_o})^2}{\theta_o}\,\,\, \rightarrow \,\,\,\, |P_{D}| <  0.725.
\er
By comparison of the relations (\ref{polm1}) and (\ref{pold1}) it can be shown, for any value of the parameter $\theta_o$, that the quantity $|P_{D}|$ is less than the Majorana polarization quantity $|P_M|$ (i.e. $|P_D| < |P_M|$). 
 
\subsection{Zero-mode bound states for $V \neq 0 $}
\label{sec:A1n0}

In this case one can derive parameter relationships by substituting (\ref{tr0})-(\ref{tl0}) in terms of the tau functions  (\ref{tau01})-(\ref{taurl})  into   (\ref{id10}). So, one has the relationships
\br
\label{k11}
\kappa &=&  - \frac{1}{32} \epsilon \, \Big\{ \b^2  \(\zeta_1^2+\zeta_2^2\) + \b^{3/2}\sqrt{\b \(\zeta_1^2+\zeta_2^2\)^2 + 256 A_1}\Big\},\,\,\,\b > 0,\\
\label{A110}
A_1 & = & \epsilon  \frac{\kappa}{4 \b^3} \Big[ 16  \epsilon  \kappa + \b^2 (\zeta_1^2+\zeta_2^2)\Big],\\
\label{A23}
A_2 & = & A_3 = 0.
\er
Notice that for  $A_1=0$ the relationship (\ref{k11}) for $\kappa$ reduces to the previous one in (\ref{k10}). Taking into account (\ref{km1}) and (\ref{k11}) one can get the expression for $A_1$ 
\br
\label{A11}
A_1 = \frac{M}{4 \b^3} \Big[ 16 M - \b^2 (\zeta_1^2+\zeta_2^2)\Big].
\er  
The equation (\ref{id10}) is satisfied for the values in (\ref{k11})-(\ref{A23}) without any additional restrictions for the parameters, except that the phase angle $\a_0$ must be $\epsilon \frac{\pi}{2}$ as in (\ref{a0}). So, one has the potential
\br
\label{potMZM}
V = A_1 \cos{(\b \Phi)}.
\er
So, the kinks of type (\ref{k10i}) host the zero-mode bound states (\ref{mr0})-(\ref{ml0}) in the model (\ref{atm1}) for the sine-Gordon potential $V$ provided in (\ref{potMZM}). Recently, for an external (without back-reaction) sine-Gordon kink which interacts weakly with fermionic iso-dublets through a Yukawa interaction term it has been discussed the appearance of Majorana zero-modes, which do not exert back-reactions on the kink  in this approximation, i.e.  the fermion-kink interaction term in the Majorana sector vanishes  \cite{loginov}.  Whereas, the non-vanishing back-reaction of the Majorana zero-modes on the kink of our model has been discussed in sec. 7.1.2. Moreover,  one can  show the vanishing of the potential (\ref{potMZM}) provided that the Majorana condition (\ref{zeta122}) is imposed on the Dirac fermion zero-mode parameters, viz., $A_1 = 0$ in (\ref{A11}) taking into account  (\ref{zeta122}). 

\subsubsection{Atiyah-Patodi-Singer-type formula and the zero modes}   
\label{sec:topozmV}

Following analogous construction to the one in sec. 6, let us examine the currents relationship (\ref{currents1}) for the modified ATM model, i.e.  for $V\neq 0$, and for spinor bound states with zero eingenvalue, i.e.  $E=0$. Let us define the potential  $\Theta$ associated to the current component ${\cal J}^{0}_{nonl}$ in (\ref{nonl}), such that 
\br
\label{nonl1110}
\pa_x \Theta \equiv {\cal J}^{0}_{nonl} (x).
\er
Therefore, in the soliton sector with topological charges $\pm 1$ provided in (\ref{k10i}) and the potential $V$ of type (\ref{potMZM})  one can write 
\br
\label{theta11}
\Theta(x) &=& \epsilon \frac{A_1}{2 \kappa^2}[ \b \Phi(x) - \epsilon \pi],\,\,\,\,\,\, \epsilon = \pm 1, 
\er
where $A_1$ is provided in (\ref{A11}). Therefore,  one has
\br
\label{del101}
 \D \Theta &\equiv & \Theta(+\infty)-\Theta(-\infty) \\
\label{del201}
          &=& \epsilon \frac{\b A_1}{2 \kappa^2}  [\Phi(+\infty)-\Phi(-\infty)]\\
           &=&   \frac{\epsilon \pi A_1}{\kappa^2}\,Q_{topol},\label{del301}
\er
where the relationship (\ref{theta11}) and the definition (\ref{topol11}) have been used. Notice that $A_1$ in (\ref{A11}) depends on the parameters $M, \b,\zeta_1, \zeta_2$. So, one concludes that $\D \Theta $ will provide a correction to the spinor charge due to the  combined effects of the potential $V$ and the spinor parameters. In order to see this, let us rewrite the charge densities equivalence relationship (\ref{currents1}) as
\br
\label{j00}
J^{0}(x) = - \frac{2}{\b} \pa_x \Phi(x) + \pa_{x} \Theta(x),
\er
where $j^0_{top}(x)$ in  (\ref{topological}) and ${\cal J}^{0}_{nonl} (x)$ in (\ref{nonl1110}) have been used in order to write the first and second terms in the r.h.s. of the equation (\ref{j00}), respectively.  So, from (\ref{j00}) the Dirac spinor charge becomes
\br
\label{equiv110}
{\cal Q} &\equiv & \int_{-\infty}^{+\infty}  dx\, J^{0}(x)\\
               &=&- \frac{2}{\b}  [\Phi(+\infty)-\Phi(-\infty)] + [\Theta(+\infty)-\Theta(-\infty)]\\
 \label{equiv1101}               &=&  \frac{2\pi}{\b} \(\frac{2}{\b} - \frac{\epsilon \pi \b A_1}{2\kappa^2} \)  |Q_{topol}|,
\er
where the relationships (\ref{del101})-(\ref{del301}) have been used. In this case also one can argue that the modified ATM model presents the equivalence between the Noether and topological charges provided that the factor of proportionality is changed to $ \(\frac{2}{\b} -  \frac{\epsilon \pi \b A_1}{2\kappa^2} \) $, such that the effect of the potential is encoded into the parameter $A_1$. So, we have verified analytically  a classical version of a formula of the Atiyah-Patodi-Singer-type (\ref{j00}) in the zero-mode bound state sector of the the modified ATM model which incorporates the effect of the potential $V$.

As a particular  case in  (\ref{A11}) one can see that the parameter $A_1$ vanishes if the parameters satisfy 
\br
\label{A1eq0}
 \zeta_1^2+\zeta_2^2 = \frac{16M}{\b^2}\,\,\,\,\,  \rightarrow\,\,\,\,\,\, A_1 =0.  
\er  
So, setting $A_1=0$ into (\ref{equiv1101}) one recovers the ATM model relationship ${\cal Q} = -\frac{4\pi}{\b^2} Q_{topol}$ for the special case (\ref{A1eq0}). Remarkably,  the special case $\zeta_1^2 = \zeta_1^2=\frac{8M}{\b^2}$, which is the Majorana fermion condition defined in (\ref{zeta122}), satisfies  the condition (\ref{A1eq0}), and so $A_1=0$, which amounts to the vanishing of the potential, $V=0$. So, one can write the next relationship for the Majorana state quantity ${\cal D}$ and the topological charge $Q_{topol}$ 
\br
\label{dentop1}
{\cal D} &=& \int_{-\infty}^{+\infty}  dx [\xi_1^2+ \xi_2^2]\\
 &=&  \frac{2\pi}{\b^2}  |Q_{topol}|.\label{dentop2}
\er  
As mentioned above, in the real field Majorana fermion case the equivalence  (\ref{topden}), which holds in the case $V=0$, can be regarded as an equivalence between the Majorana state density $(\xi_1^2+ \xi_2^2)$ and the topological charge density $j^0_{topol}$, which implies  the relationship (\ref{dentop2}). One can conclude that the kink (\ref{k10i}), with topological charges $\pm 1$, may host a Majorana fermion for the vanishing potential case only, viz., $V=0$, such that the non-local contribution to the Atiyah-Patodi-Singer-type formula carried by the potential $\Theta$ in the r.h.s. of (\ref{j00}) vanishes. 

\section{Numerical simulations}
\label{sec:numer1}

We have shown above that the system of equations (\ref{5211})-(\ref{phizetas1}) can be solved analytically and obtained various types of  kinks  which interpolate the vacua $\Phi(\pm \infty) = \pm \frac{\pi}{3}, \pm \frac{2\pi}{3}, \pm \pi$, respectively. The analytic kinks and their relevant spinor bound states, as well as the associated eigenvalues $E$ have been found. However, it is desirable to get analogous solutions, as well as for the eigenvalues $E$, such that  the parameter $\theta_0$ ($\Phi(\pm \infty) \equiv \pm \theta_0$) belongs to a continuous interval. A general analytical method is not known for arbitrary values of that parameter. Below we will use the so-called relaxation method  to numerically solve the system of equations; this method proceeds to find the solution by starting with a guess solution and searching for the true solution. In order to achieve a fast convergence, it is needed a good initial trial guess solution satisfying the same boundary conditions for the kink and the bound states as the real solutions. Thus, as suitable guess initial solutions we will choose the form of the analytic kink and fermionic bound states in (\ref{spinors12}) and (\ref{kink1}), provided that certain values of the parameters are assumed.

\begin{figure}
\centering
\includegraphics[width=2cm,scale=1, angle=0,height=6cm]{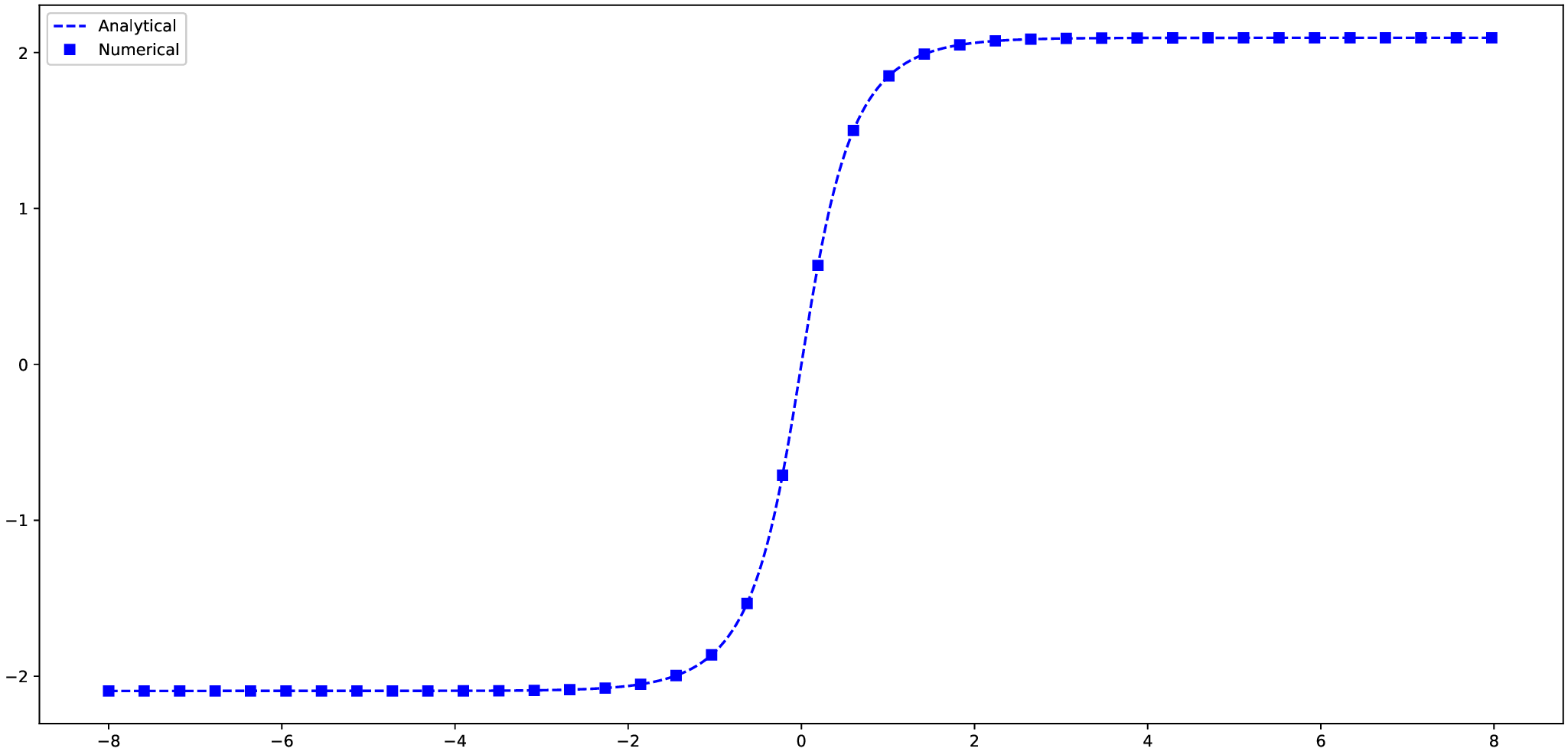}
\parbox{5in}{\caption{The comparison of the analytical (\ref{kink1}) and numerically generated kink $\Phi(x)$ for the parameters $\theta_0 = \frac{2\pi}{3},\, \b =1,\, \kappa = 1, M=1$.}}
\end{figure}

\begin{figure}
\centering
\includegraphics[width=2cm,scale=5, angle=0,height=7cm]{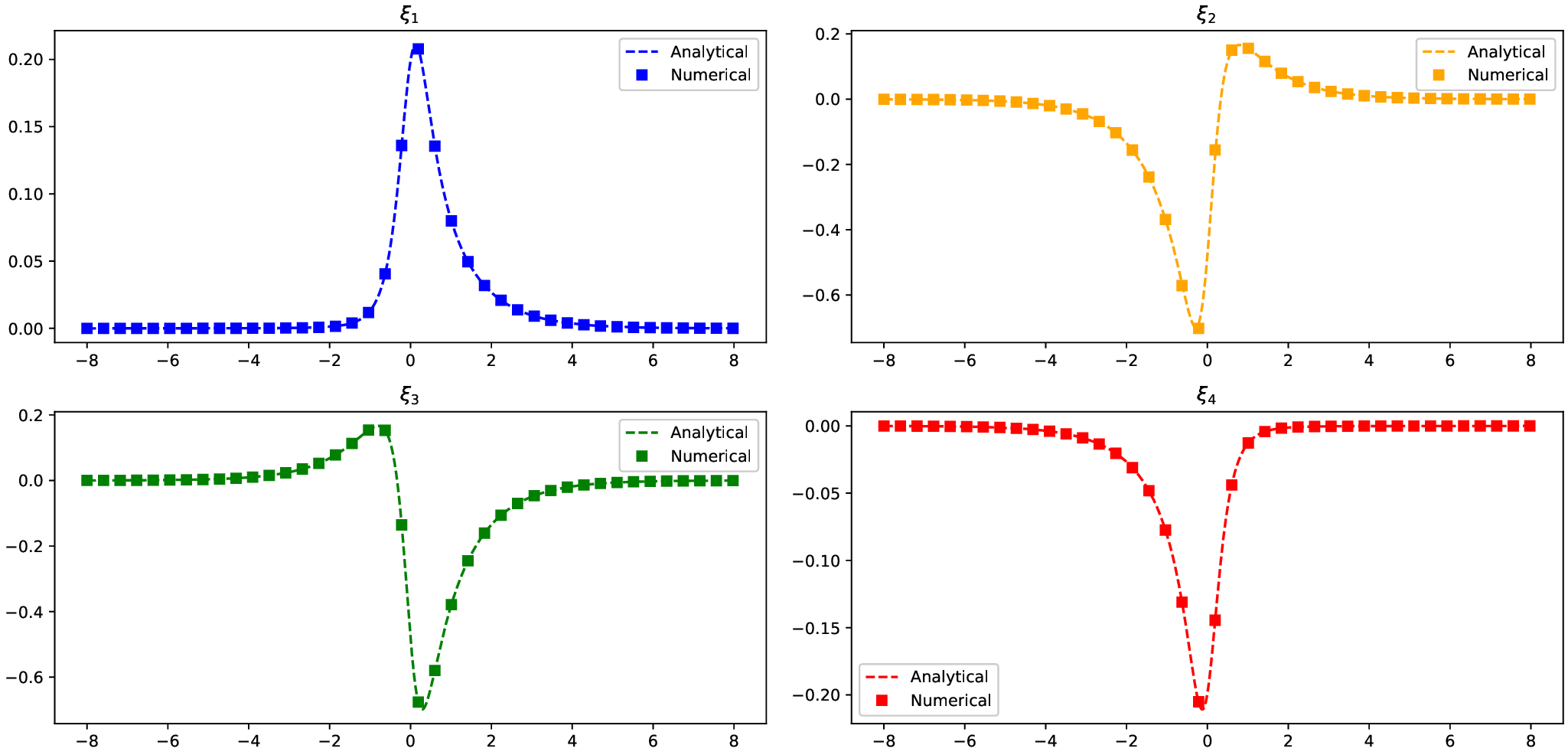}
\parbox{5in}{\caption{The comparison of the analytical bound state components of the fermion (\ref{spinors12}) and the numerically simulated ones as a function of $x$ for {\bf positive parity} $\s=+1$, It is hosted by the kink of Fig. 15. Plotted for $\rho_1 = 0.50, \rho_2 = 2.2,\, \theta_1 = 1.54,\, \theta_2 = 1.55,\,\kappa =1, \theta_0 = 2 \pi/3, M=1$. Notice that the spinor components satisfy the parity condition $\xi_{1}(-x)= - \xi_{4}(x)$ and $\xi_{2}(-x)= + \xi_{3}(x)$.}}
\end{figure}

\begin{figure}
\centering
\includegraphics[width=2cm,scale=1, angle=0,height=8cm]{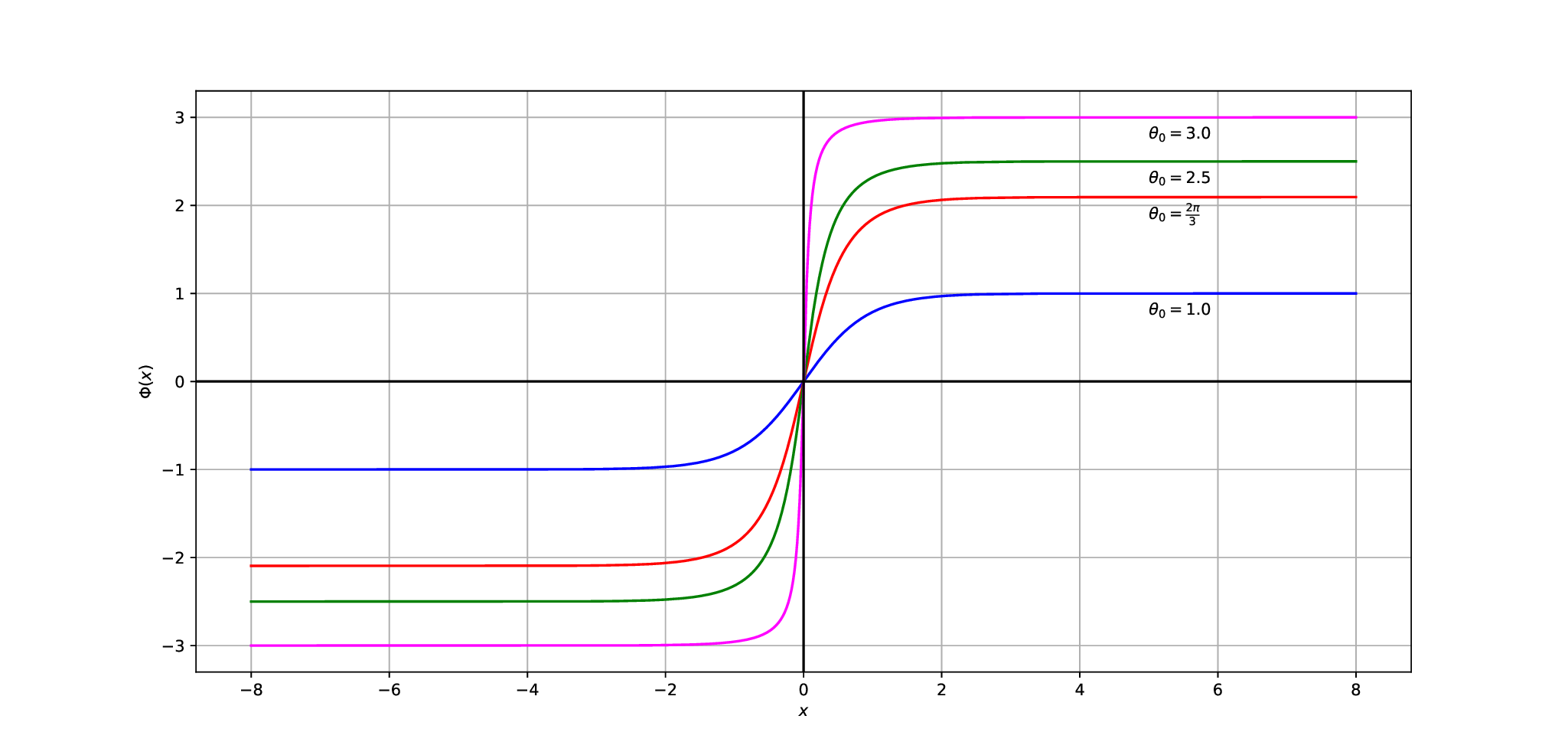}
\parbox{5in}{\caption{Numerically simulated kinks $\Phi(x)$ for the set of parameters $\theta_0 = \{ 1,\, \frac{2\pi}{3}, \,2.5,\, 3\}$, with blue, red, green and purple lines, respectively.}}
\end{figure}

\begin{figure}
\centering
\includegraphics[width=2cm,scale=5, angle=0,height=8cm]{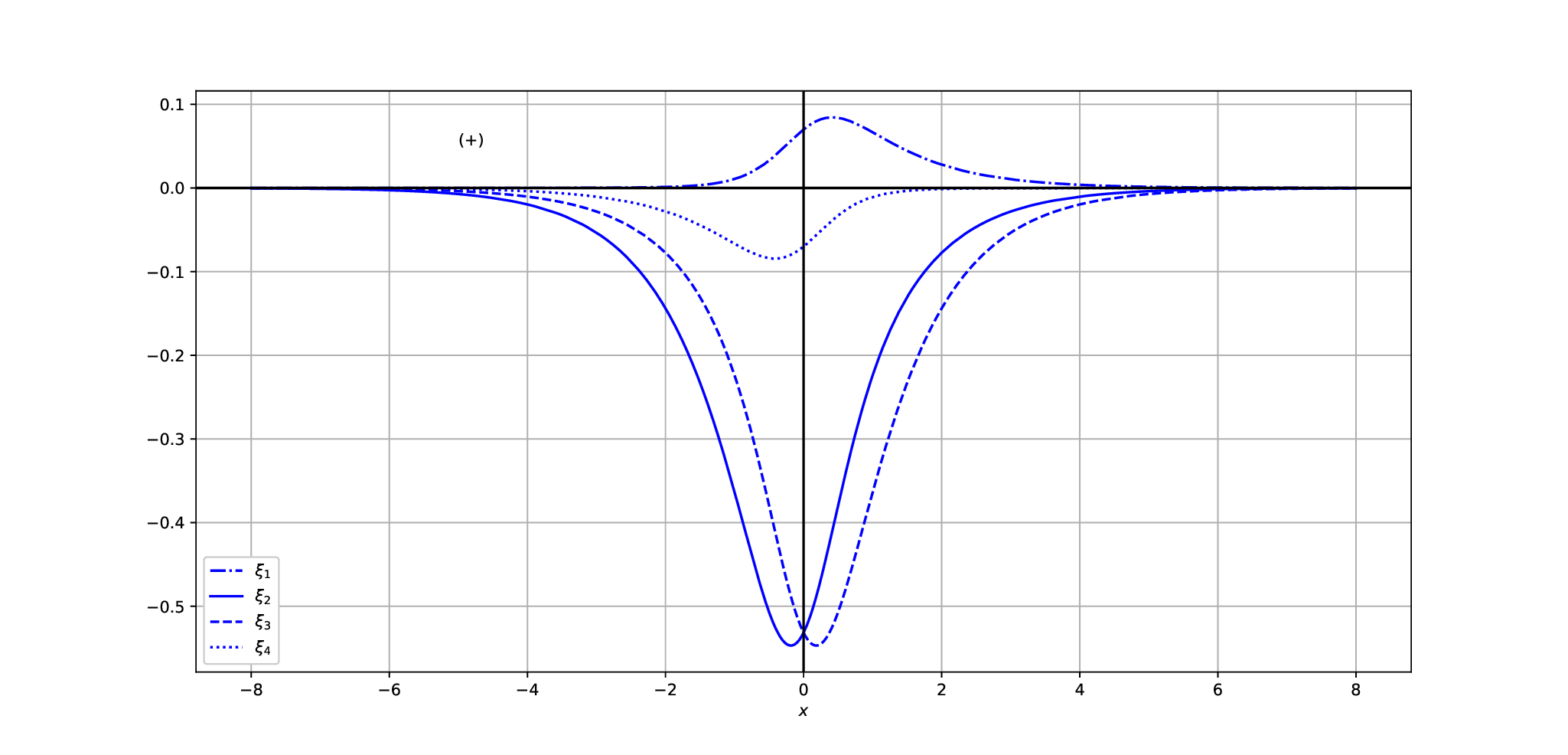}
\includegraphics[width=2cm,scale=5, angle=0,height=8cm]{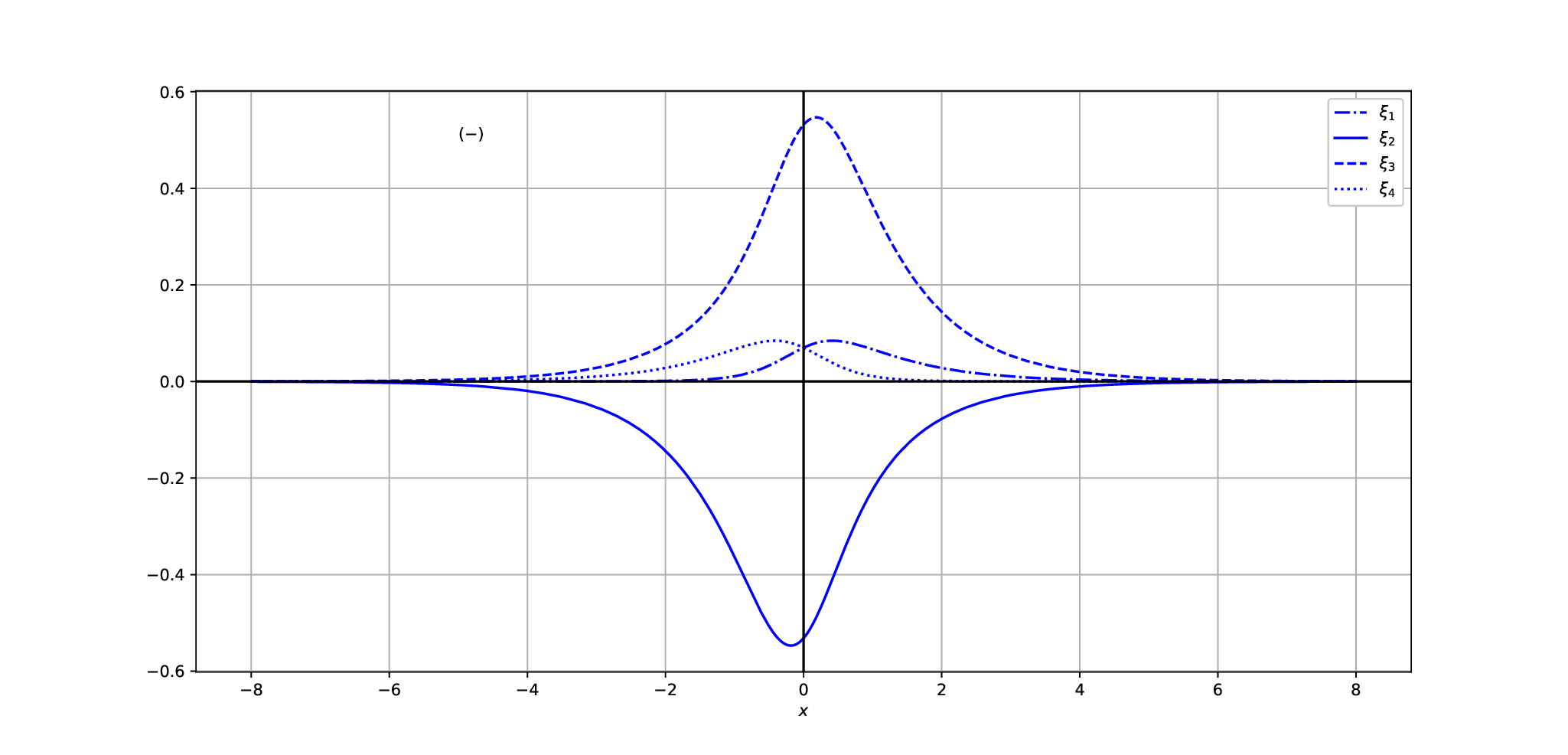} 
\parbox{5in}{\caption{Numerically simulated bound state components  as functions of $x$ for {\bf positive parity} $\s=+1$ (Top Fig.) and  {\bf negative parity} $\s=-1$ (Bottom Fig.). They are hosted by the kink with $\theta_0 =1.0$ (blue kink in the Fig. 17).}}
\end{figure}

\begin{figure}
\centering
\includegraphics[width=2cm,scale=5, angle=0,height=8cm]{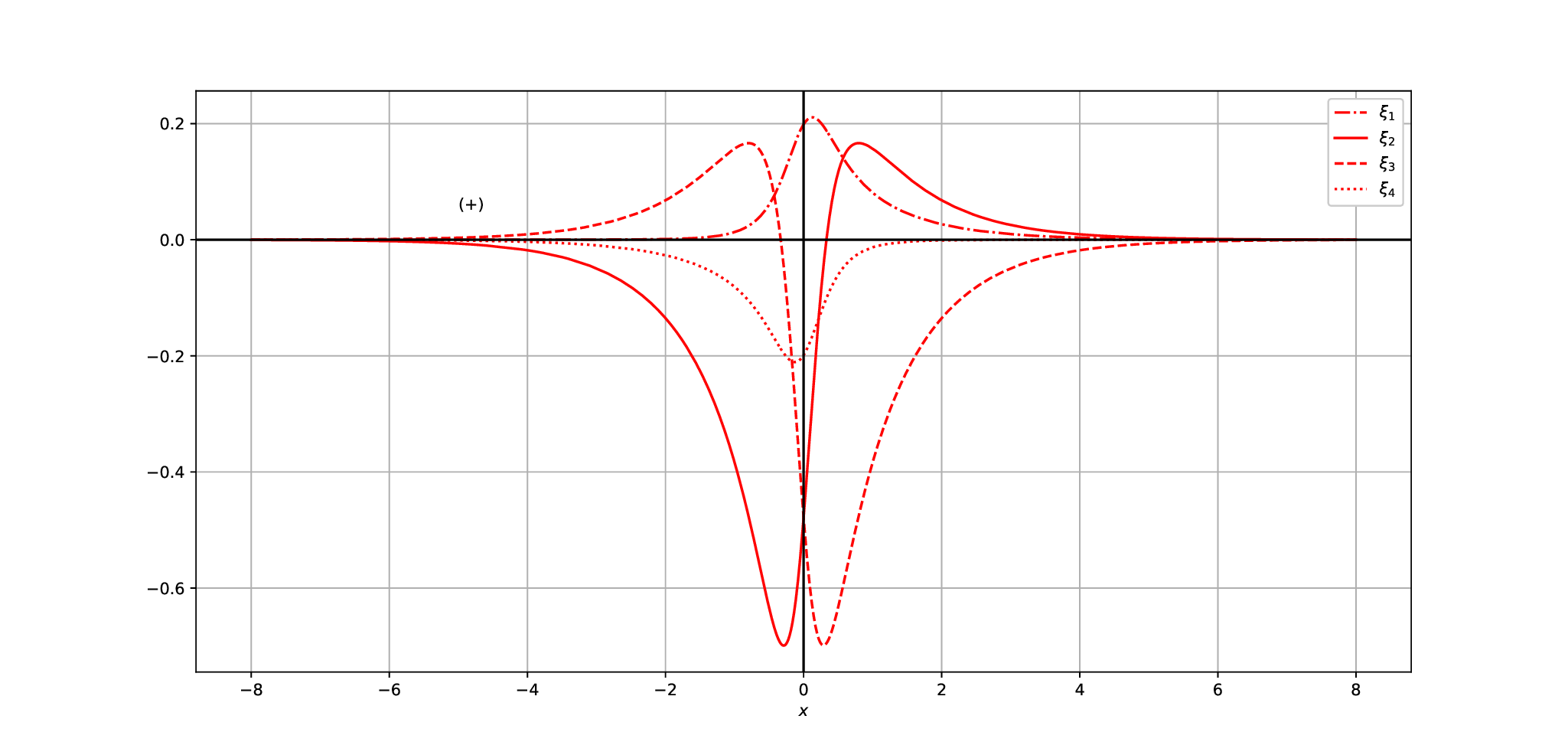}
\includegraphics[width=2cm,scale=5, angle=0,height=8cm]{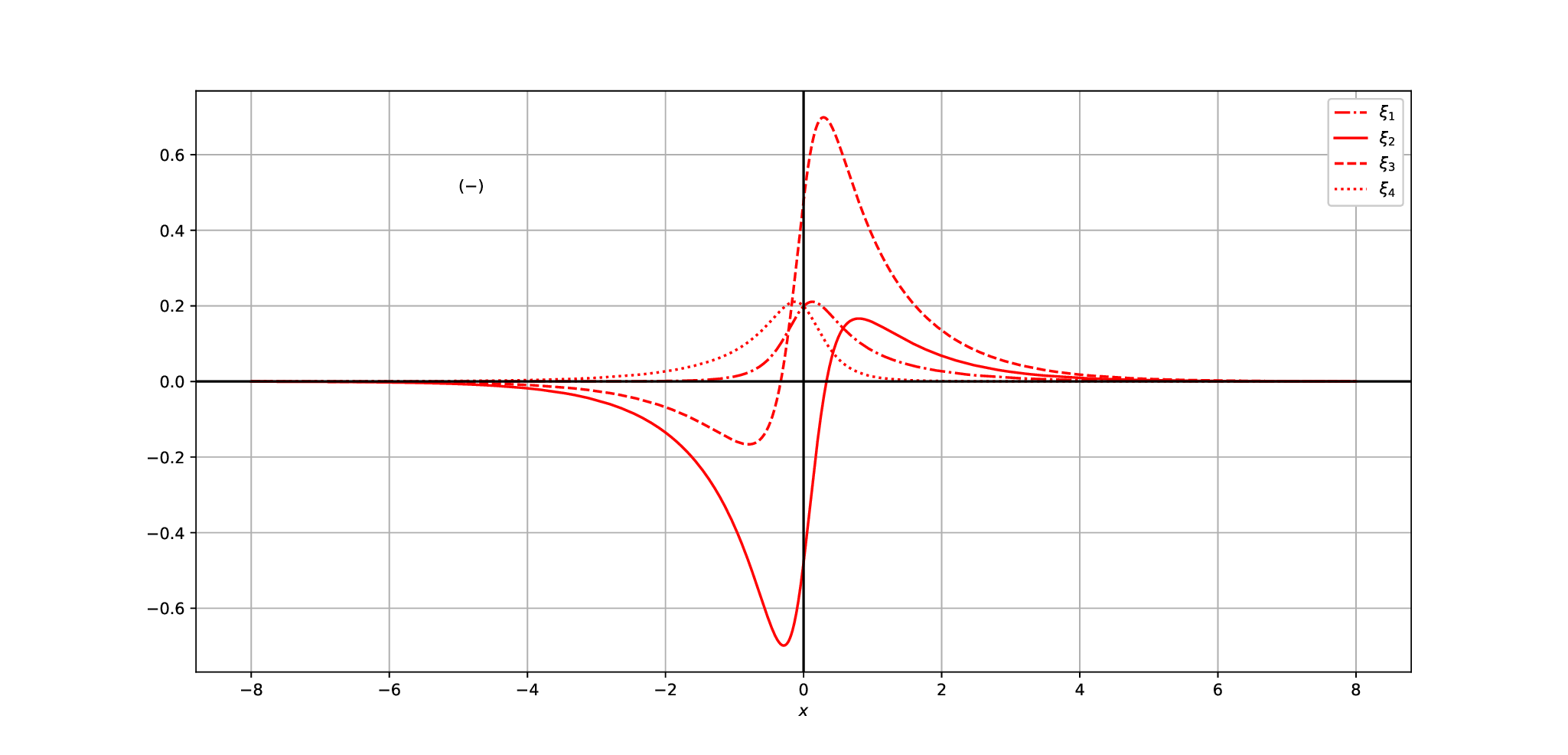} 
\parbox{5in}{\caption{Numerically simulated bound state components  as functions of $x$ for {\bf positive parity} $\s=+1$ (Top Fig.) and  {\bf negative parity} $\s=-1$ (Bottom Fig.). They are hosted by the kink with $\theta_0 = \frac{2 \pi}{3}$ (red kink in the Fig. 17).}}
\end{figure}

\begin{figure}
\centering
\includegraphics[width=2cm,scale=5, angle=0,height=8cm]{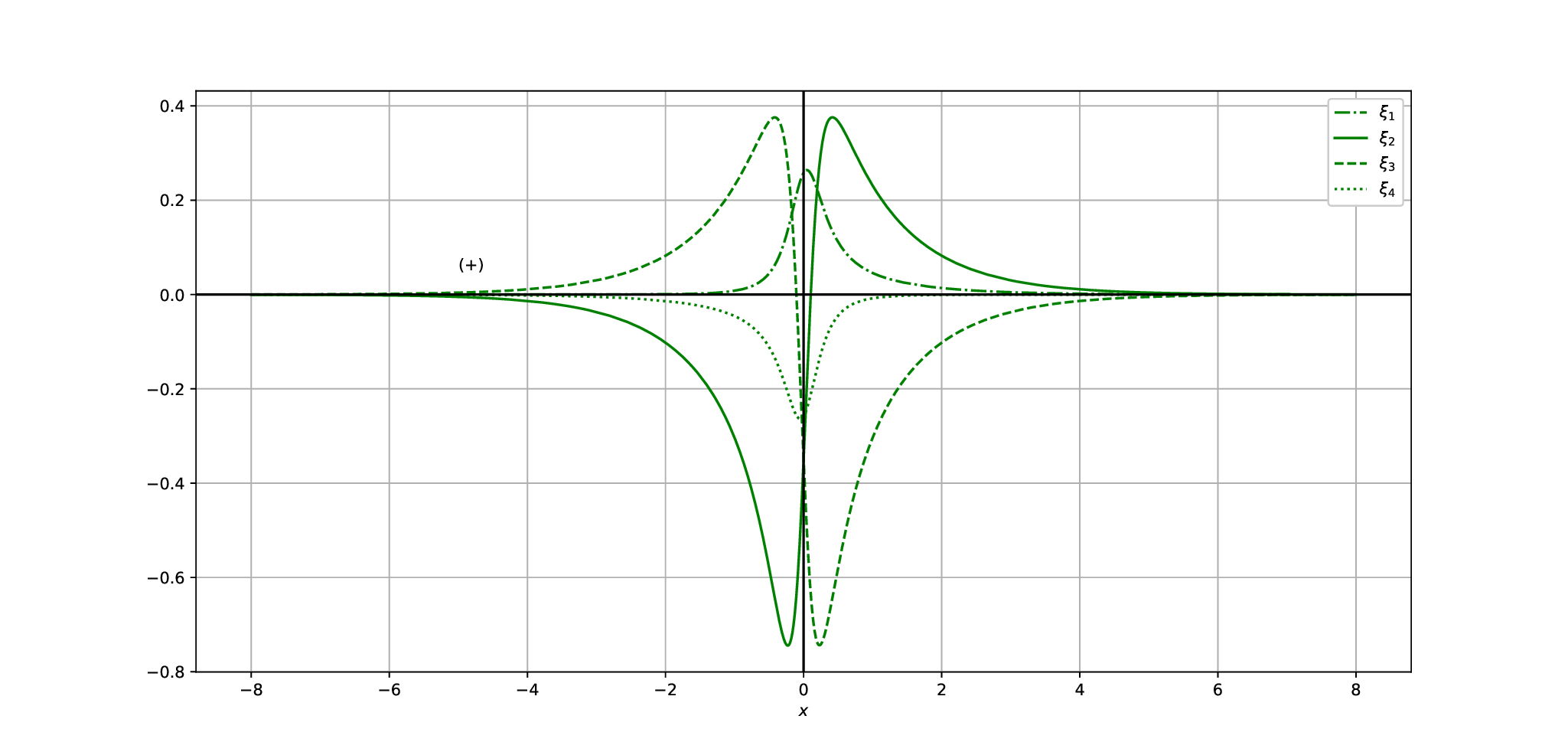}
\includegraphics[width=2cm,scale=5, angle=0,height=8cm]{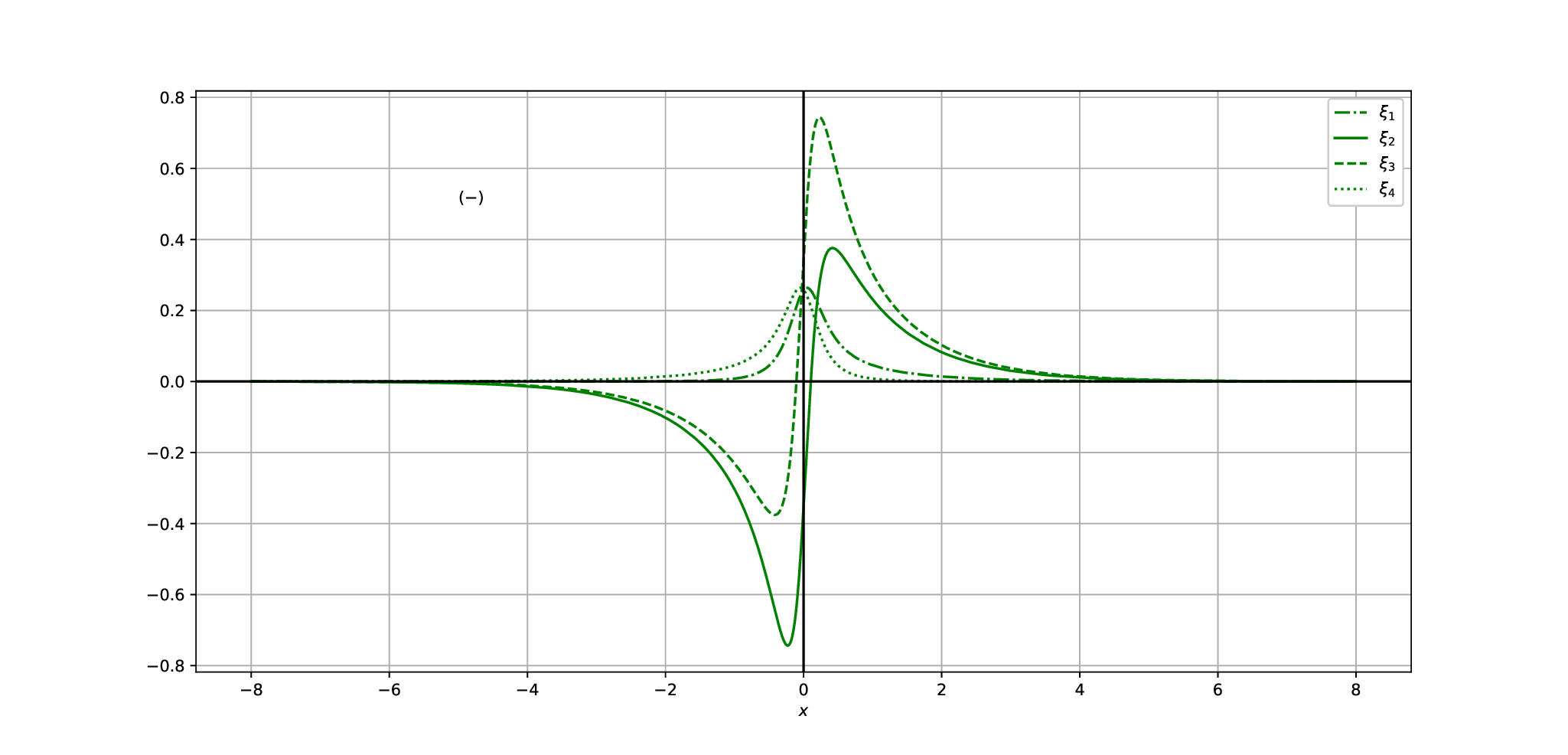} 
\parbox{5in}{\caption{Numerically simulated bound state components  as functions of $x$ for {\bf positive parity} $\s=+1$ (Top Fig.) and  {\bf negative parity} $\s=-1$ (Bottom Fig.). They are hosted by the kink with $\theta_0 =2.5$ (green kink in the Fig. 17).}}
\end{figure}

\begin{figure}
\centering
\includegraphics[width=2cm,scale=10, angle=0,height=8.5cm]{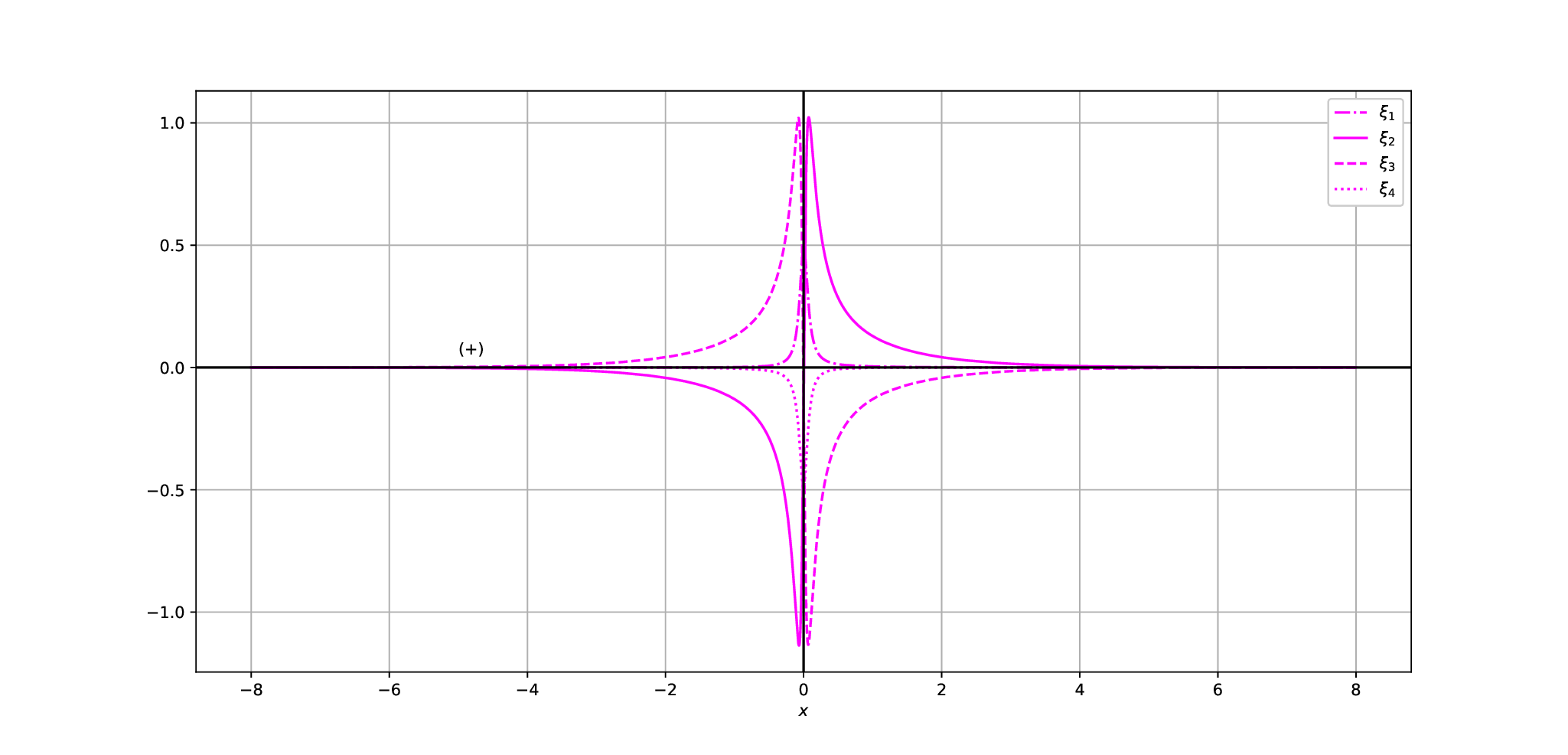}
\includegraphics[width=2cm,scale=10, angle=0,height=8.5cm]{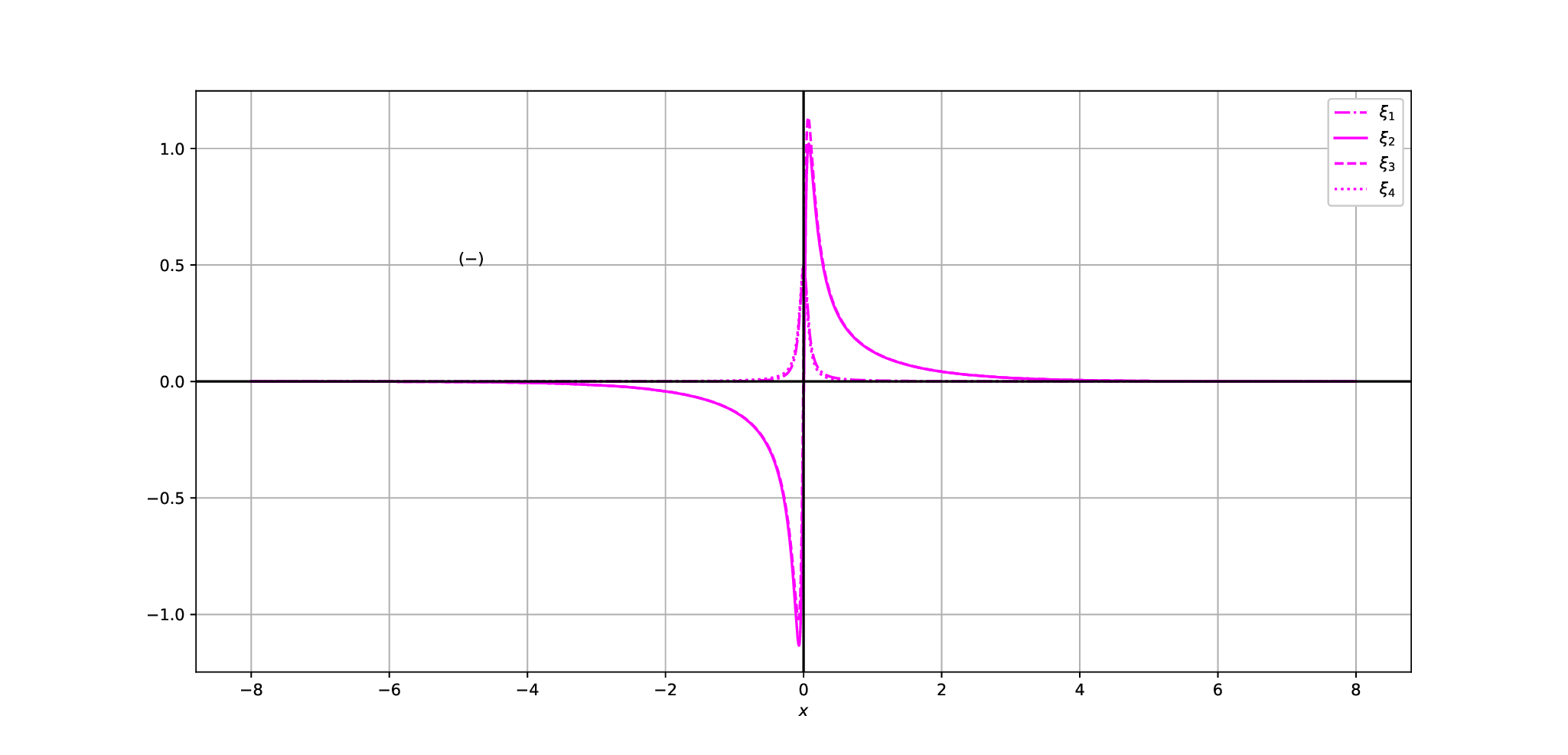} 
\parbox{5in}{\caption{Numerically simulated bound state components  as functions of $x$ for {\bf positive parity} $\s=+1$ (Top Fig.) and  {\bf negative parity} $\s=-1$ (Bottom Fig.). They are hosted by the kink with $\theta_0 =3.0$ (purple kink in the Fig. 17).}}
\end{figure}

Our aim is to obtain numerical solutions for the static version of the system of eqs. (\ref{5211})-(\ref{phizetas1}) corresponding  to the potential of type $V_1(\Phi)$ in (\ref{pot12}), such that nontrivial topological configurations can be achieved by the field $\Phi$, interacting with the bound states formed by the fermion field $\psi$. It is expected that the spinor bilinears $\bar{\psi}\psi$ and $\bar{\psi}\g_5 \psi$ will develop some localized bound states when the scalar field $\Phi$ becomes a kink type soliton. In the appendix \ref{sec:apprelax} we will consider the static version of the eqs. of motion (\ref{5211})-(\ref{phizetas1}) in order to implement their relevant discretizations .   

We will search for static solutions for $\Phi$ representing a kink type soliton satisfying the parity and boundary conditions (\ref{parity0})-(\ref{infty0}) and solving the system (\ref{5211})-(\ref{phizetas1}) such  that the bound states  are
invariant under the parity symmetry (\ref{par1}) and (\ref{parity}). Under these assumptions we will search for solutions to the system to be eigen-functions of the parity operator ${\cal P}_x$, with $\s=\pm 1$ eigenvalues. 
  
\begin{figure}
\centering
\includegraphics[width=5cm,scale=1, angle=0,height=8.5cm]{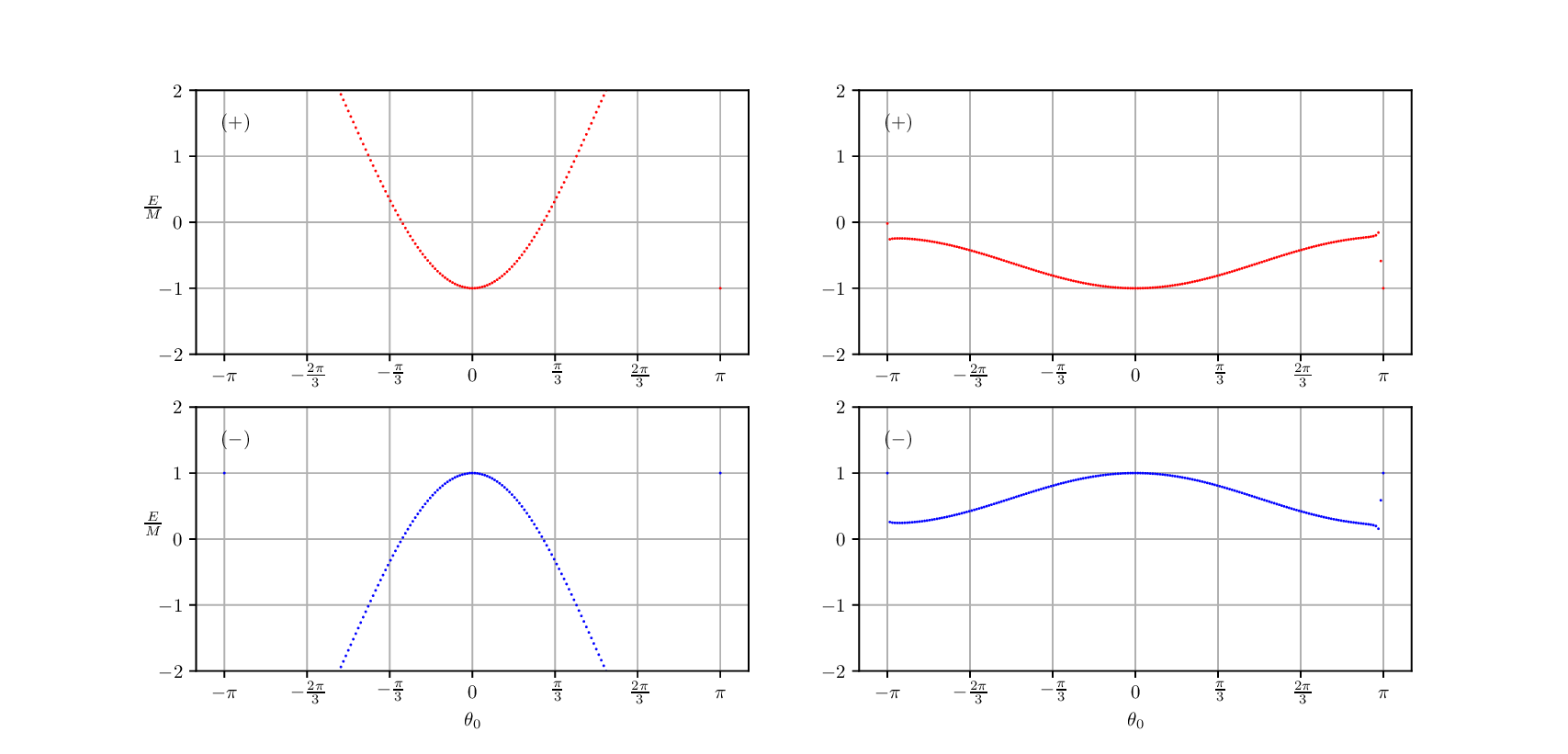}
\parbox{5in}{\caption{ $\frac{E}{M}$ vs $\theta_0$. The parities of the bound states are indicated as $(\pm)$. Notice the BIC states in the left column such that $|\frac{E}{M}| >1$ in some regions of $\theta_0$.}}
\end{figure}
In the appendix \ref{sec:apprelax} we present the relevant developments of the numerical simulation in the context of the relaxation method. At the final step of each iteration, following the algorithm in this method, one gets the relevant eingenvalue $E$ defined as a constant independent of the variable $x$.

In the Figs.15-25 we present our numerical results. In Fig. 15  we perform a comparison of the analytical (\ref{kink1}) and numerically generated kinks $\Phi(x)$ for the parameter values $\theta_0 = \frac{2\pi}{3},\, \b =1,\, \kappa = 1$. In Fig. 16 we present the comparison of the analytical bound state components (\ref{spinors12}) and the numerically simulated ones as a function of $x$ for {\sl positive parity} $\s=+1$. This bound state is hosted by the kink of Fig. 15 and plotted  for $\rho_1 = 0.50, \rho_2 = 2.2,\, \theta_1 = 1.54,\, \theta_2 = 1.55,\,\kappa =1, \theta_0 = 2 \pi/3$. Notice that the spinor components satisfy the parity symmetry  $\xi_{1}(-x)= - \xi_{4}(x)$ and $\xi_{2}(-x)= + \xi_{3}(x)$. In order to better illustrate how small the discrepancy is, we have inserted in the Figs. 15 and 16 some color filled squares on a much greater scale corresponding to the points of our numerical simulations. Notice that there is hardly any discrepancy between the numerical and analytical values. Moreover, we have checked numerically the eigenvalues for these field configurations. So, the analytical BIC eigenvalues $E = +1.01$ and $-1.1$ presented in  the table of the Fig. 6, corresponding to the $ \theta_0 = 2 \pi/3$ and $\s = +1$ values have been checked. Our numerical results reproduce these values within numerical accuracy with $\sim 1 \%$ error in the both calculations.    

In the Figs. 17-21  we plot a set of numerically simulated kinks and their hosted bound states for positive and negative parities. In Fig. 17 we numerically simulate the set of kinks $\Phi(x)$ for the parameters $\theta_0 = \{ 1,\, \frac{2\pi}{3}, \,2.5,\, 3\}$ with blue, red, green and purple lines, respectively. The numerically simulated bound state components  as functions of $x$ for {\sl positive parity} $\s=+1$ (Top Figs.) and  {\sl negative parity} $\s=-1$ (Bottom Figs.) are provided in the Figs. 18-21, respectively. The hosted bound state components are plotted with the same color as the one of  the relevant  kink in the Fig. 17 (i.e. blue, red, green or purple). These plots reproduce qualitatively all our expectations regarding their asymptotic behavior and parity symmetries. Namely,  they vanish at $x = \pm \infty$ and satisfy $\xi_{1}(-x)= - \s \xi_{4}(x)$ and $\xi_{2}(-x)= \s  \xi_{3}(x)$, with $\s = \pm 1$. Notice that the profiles of the spinor components are mainly concentrated on the region where the kink's  slope is appreciable. In fact, the slope of a kink reaches its maximum value at the origin. In addition, the kinks in the Fig. 17 show greater slope at the origin as the relevant parameter $\theta_0$ increases, so they reproduce qualitatively the analytical result (\ref{slope0}). 

So, as emphasized above, the properties of the numerical kinks and their hosted spinor bound state components, shown in the Figs. 17-21 and described above, resemble to the ones of the analytical solutions obtained in sec. 4; and so, allow us to argue, at least qualitatively, that the numerical solutions will satisfy the formula of the Atiyah-Patodi-Singer-type for the charge densities as presented in  (\ref{jj00}). Actually, its exact validity has been examined in (\ref{equivch1})-(\ref{equivch2}) for the analytical solutions, and it deserves further numerical computations for other set of values in the parameter space.       
 
In the Fig. 22 it is shown the numerical plot  $E$ vs $\theta_0$ for the interval $[-\pi, \pi]$. In this case the initial trial kink and bound state functions have been defined for the analytical solutions (\ref{kink1}) and (\ref{spinors12}) for $\kappa = 1, M=1,  \beta=1, \theta_1 = 1.548, \theta_2 = 1.552, A_3 = 10^5, z=4.0, \rho_1 = 0.5, \rho_2 = 2.19$. So, it is numerically simulated the behavior of $E$  as the angle $\theta_0$ varies. For a given $\theta_0$ one gets the two energies $\pm E$ corresponding to the conjugation symmetry and to the positive and negative parities. Notice that the plots in the Figs. 22  approximately approach the values $E = \pm 0.33, \pm 0.83$, within numerical accuracy. In fact, these values correspond to the angle $\theta_0 = \pi/3$ and $\kappa/M=1$ of the analytical results presented in the Figs.4 and 5. In addition, the both Figs.22 show qualitatively that the eigenvalue $E$ as a function of $\theta_0$ is an even function, i.e. $E(-\theta_0) = E(\theta_0)$. Notice that the analytical result  for $E$  (\ref{esp2}) exhibits this symmetry.    

Moreover, the two  plots in the right column of Fig.22  show that the spectra belongs entirely to the in-gap states,  i.e. $|\frac{E}{M}| <1$. Remarkably, the left column  of Fig. 22 exhibits the appearance of the bound states in the continuum (BIC) such that  $|\frac{E}{M}| >1$ for certain continuous regions of the parameter; i.e. $\frac{\pi}{3}<\theta_0 < \frac{2\pi}{3}$ and $-\frac{2\pi}{3} < \theta_0 < -\frac{\pi}{3}$.

Next, in the Fig.23 the initial trial kink and bound state functions in (\ref{kink1}) and (\ref{spinors12}) for $\theta_0 = \frac{\pi}{3}, M=1,  \beta=1, \theta_1 = 1.548, \theta_2 = 1.552, A_3 = 10^5, z=4.0, \rho_1 = 0.5, \rho_2 = 2.19$ are used to simulate the behavior of $E$ under the variation of $\kappa$ in the interval $\frac{\kappa}{M} \in [-1.5\,,\,1.5]$ for the fixed value of $\theta_0 = \frac{\pi}{3}$. Notice that all the states are confined to the in-gap region of the spectra.  The plots  approach the values $E = \pm 0.33$ for $\kappa/M=1$, within numerical accuracy. In fact, these values correspond to the angle $\theta_0 = \pi/3$ and $\kappa/M=1$ of the analytical results presented in the Figs.4 and 5. In addition, the Figs. 23 show qualitatively that the eigenvalue $E$ changes sign as $E \rightarrow -E$ under the transformation:  $\kappa\rightarrow -\kappa$ and $\s \rightarrow -\s$. This symmetry also appears in the analytical results, as stated in the eq. (\ref{Eksign}). 
 
Next, in the Fig. 24 we present $E$ vs $\theta_0$ for the interval  $[-\pi, \pi]$ for another set of initial trial parameters. In fact,  the initial trial kink and bound state functions have been assumed for the analytical solutions (\ref{kink1}) and (\ref{spinors12}) as $\kappa = -1, M=1,  \beta=1, \theta_1 = 1.1, \theta_2 = 0.9, A_3 = 10^5, z=4.0, \rho_1 = 1.1, \rho_2 = 0.9$. Notice that the plots in the right column of the Fig. 24  approximately approach the value $E = \pm 1.1$. In fact, this value corresponds to the angle $\theta_0 = - 2\pi/3$ and $\kappa/M=-1$ of the analytical results presented in the Fig.6.  

Finally, in the Figs. 25 the initial trial kink and bound state functions used  in the Fig.24 are used to simulate the behavior of $E$ under the variation of $\kappa$ in the interval $\frac{\kappa}{M} \in [-1.5\,,\,1.5]$ for the fixed value of $\theta_0 = \frac{2\pi}{3}$. Notice that all the states are confined to the in-gap region of the spectra in this case. The plots approach the values $E = \mp 0.06$ for $\kappa/M=\pm 1$, within numerical accuracy. In fact, these values correspond to the angle $\theta_0 = 2\pi/3$ and $\kappa/M= \mp 1$ of the analytical results presented in the Figs.6 and 7. In addition, the Figs. 25 show qualitatively that the eigenvalue $E$ changes sign as $E \rightarrow -E$ under the transformation:  $\kappa\rightarrow -\kappa$ and $\s \rightarrow -\s$. This symmetry also appears in the analytical results, as stated in the eq. (\ref{Eksign}). 
 
So, the exact values of the set $\{ E , \theta_0, \s, M \neq 0\}$ of our analytical results provided in the tables and figures 4, 5, 6, and 7 for the set  $\theta_0 = \pm \frac{\pi}{3}, \pm \frac{2\pi}{3}$, can be approached numerically provided that a convenient set of initial trial  parameters are assumed for each simulation.

\begin{figure}
\centering
\includegraphics[width=5cm,scale=1, angle=0,height=8.5cm]{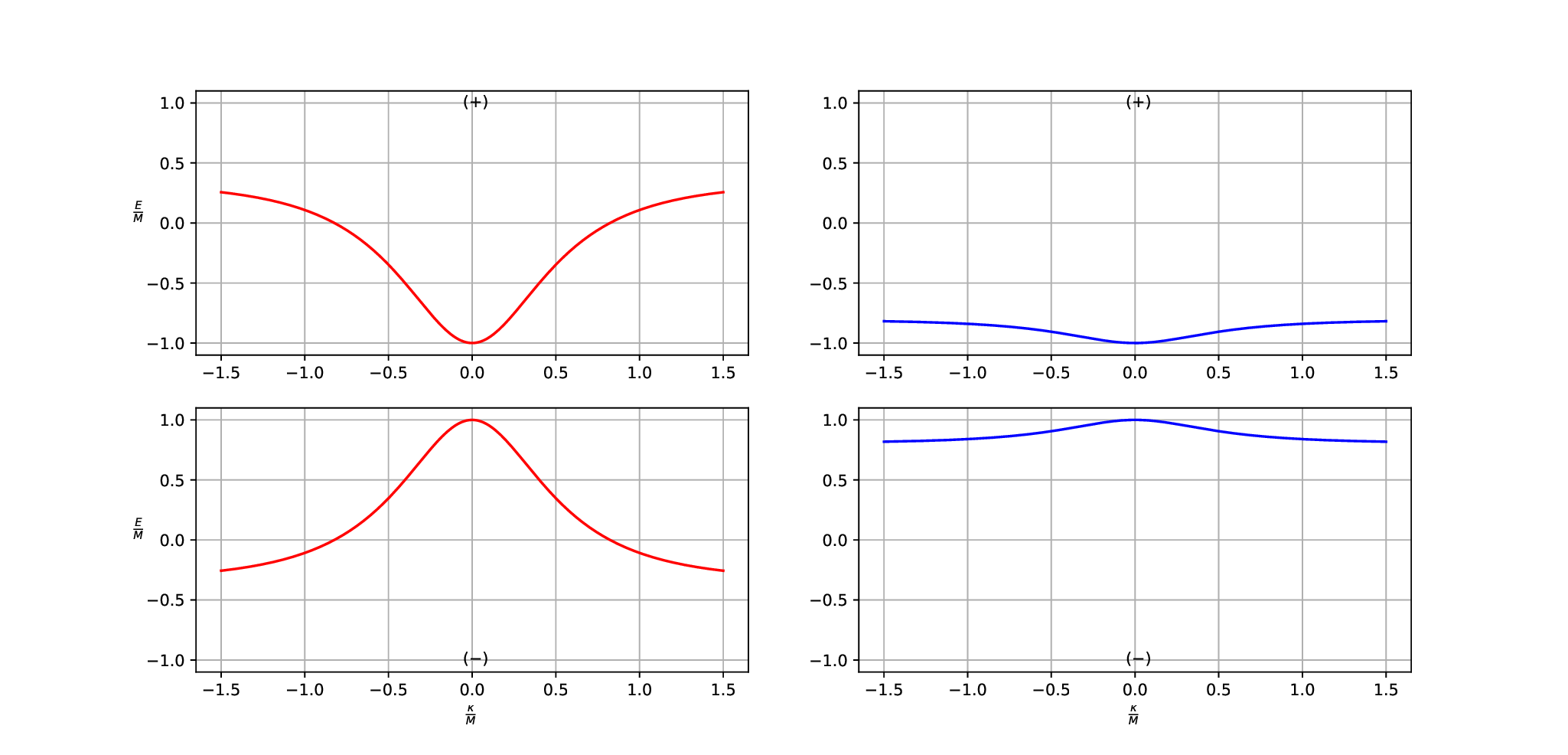}
\parbox{5in}{\caption{ $\frac{E}{M}$ vs $\frac{\kappa}{M}$ for $\theta_0=\pi/3$. The parities of the bound states are indicated as $(\pm)$. These plots show in-gap states ($|\frac{E}{M}| < 1$).}}
\end{figure}

\begin{figure}
\centering
\includegraphics[width=5cm,scale=1, angle=0,height=8.5cm]{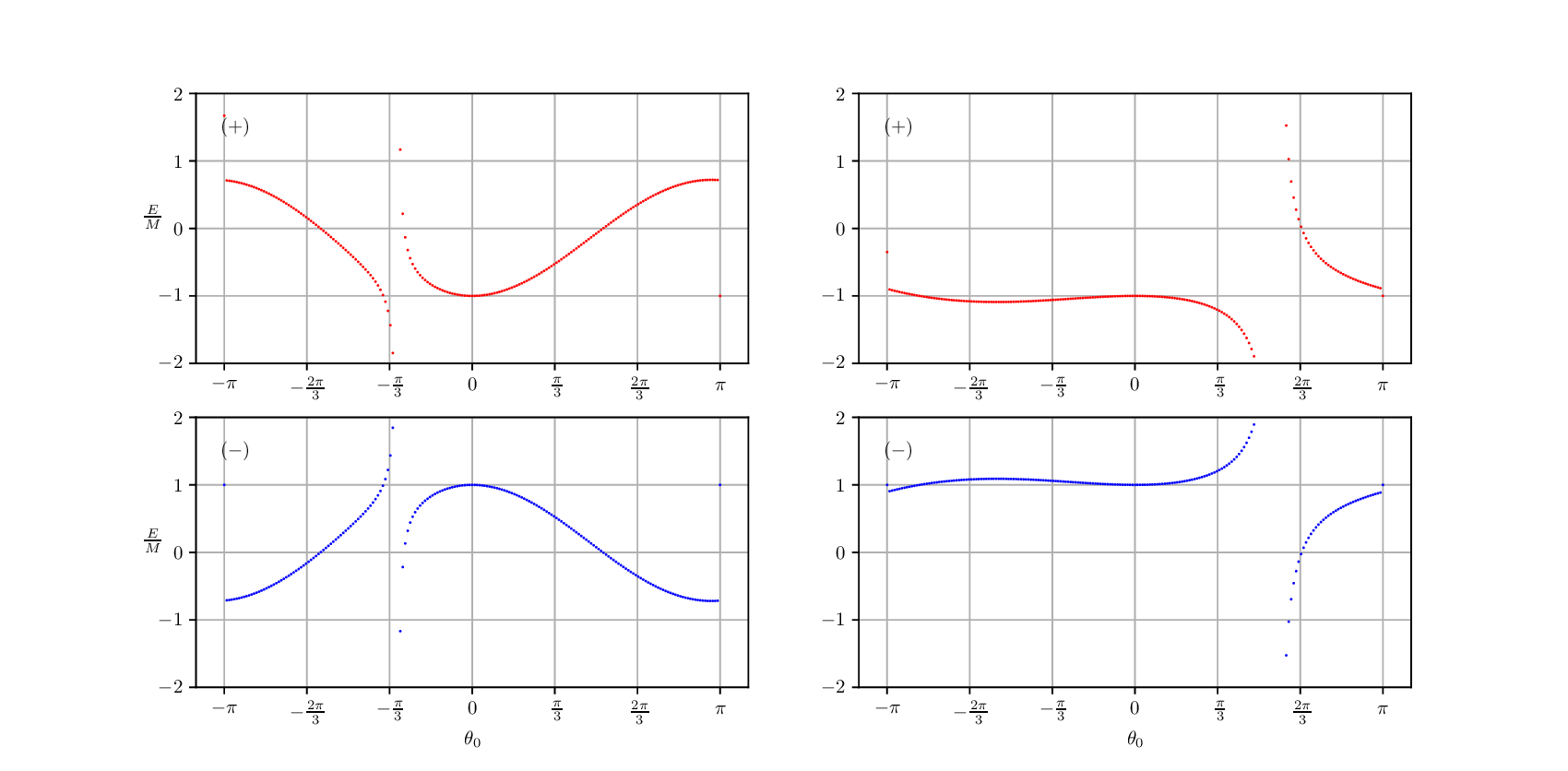}
\parbox{5in}{\caption{ $\frac{E}{M}$ vs $\theta_0$. The parities of the bound states are indicated as $(\pm)$. Notice the BIC states in the both Figs. such that $|\frac{E}{M}| >1$ for certain discrete and continuous regions of $\theta_0$.}}
\end{figure}

\begin{figure}
\centering
\includegraphics[width=5cm,scale=1, angle=0,height=8.5cm]{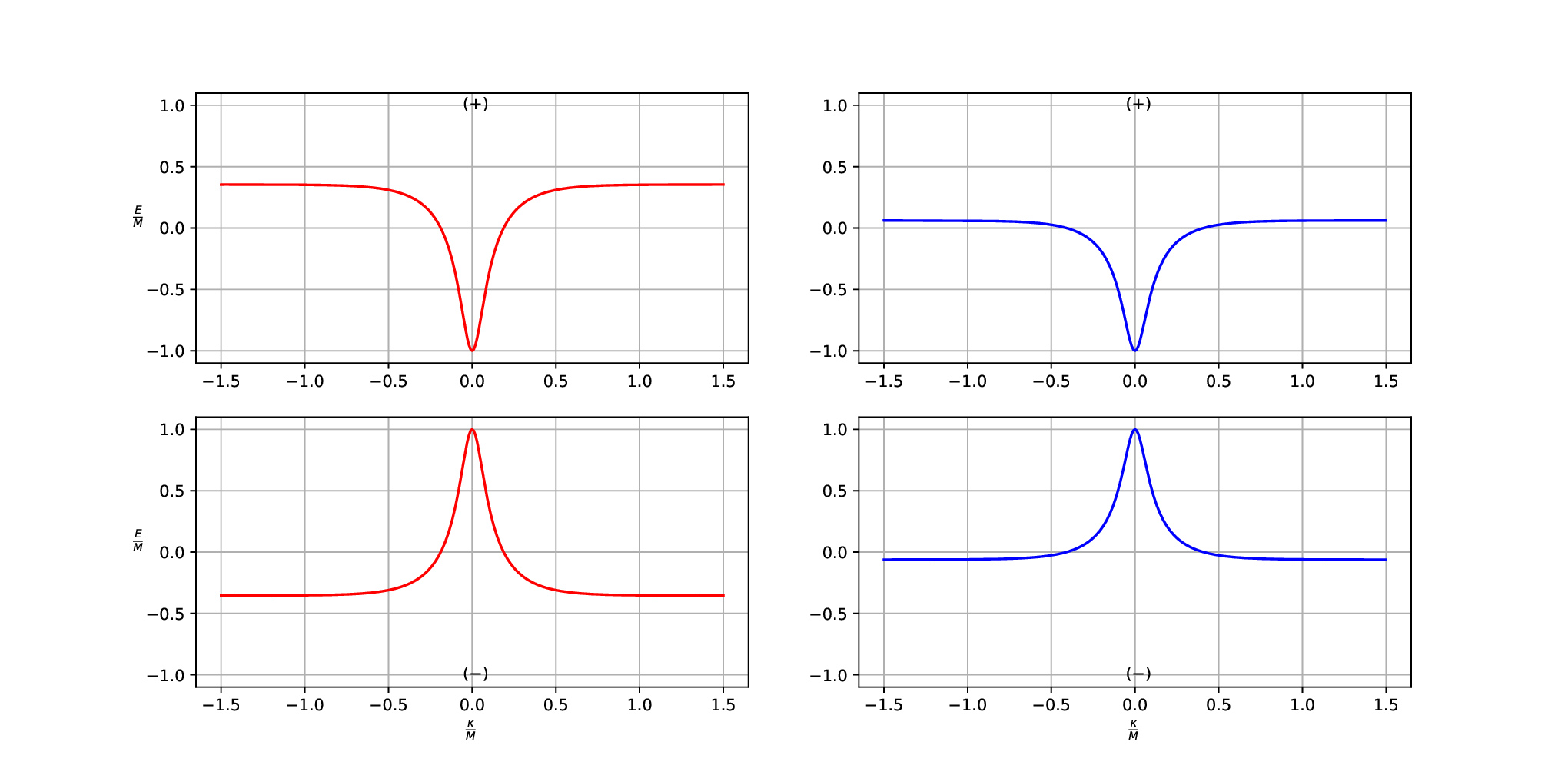}
\parbox{5in}{\caption{ $\frac{E}{M}$ vs $\frac{\kappa}{M}$ for $\theta_0=2\pi/3$. The parities of the bound states are indicated as $(\pm)$. These plots show in-gap states ($|\frac{E}{M}| < 1$).}}
\end{figure}     

Moreover, inspecting the Figs. 22-25 one can see the appearance of certain zero-modes for some values  of the set of parameters $\{\theta_0, \kappa, \s,\, M \neq 0\}$. In fact, the left columns of Figs. 22-23 and the both Figs. 24-25  exhibit the presence of zero-modes. They are solitons with topological charge $\frac{\theta_0}{\pi}$ hosting exotic bound state fermions. It would be interesting to examine numerically their properties as compared  to  the ones of the zero-mode Majorana fermions; e.g.  the degree of divergence from the Majorana component relationships  (\ref{majo1})-(\ref{majo2}) and their relevant polarization $P_D$ as compared to the ones for the Majorana fermion $P_M$ as in (\ref{pold}) and (\ref{polm}), respectively. So, the numerical solutions satisfying those properties deserve a further examination and we postpone it for a future work.      

\section{Discussion}
\label{sec:discussion}

We have considered a modified Toda model coupled to matter providing the analytical and numerical solutions as bound states of the fermion and solitons of the scalar field. A deformation of the ATM model was performed by adding a multi-frequency scalar potential. The solutions related to a special set of co-prime integers $\nu_j=j \,(j=1,2,3)$ for the self-couplings $\frac{\b_j}{\b}=\nu_j$ (\ref{betas123r}) were investigated. Special analytical solutions of the static version of the system were obtained by using the tau function approach. The self-consistent solutions can be classified by the topological charge associated to the asymptotic values of the scalar field $\Phi(\pm \infty) = \pm \theta_0$, and  the spinor bound states are found analytically for each topological configuration of the background scalar field. In these developments the soliton shape depends crucially on the spinor bound state parameters, i.e. it depends on the fermionic state to which it is coupled. It is worth mentioning that the back-reaction of the spinor on the soliton has been described nonperturbatively and exactly through our analytical soliton and bound state solutions; so, this fact may be contrasted to the previous results in the literature in which the spinor back-reaction has been treated in the weak coupling regime or neglected altogether.  
 
The bound state energy eigenvalue $E$, in each topological sector, satisfies a second order algebraic equation  with coefficients depending  on the set of  parameters $\{\kappa, M, \s, \theta_0\}$ as in (\ref{esp2}). We provided analytical solutions for the topological sectors with charges $
Q_{topol} =\pm \frac{1}{3},\, \pm \frac{2}{3}$ and  $\pm 1$ in (\ref{topcharges1}) and (\ref{topo11}), respectively. The charges $Q_{topol} =\pm 1$ are special in the sense that the relevant solitons host the Majorana zero-modes, provided that the parameter relationships in section \ref{sec:majoV0} hold for the cases I and II, respectively . 

The parameter $\kappa$, which defines the soliton width $\sim \frac{1}{\kappa}$, was shown to depend on $\theta_0$ and $\s$ only (\ref{mk2}). The solutions for $\theta_0$ belongs to the discrete set $\{\pm \frac{\pi}{3},\,\pm \frac{2\pi}{3}\}$ in (\ref{thetas0}); so, the values of $\kappa$ and the relevant bound state energies $E$ were provided in the Figs. 4-7. The values of $E$ belong to the in-gap ($|\frac{E}{M}| < 1$) and the continuum ($|\frac{E}{M}| > 1$) bound states. In this analytical approach the so-called  bound states in the continuum  (BIC) emerge for the values $\theta_0=\pm \frac{2\pi}{3},\,\s= \pm 1$,  see Figs. 6-7.

The back-reaction of the spinor field on the soliton has been studied analytically by constructing a mapping of the spinor components  as functionals of the scalar field (\ref{map11})-(\ref{map14}) for the soliton and spinor analytical solutions in (\ref{kink1}) and (\ref{spinors12}). We have computed the spinor-soliton coupling terms of the equations of motion (\ref{id2}); such that one is left with an effective potential associated to the so-called double sine-Gordon model (DSG) (\ref{Veff}) which incorporates the back-reaction of the spinor. So, it has been shown the weak-strong coupling mapping such that the strong coupling sector is described by a DSG model with potential (\ref{Veff}) with  kink solution possessing a finite slope $\mu^A$ (\ref{strong221}) in the strong coupling limit, i.e. as $\b \rightarrow$ large. The degree of this  back-reaction can be measured by computing the ratio in (\ref{ratio1}) between the slope $\mu^{A}$ of the kink coupled to the fermion and the slope $\mu$ of the DSG kink of the strong coupling sector. Notice that this ratio depends on the spinor parameters and $\theta_0$. Their qualitative behavior can be seen in the Fig. 12 for weak and strong coupling constant $\b$.

The weak coupling sector of the soliton-particle duality can be examined in a similar manner by implementing the inverse mapping, such that the expressions $\cos{\b \Phi}$ and $\sin{\b \Phi}$ for the scalar field can be written as functionals of the spinor components from  (\ref{map11})-(\ref{map14}) which hold for the kink(anti-kink) and spinor solutions of the model (\ref{kink1}) and (\ref{spinors12}), respectively. The relevant fermion model and its equations of motion in the weak coupling limit of the modified ATM model (\ref{atm1}) is under current investigation and will be reported elsewhere.    

Similarly, through the tau function formalism we have obtained the sine-Gordon type kink (anti-kink) (\ref{k10i}) and spinor component bound states (\ref{mr0})-(\ref{ml0}) as the relevant zero-modes. So, the SG-type kink hosts the Majorana zero-mode for a convenient set of parameters provided in sec. \ref{sec:majoV0}. In order to compute the back-reaction of the spinor on the kink we followed analogous procedure to the case for $E\neq 0$. So, using the mappings (\ref{ms2})-(\ref{ms22}) and  (\ref{ms1})-(\ref{ms11})  it has been written the interacting terms as in the r.h.s. of  (\ref{compot1}), which is related to the double sine-Gordon type potential  (\ref{potdsg1}). The SG-type kink hosting the Majorana fermion and the DSG kink of the strong coupling sector of the model interpolate the relevant vacua $\pm \pi$ and $\pm \frac{2\pi}{3}$, respectively. However, the ratio of their  relevant slopes at $x=0$ has been computed in (\ref{ratio2}) such that the DSG kink slope is greater than the one of the SG kink by a factor $1.5$. So, one can argue that, in the strong coupling limit of the model, the effect of the back-reaction of the Majorana zero mode on the kink is to catalyze the appearance of the DSG kink with lower topological charge and higher slope at the center.   

The weak coupling sector of the soliton-particle(Majorana) duality  was uncovered by using the inverse mapping, i.e. by written the scalar as functional of the spinor components (\ref{fermioni}). Then, one arrives at the real field Thirring-type equations of motion for the spinor components (\ref{maeq1})-(\ref{maeq2}). So, one can argue that for the fermion zero mode, the weak coupling sector of the ATM model is described by the Majorana fermion system (\ref{maeq1})-(\ref{maeq2}) and the strong coupling sector by the DSG model (\ref{sec201}). The relativistic formulation and the bosonization of the spinor sector deserve a further study.

Moreover, the Majorana and Dirac field polarization quantities $P_M$ and $P_D$ in (\ref{polm1}) and (\ref{pold1}), respectively, have been computed and it has been shown that $|P_D|< |P_M|$ for the relevant analytical solutions.  

We have examined the relationship of the Atiyah-Patodi-Singer-type for the modified ATM model (\ref{currents1}) which incorporates  the effect of the potential $V$ into the dynamics of the system. The contributions of the topological and non-local charges have been computed in  (\ref{jj00})-(\ref{equivch2}) and (\ref{j00})-(\ref{equiv1101}) using the analytical kinks and their relevant non-zero and zero mode bound states, respectively. The outcome in each case was the appearance of a new proportionality constant factor, which replaces the relevant factor of the equivalence between the Noether and topological charges in the undeformed ATM model. In each case, the new constant factor depends on the potential and spinor parameters. Therefore, the equivalence between the Noether and topological charges still holds in the soliton sector of the modified ATM model (\ref{atm1}).  

In sec. \ref{sec:numer1} we have checked our analytical results through numerical simulations in the context of the relaxation method. The trial solutions required in the method have been assumed to be the analytical kink and bound state spinor solutions (\ref{kink1}) and (\ref{spinors12}), respectively,  for some set of parameters. So, the exact values for $\{ \frac{E}{M},  \s,\}$ provided in the figures 4, 5, 6, and 7 for the relevant solutions $\theta_0 = \pm \frac{\pi}{3}, \pm \frac{2\pi}{3}$ can be approached numerically for some trial solutions with a convenient set of initial parameters for each simulation. Although the numerical solutions can be constructed for arbitrary real values of the set $\nu_j \in \IR$ in the potential $V$, the analytical solutions would be formulated for any set of integers $\nu_j$ in the context of the tau-function formalism. 

The zero-modes appearing in our simulations deserve a further analysis regarding the Majorana condition and polarization quantities. In addition, the numerical simulations assuming as trial solution the sine-Gordon type kink (\ref{k10i}) and spinor solutions in (\ref{mr0})-(\ref{ml0}), for some set of parameters, deserve a proper treatment in order to uncover exotic zero-modes which would possess different properties from the ones for the Majorana fermions.    
 
Some analytical and numerical extensions of this work deserve to be considered, such as the kink-antikink scattering in order to understand  the behavior of the fermions and host bound states, as well as the extensions for more than one species of chiral fermions and scalars in the scalar-fermion interaction term. Those issues are currently performed in the context of the deformations of the  $\hat{sl}(n)$ affine Toda model coupled to matter  \cite{matter, jhep11, jhep2, jmp1} and will be reported elsewhere.  

Finally, it would be an interesting issue to study the dynamics of kinks and bound state solutions of the model (\ref{atm1})  in the context of quasi-integrability, which would be relevant to the study of the quasi-conservation laws such as (\ref{quasi1}), and the relevant quasi-conserved quantities in the so-called (quasi-)integrable approach \cite{jhep1, arxiv2, jhep20, jhep4}. So far, to our knowledge, a model possessing a strong-weak duality sectors had not been studied in the quasi-integrability context. Besides, it could be relevant in the context of one-dimensional topological superconductors in order to discuss how the interactions influence the shape and the lifetime of the bound states in view of the recent proposal to identify the Majorana modes by means of local integrals of motion in interacting systems \cite{prl2}. 
  
\vspace{1cm}

\noindent {\bf Acknowledgements}

The authors are thankful to the anonymous referee for relevant comments and suggestions.  
HB thanks IMCA at FC-UNI (Lima-Per\'u) for hospitality during the initial stage of the work. RQB thanks the Peruvian agency Concytec for financial support and IF-UFMT (Cuiab\'a-MT, Brazil) for hospitality during his visit (2018-2019).  The authors thank A. C. R. do Bomfim, H. F. Callisaya, L.F. dos Santos and M.J. B. F. da Silva for discussions.  

\appendix

\section{Spinor equations in terms of the tau functions}
\label{app1}
The first four eqs.  (\ref{5211})-(\ref{5241}) in terms of the tau functions become 
\br
\nonumber
2\tau_1^2 e^{-i \theta_1} (\tau_0 \tau_{\xi ,1}^{'}-\tau_0^{'} \tau_{\xi ,1}) + 2\tau_0^2 e^{i \theta_1} (\tau_1  \widetilde{\tau}_{\xi ,1}^{'}- \tau_1^{'} \widetilde{\tau}_{\xi ,1}) + e^{i \theta_0} M \tau^3_0 (e^{i \theta_3} \widetilde{\tau}_{\xi ,3} -e^{i \theta_4} \widetilde{\tau}_{\xi ,4} ) +\\
e^{-i \theta_0} M \tau^3_1 (e^{-i \theta_3} \tau_{\xi ,3} -e^{-i \theta_4} \tau_{\xi ,4} ) + \tau_0 \tau_1 [\tau_0 e^{i\theta_0}(M e^{-i\theta_3}  \tau_{\xi ,3} + M e^{-i\theta_4}  \tau_{\xi ,4} + 2 i E  e^{i(\theta_2-\theta_0)} \widetilde{\tau}_{\xi ,2} )+ \label{exi1}\\
\tau_1 e^{-i\theta_0}(M e^{i\theta_3}  \widetilde{\tau}_{\xi ,3} + M e^{i\theta_4}  \widetilde{\tau}_{\xi ,4} - 2 i E  e^{-i(\theta_2-\theta_0)} \tau_{\xi ,2} )]=0,
\nonumber
\er
\br
\nonumber
2i \tau_1^2 e^{-i \theta_2} (\tau'_0 \tau_{\xi ,2}-\tau_0 \tau'_{\xi ,2}) - 2i\tau_0^2 e^{i \theta_2} (\tau'_1  \widetilde{\tau}_{\xi ,2}- \tau_1 \widetilde{\tau}'_{\xi ,2}) + i e^{i \theta_0} M \tau^3_0 (e^{i \theta_3} \widetilde{\tau}_{\xi ,3} -e^{i \theta_4} \widetilde{\tau}_{\xi ,4} ) -\\
i e^{-i \theta_0} M \tau^3_1 (e^{-i \theta_3} \tau_{\xi ,3} -e^{-i \theta_4} \tau_{\xi ,4} ) + \tau_0 \tau_1 [\tau_0 e^{-i\theta_0} (e^{-i\theta_3}M   \tau_{\xi ,3} + e^{-i\theta_4} M  \tau_{\xi ,4} + 2 i E  e^{i(\theta_1-\theta_0)} \widetilde{\tau}_{\xi ,1} )+\label{exi2}\\
\tau_1 e^{-i\theta_0}(Me^{i\theta_3}  \widetilde{\tau}_{\xi ,3} +M e^{i\theta_4}  \widetilde{\tau}_{\xi ,4} - 2 i E  e^{-i(\theta_1-\theta_0)} \tau_{\xi ,1} )]=0,
\nonumber
\er
\br
\nonumber
2 \tau_1^2 e^{-i \theta_3} (\tau_0 \tau'_{\xi ,3}-\tau'_0 \tau_{\xi ,3}) - 2\tau_0^2 e^{i \theta_3} (\tau'_1  \widetilde{\tau}_{\xi ,3}- \tau_1 \widetilde{\tau}'_{\xi ,3}) +  e^{-i \theta_0} M \tau^3_1 (e^{-i \theta_1} \tau_{\xi ,1} -e^{-i \theta_2} \tau_{\xi ,2} ) +\\
 e^{i \theta_0} M \tau^3_0 (e^{i \theta_1} \widetilde{\tau}_{\xi ,1}-e^{i \theta_2} \widetilde{\tau}_{\xi ,2}) + \tau_0 \tau_1 [\tau_0 e^{i\theta_0} (e^{-i\theta_1}M   \tau_{\xi ,1} + e^{-i\theta_2} M  \tau_{\xi ,2} + 2 i E  e^{i(\theta_4-\theta_0)} \widetilde{\tau}_{\xi ,4} )+\label{exi3}\\
\tau_1 e^{-i\theta_0}(Me^{i\theta_1}  \widetilde{\tau}_{\xi ,1} +M e^{i\theta_2}  \widetilde{\tau}_{\xi ,2} - 2 i E  e^{-i(\theta_4-\theta_0)} \tau_{\xi ,4} )]=0,
\nonumber
\er
and
\br
\nonumber
2i \tau_1^2 e^{-i \theta_4} (\tau_0 \tau'_{\xi ,4}-\tau'_0 \tau_{\xi ,4}) - 2i\tau_0^2 e^{i \theta_4} (\tau_1  \widetilde{\tau}'_{\xi ,4}- \tau'_1 \widetilde{\tau}_{\xi ,4}) + i  M e^{-i \theta_0} \tau^3_1 (e^{-i \theta_1} \tau_{\xi ,1} -e^{-i \theta_2} \tau_{\xi ,2} ) -\\
 i M e^{i \theta_0} \tau^3_0 (e^{i \theta_1} \widetilde{\tau}_{\xi ,1} - e^{i \theta_2} \widetilde{\tau}_{\xi ,2}) + \tau_0 \tau_1 [-i \tau_0 e^{i\theta_0} (M e^{-i\theta_1}  \tau_{\xi ,1} + M e^{-i\theta_2}  \tau_{\xi ,2} +2 i E  e^{i(\theta_3-\theta_0)} \widetilde{\tau}_{\xi ,3} )+\label{exi4}\\
i \tau_1 e^{-i\theta_0}(Me^{i\theta_1}  \widetilde{\tau}_{\xi ,1} +M e^{i\theta_2}  \widetilde{\tau}_{\xi ,2} - 2 i E  e^{-i(\theta_3-\theta_0)} \tau_{\xi ,3} )]=0.
\nonumber
\er

\section{Scalar field equation in terms of the tau functions}
\label{app2}

The eq. (\ref{phizetas1}) written in terms of the tau functions becomes 
\br
\nonumber
\frac{2i}{\b} \tau_0 \tau_1 [\tau^2_1 ((\tau'_0)^2 - \tau_0 \tau''_0) - \tau^2_0 ((\tau'_1)^2 - \tau_1 \tau''_1)]+ 3i A_3 \b [e^{-3i\theta_0}  \tau^6_1 - e^{3i\theta_0}  \tau^6_0 ] +\\
\nonumber
 2i\tau_0 \tau_1 \b A_2 [e^{-2i\theta_0}  \tau^4_1 -  e^{2i\theta_0}  \tau^4_0]+ i\tau^2_0 \tau^2_1 \b A_1 [e^{-i\theta_0}  \tau^2_1 -  e^{i\theta_0}  \tau^2_0]+ \\
\nonumber
2 M \b \tau^4_1 e^{i\theta_0} (e^{-i\theta_1}  \tau_{\xi ,1} - e^{-i\theta_2}  \tau_{\xi ,2} )(e^{-i\theta_3}  \tau_{\xi ,3} - e^{-i\theta_4}  \tau_{\xi ,4})+\\
\nonumber
2 M \b \tau^4_0 e^{-i\theta_0} (e^{i\theta_1}  \widetilde{\tau}_{\xi ,1} - e^{i\theta_2}  \widetilde{\tau}_{\xi ,2} )(e^{i\theta_3}  \widetilde{\tau}_{\xi ,3} - e^{i\theta_4}  \widetilde{\tau}_{\xi ,4})+ \\
\label{secorder}
2 M \b  \tau^2_0 \tau^2_1 \Big[(e^{i(\theta_0-\theta_1-\theta_3)} \tau_{\xi ,1} \tau_{\xi ,3}+ c.c) + (e^{i(\theta_0-\theta_2-\theta_3)} \tau_{\xi ,2} \tau_{\xi ,3}+ c.c) +\\
\nonumber
 (e^{i(\theta_0-\theta_1-\theta_4)} \tau_{\xi ,1} \tau_{\xi ,4}+ c.c) + (e^{i(\theta_0-\theta_2-\theta_4)} \tau_{\xi ,2} \tau_{\xi ,4}+ c.c) \Big]+\\
\nonumber
2 M \b  \tau_0 \tau_1 e^{i\theta_0} \Big[ ( e^{i(\theta_1- \theta_3)}  \tau^2_0 \widetilde{\tau}_{\xi ,1} \tau_{\xi ,3} + c.c. )+( e^{i(\theta_1- \theta_4)}  \tau^2_0 \widetilde{\tau}_{\xi ,1} \tau_{\xi ,4} + c.c. )+( e^{i(\theta_2 -\theta_3)}  \tau^2_0 \widetilde{\tau}_{\xi ,2} \tau_{\xi ,3} + c.c. )+\\
\nonumber
( e^{i(\theta_2- \theta_4)}  \tau^2_0 \widetilde{\tau}_{\xi ,2} \tau_{\xi ,4} + c.c. )+( e^{-i(\theta_1- \theta_3)}  \tau^2_0 \tau_{\xi ,1} \widetilde{\tau}_{\xi ,3} + c.c. )+( e^{-i(\theta_2- \theta_3)}  \tau^2_0 \tau_{\xi ,2} \widetilde{\tau}_{\xi ,3} + c.c. )+\\
( e^{-i(\theta_1- \theta_4)}  \tau^2_0 \tau_{\xi ,1} \widetilde{\tau}_{\xi ,3} + c.c. )+ ( e^{-i(\theta_2- \theta_4)}  \tau^2_0 \tau_{\xi ,2} \widetilde{\tau}_{\xi ,4} + c.c. )
\Big] =0,\nonumber
\er
where $c.c.$ inside the parentheses $(...)$ in the eqs. above stands for complex conjugation of the preceding term.

\section{Consistency conditions and tau functions}
\label{app3}

The  eqs. (\ref{consist1}) (for $c_1 =0$)  and (\ref{consist2}) in terms of the tau functions become, respectively
\br
\nonumber
 \{\tau^2_1 [ e^{-2i(\theta_1)} \tau_{\xi ,1}^2-  e^{-2i(\theta_2)} \tau_{\xi ,2}^2- e^{-2i(\theta_3)} \tau_{\xi ,3}^2 + e^{-2i(\theta_4)} \tau_{\xi ,4}^2 ] + c.c. \}+\\
2 \tau_1 \tau_2 [\tau_{\xi ,1} \widetilde{\tau}_{\xi ,1}+ \tau_{\xi ,2} \widetilde{\tau}_{\xi ,2}-\tau_{\xi ,3} \widetilde{\tau}_{\xi ,3}-\tau_{\xi ,4} \widetilde{\tau}_{\xi ,4}] =0,\label{consist11}
\er
and
\br
\nonumber
\Big\{48 M e^{i \theta_0}  \tau^4_0 \Big[ e^{i \theta_1}  \widetilde{\tau}_{\xi ,1} - e^{i \theta_2}  \widetilde{\tau}_{\xi ,2} \Big]\Big[  e^{i \theta_3}  \widetilde{\tau}_{\xi ,3} - e^{i \theta_4}  \widetilde{\tau}_{\xi ,4} \Big] + c.c.\Big\} &+&\\
\nonumber
8 M \tau^2_0 \tau^2_1 \Big\{ \Big[ e^{i \theta_0 -i \theta_1-i \theta_3} \tau_{\xi ,1}  \tau_{\xi ,3} + c.c. \Big] +  \Big[ e^{i \theta_0 -i \theta_1-i \theta_4} \tau_{\xi ,1}  \tau_{\xi ,4} + c.c.\Big]  &+&\\ 
\nonumber   \Big[ e^{i \theta_0 -i \theta_2-i \theta_3} \tau_{\xi ,2}  \tau_{\xi ,3} + c.c. \Big] +  \Big[ e^{i \theta_0 -i \theta_2-i \theta_4} \tau_{\xi ,2}  \tau_{\xi ,4} + c.c. \Big]  \Big\} -\\
\label{consist22} 8 M \tau_0 \tau_1 \Big\{ \tau_{\xi ,1}   \widetilde{\tau}_{\xi ,1}   + \tau_{\xi ,2}   \widetilde{\tau}_{\xi ,2}  +\tau_{\xi ,3}   \widetilde{\tau}_{\xi ,3}    +  \tau_{\xi ,4}   \widetilde{\tau}_{\xi ,4}  \Big\}  \frac{d}{dx} [ \tau_0 \tau_1]+\\
8 M \tau_0^2 \tau_1^2    \frac{d}{dx} \Big\{ \tau_{\xi ,1}   \widetilde{\tau}_{\xi ,1}+ \tau_{\xi ,2}   \widetilde{\tau}_{\xi ,2}+  \tau_{\xi ,3}   \widetilde{\tau}_{\xi ,3}+  \tau_{\xi ,4}   \widetilde{\tau}_{\xi ,4} \Big\}+\nonumber \\
12 M \tau_0 \tau_1 \Big[e^{i \theta_0} \tau_0^2 +e^{-i \theta_0} \tau_1^2 \Big] \Big\{ \Big[ e^{-i (\theta_1- \theta_3)}  \tau_{\xi ,1}   \widetilde{\tau}_{\xi ,3} + c.c.\Big ]  +  \Big[ e^{-i (\theta_1- \theta_3)}  \tau_{\xi ,2}   \widetilde{\tau}_{\xi ,4} + c.c \Big] \Big\} + 
\nonumber\\
12 M \tau_0 \tau_1 \Big[e^{i \theta_0} \tau_0^2 - e^{-i \theta_0} \tau_1^2 \Big] \Big\{
  \Big[ e^{-i (\theta_1- \theta_4)}  \tau_{\xi ,1}   \widetilde{\tau}_{\xi ,4} + c.c.\Big ] + \Big[ e^{-i (\theta_2- \theta_3)}  \tau_{\xi ,2}   \widetilde{\tau}_{\xi ,3} + c.c.\Big ] \Big\} = 0,
\nonumber
\er
where $c.c.$ in the eqs. above stands for complex conjugation of all the preceding terms inside each bracket  of type $\{...\}$ or $[...]$.

\section{Coefficients of  the polynomial in (\ref{consist2i})}
\label{app:b0b1}

Coefficients in the polynomial $b_0 + b_1 \cosh{(2 \kappa x)} $ in the r.h.s. of  the identity (\ref{consist2i})
\br
\nonumber
b_0 \equiv \frac{1}{2 \kappa} \rho_1^2 \Big\{8 \kappa \cos{(\theta_1-\theta_0)} \cos{\theta_1}  - \s M \Big[3+ 4 \cos{(2 \theta_1- \theta_0)} \cos{\theta_0} +  \cos{(2 \theta_0)} \sin{\theta_0}\Big]\Big\} +\\
\nonumber
\frac{1}{2 \kappa} \rho_2^2 \Big\{8 \kappa \sin{(\theta_2-\theta_0)} \sin{\theta_2}  - \s M \Big[3 -  4 \cos{(2 \theta_2- \theta_0)} \cos{\theta_0} +  \cos{(2 \theta_0)} \sin{\theta_0}\Big]\Big\}+\\
\frac{1}{\kappa} M \s \rho_2  \rho_1  \cos{(\theta_1-\theta_2)} [3 + \cos{(2\theta_0)}] \sin{\theta_0} .
\label{h00}
\er
and
\br
\nonumber
b_1 \equiv \frac{1}{\kappa} \rho_1^2 \Big[\kappa + \kappa \cos{(2\theta_1-\theta_0)} \cos{\theta_0}  - 2 \s M \cos{(\theta_1- \theta_0)} \cos{\theta_1} \sin{\theta_0}\Big]+\\
\nonumber
\frac{1}{\kappa} \rho_2^2 \Big[\kappa - \kappa \cos{(2\theta_2-\theta_0)} \cos{\theta_0}  - 2 \s M \sin{(\theta_2- \theta_0)} \sin{\theta_2} \sin{\theta_0}\Big]+\\
\frac{1}{\kappa} M \s \rho_2  \rho_1  \cos{(\theta_1-\theta_2)} \sin{(2\theta_0)}.
\label{h11}
\er

\section{Coefficients of  the polynomial in  (\ref{c01})}
\label{app:curr22}

The coefficients in the polynomial (\ref{c11}) become
\br
c_0 = \Big[\frac{a_2 \cos{(2\theta_0)} +2 (a_2+ 8 b_0 \kappa - a_0)}{4 \kappa} + \frac{2 \kappa \sin{(2 \theta_0)}}{\b^2} \Big]\label{c00} 
\er
\br
\nonumber
= \frac{1}{8 \kappa} \rho_1^2 \Big\{ 32 \kappa \cos{(\theta_1-\theta_0)} \cos{\theta_1}  - \s M \Big[  \frac{14 + \cos{(4 \theta_0)} (2+\log{4}) +  \log{64} +  \cos{(2 \theta_0)}  (11+\log{1024})}{\sin{\theta_0}} + \nonumber\\
8  \cos{(2 \theta_1-\theta_0)}  \sin{(2 \theta_0)} \Big] \Big\} + \nonumber\\
\nonumber
\frac{1}{8 \kappa} \rho_2^2 \Big\{16 \kappa \cos{\theta_0} - \s M \Big[ \frac{14 + \cos{(4 \theta_0)} (2+\log{4}) +  \log{64} +  \cos{(2 \theta_0)}  (11+\log{1024})}{\sin{\theta_0}}\Big] - \\
\nonumber
8  \cos{(2 \theta_2-\theta_0)} [ 2\kappa - M \s \sin{(2 \theta_0)} ] \Big\}  +\\
\nonumber
\frac{M \s}{4 \kappa \sin{\theta_0}} \rho_1 \rho_2  \cos{(\theta_1-\theta_2)} \Big[14 + \cos{(4 \theta_0)} (2+\log{4}) +  \log{64} +  \cos{(2 \theta_0)}  (11+\log{1024})\Big].
\er
and
\br
c_1  = \Big[\frac{16 \kappa^2  \sin{\theta_0} + \b^2 (4 a_2 \cos{\theta_0} + 16 b_1 \kappa - a_1)}{4 \kappa \b^2 } \Big]   \label{c11}
\er
\br
=\nonumber
-2 \b^2 \rho_1^2  \Big\{  3 M \s  \cot{\theta_0} (3 + \log{4}) - 4 \cos{(2 \theta_1-\theta_0)}  (\kappa  \cos{\theta_0} - M \s \sin{\theta_0})- 4 [ \kappa + M \s (1+ \log{2}) \sin{(2\theta_0)} ]\Big\}\\ 
\nonumber
-2 \b^2 \rho_2^2  \Big\{  3 M \s  \cot{\theta_0} (3 + \log{4}) - 4 \cos{(2 \theta_2-\theta_0)}  (\kappa  \cos{\theta_0} - M \s \sin{\theta_0})- 4 [\kappa + M \s (1+ \log{2}) \sin{(2\theta_0)} ]\Big\}+\\
4 M \b^2  \rho_1\rho_2 \, \s \cot{\theta_0}  \cos{( \theta_1-\theta_2)} \Big[5 + \log{4} +  \cos{(2 \theta_0)} (4 + \log{16} )\Big].
\nonumber  
\er

\section{Relaxation method}

\label{sec:apprelax}

We perform the numerical simulation of the system using the relaxation method. In this context  we define seven coupled first-order ODEs comprising: The four first-order equations  (\ref{5211})-(\ref{5241}),  the one second-order equation (\ref{phizetas1}) rewritten as two first order equations and in addition, an equation for the eigenvalue in the form of a constancy condition, i.e. $E$ to be a constant independent of the variable $x$.

It is convenient to redefine the new quantities as
\br
\label{rees1}
\hat{\Phi} = \b \Phi,\,\,\hat{x} = M x,\,\, \hat{E} = E/M, \,\, \l_1 = \frac{2 \b^2}{M},\,\, \l_2 = \frac{\b^2}{M^2},\,
\hat{\b}_i = \frac{\b_i}{\b}, \, i =1,2,3.
\er
We will find the solutions which are eigenstates
of the parity operator ${\cal P}_x$; so,  it is enough to solve the system of equations in the interval $x \geq 0$. For simplicity, we map the interval  $x \in [0, \infty>$ to $ X \in [0, 1]$ through  $X = tanh(\hat{x})$. So, the static version of the system of eqs.  (\ref{5211})-(\ref{phizetas1}) written in the new coordinate and taking into account the rescaling (\ref{rees1}) become
\br
\label{5211X}
(1-X^2) \frac{d\xi_1}{dX} + \hat{E}\, \xi_2 -  \xi_4 \sin{\hat{\Phi}}+   \xi_3 \cos{\hat{\Phi}}&=&0, \\ \label{5221X}
(1-X^2) \frac{d\xi_2}{dX}  - \hat{E}\, \xi_1+  \xi_3 \sin{ \hat{\Phi}}+  \xi_4\cos{ \hat{\Phi}}&=&0, \\ \label{5231X}
(1-X^2) \frac{d\xi_3}{dX}  - \hat{E}\, \xi_4 + \xi_2 \sin{\hat{\Phi}}+  \xi_1\cos{ \hat{\Phi}}&=&0, \\ \label{5241X}
(1-X^2) \frac{d\xi_4}{dX}  + \hat{E}\,  \xi_3 -  \xi_1\sin{\hat{\Phi}}+  \xi_2\cos{\hat{\Phi}}&=&0, \\
-(1-X^2)^2  \frac{d Z }{dX}  + 2 X (1-X^2)  Z + \l_1  \Big[ (\xi_1\xi_3+\xi_2\xi_4)\cos{\hat{\Phi}}- (\xi_1\xi_4-\xi_2\xi_3)\sin{\hat{\Phi}}\Big] +  \frac{d \hat{V}(\hat{\Phi})}{d\hat{\Phi}} &=&0,\label{phizetas1X} \\
\frac{d\hat{\Phi} }{dX} - Z &=& 0,
\label{Z11}
\er
where 
\br
\hat{V}(\hat{\Phi}) = \hat{A}_1 \cos{(\hat{\beta}_1 \hat{\Phi})} + \hat{A}_2 \cos{(\hat{\beta}_2 \hat{\Phi}))} + \hat{A}_3 \cos{(\hat{\beta}_3 \hat{\Phi})} ,\,\, \,\,\,\,\hat{A}_i = \l_2 A_i,\,\, i=1,2,3.
\er
As mentioned above, it will be imposed the next equation
\br
\label{ec1}
\frac{d}{dX} \hat{E}(X)=0.
\er
 
Therefore, in the framework of the relaxation method we must impose seven conditions at the two boundaries $X = 0$ and $X = 1$. At $X = 0$ we impose three conditions: one
parity condition for the spinor (e.g. $\xi_1(0) = - \s \xi_4(0)$),  $\hat{\Phi}(0) = 0$ and another one setting  a value for one of the $\xi_i (0)'s$, say $\xi_1(0) = \xi_0$. This value can be adjusted in order to have a normalization of $\psi$ as $\int_{-\infty}^{+\infty} \psi^{\dagger} \psi =1$. At $X = 1$ we choose the remaining four conditions: three of the $\xi_i (1)'s$
are set to zero (i.e. $\xi_i(1) =0\,\, \mbox{for} \,\, i=2,3,4$) and $\Phi(1) = \phi_0$. Notice that these last conditions specify the asymptotic values of the corresponding fields at $x= +\infty$. 

Let us rewrite one representative equation of the system (\ref{5211X})-(\ref{5241X}), e.g. the first equation as 
\br
\label{eq11r}
\frac{d}{dx} y(x)+ \hat{E}(x)\,  z(x)  - h(x) =0,
\er
where the functions $y(x) \equiv \xi_1(x)$, $h(x) \equiv  (\xi_4 \sin{\hat{\Phi}} -   \xi_3 \cos{\hat{\Phi}})/(1-x^2)$   and  $z\equiv  \xi_2(x)/(1-x^2)$ have been introduced. The variable $X$ was relabeled  as $x$.
  
Next, we solve numerically the above differential eq. (\ref{eq11r}) in the context of the relaxation method. In particular, we need to solve the eq. (\ref{eq11r}) for the b.c. described above. In that process we will be able to find numerically the eingenvalues $\hat{E}$.  These values will depend on the parameters  $\{ \phi_0, \s, M\}$, as it is clear from the exact result  for $E$  in (\ref{esp2}). The parameter $\phi_0$ is used to define the asymptotic values of the kink $\Phi(\pm \infty) = \pm \phi_0$. 

The relaxation method is known to be an useful method in order to tackle numerically this type of b.c. In fact, as we will show below, the relaxation method assumes a suitable initial trial function (a guess function), and then obtains successively a set of approximate solutions until some number of iterations, say $M_k$, the numerical solution converges to the true solution.  We will check the accuracy of our numerical solutions by comparing them to the exact analytical results obtained  in sec. \ref{sec:tau} for the particular values of $\phi_0 \equiv \theta_0 = \{ \frac{\pi}{3}, \frac{2\pi}{3}\}$.    
  
So, let us  define the following  discretizations
\br
x_n &=& x_i + (n-2) dx,\,\,\,\,\,\,dx = \frac{1}{M},\,\,\, n =1,2,3,....M,   \\
y_n &=&  y(x_n),\,\,\,\,\,\,z_n =  z(x_n),\,\,\,\,\,\,h_n =  h(x_n),\\
 \hat{E}_n &=&  \hat{E}(x_n),
\er
where $M$ ($M >> 2$) is an integer and $dx$ defines the grade spacing $dx = x_{n+1}-x_n$.  We use the  symmetric difference quotient in order to write the next first derivative approximation  
\br
\label{diff}
y'  \approx   \frac{y_{n+1}-y_{n-1}}{2 \,dx}.
\er
Next, let us approximate  the  differential eqs.  (\ref{ec1}) and (\ref{eq11r}) through the finite difference method. So, making use of the approximation  (\ref{diff}) one has 
\br
\frac{ \hat{E}_{n+1}- \hat{E}_{n-1}}{2 \,dx} &=& 0,\label{lapr}\\
\label{eq1apr}
  \frac{y_{n+1}-y_{n-1}}{2 \,dx} + \hat{E}_n \, z_n -  h_n &=& 0.
\er
From (\ref{eq1apr}) one can get
\br
\label{ln}
 \hat{E}_n =    \frac{2 dx\,  h_{n}+y_{n-1} - y_{n+1}}{2 dx \, z_n }.
\er
Therefore, substituting  (\ref{ln}) into (\ref{lapr}), written in the form $\hat{E}_{n+1} = \hat{E}_{n-1}$, one can write  
\br
\label{iter}
y_n = \frac{z_{n+1} (2 dx\, h_{n-1}+ y_{n-2} )  - z_{n-1} (2 dx\, h_{n+1} -  y_{n+2} )  }{ z_{n+1} + z_{n-1}}
\er
Some comments are in order here.  

First,  the b.c. to be imposed on the system of eqs. (\ref{lapr})-(\ref{eq1apr})  are
$y_0 = \xi_1(0),\,\,y_M =z_0 = z_M =h_0=h_M=0$ and $\hat{E}_0 =  \hat{E}_M= const.$\, $M$ being a large integer of the order $\sim 10^4$. 

Second, the eq. (\ref{iter}) will be used below in order to perform the iterative construction of the solutions for the arrays  $y_n$, as is usual in the relaxation method.  Notice that (\ref{iter}) incorporates implicitly the constancy condition $\hat{E}=const.$ 

Third, one starts by assuming an initial guess function for $y_n \equiv  y_n^{(0)}, z_n \equiv  z_n^{(0)}, h_n \equiv  h_n^{(0)}$ and assigning  those values on the right hand side of (\ref{iter}) to the variables $y_n, z_n, h_n$ in order to construct the first iterated array $y_n \equiv  y_n^{(1)}$ on the left hand side. An initial trial solution is taken from the set of analytic kink and spinor bound state solutions (\ref{spinors12}) and (\ref{kink1}) provided that a suitable set of the parameter values are assumed. The initial set of parameters are chosen such that the normalization condition (\ref{norm0}) is satisfied.

Fourth, an analogous equation to (\ref{iter}) can be implemented for each of the first order equations of the system (\ref{5211X})-(\ref{5241X}).  In addition, for each of the first order equations (\ref{phizetas1X})-(\ref{Z11}) one can construct directly an iteration equation, since they do not depend explicitly on  the eigenvalue $\hat{E}$. So, the outcome for each iteration in the system (\ref{5211X})-(\ref{Z11}) will be a  set of $k^{th}-$interated vectors $\xi_{a,\,
n}^{(k)}\, (a=1,2,3,4),\, \Phi_n^{(k)}$ and $Z_n^{(k)}$. The normalization condition (\ref{norm0}) is checked in each iteration of the numerical simulation.

Fifth, the values of the function $E(x)$ must be computed for each order of the iteration $\hat{E} ^{(k)}$. So, the process must be iterated maintaining the constraint (\ref{norm0}) until one reaches a constant $\hat{E}_n = E/M$.  

Sixth, the process above must be repeated $M_k$ times until the set of solutions $\{\xi_{a,\,
n}^{(M_k)}\, (a=1,...,4),\, \Phi_n^{(M_k)},$ $\,Z_n^{(M_k)}\}$ converge and the  values $\hat{E} ^{(k)}$ tend  to a constant $\hat{E} ^{(M_k)}$, within numerical accuracy.

\end{document}